\documentclass[journal]{IEEEtran}
\usepackage{amsmath,amsfonts}
\usepackage{multirow}
\usepackage{threeparttable, adjustbox}
\usepackage{amsthm}
\usepackage{ifthen}
\usepackage{pgfplots}
\usepackage{pgfplotstable}
\usepackage[style=ieee,url=false,eprint=false]{biblatex}
\usepgfplotslibrary{groupplots}
\pgfplotsset{compat=newest}
\usepgfplotslibrary{fillbetween}
\usepackage{adjustbox}
\usepackage{bm}
\usepackage{caption}
\usepackage{csquotes}
\usepackage{subcaption}
\usepackage{mathtools}
\usepackage{microtype}
\usepackage{graphicx}
\usepackage{tikz}
\usetikzlibrary{arrows.meta, positioning, positioning, matrix, backgrounds, fit, external, plotmarks}
\usepackage{algorithm}
\usepackage[noEnd,indLines,commentColor=gray]{algpseudocodex}
\AddToHook{env/algorithmic/begin}{\tikzexternaldisable}
\usepackage{booktabs}
\usepackage{bbm}
\usepackage{xcolor}
\definecolor{ieeeblue}{RGB}{0,0,245}
\usepackage[colorlinks=true,citecolor=ieeeblue,linkcolor=ieeeblue]{hyperref}
\usepackage[nameinlink]{cleveref}
\crefname{equation}{}{}
\usepackage{bm}
\usepackage{glossaries-extra}
\newabbreviation{mse}{MSE}{mean-squared error}
\newabbreviation{mmse}{MMSE}{minimum-mean-squared error}
\newabbreviation{mri}{MRI}{magnetic resonance imaging}
\newabbreviation{sde}{SDE}{stochastic differential equation}
\newabbreviation{ct}{CT}{computed tomography}
\newabbreviation{awgn}{AWGN}{additive white Gaussian noise}
\newabbreviation{mcmc}{MCMC}{Markov chain Monte Carlo}
\newabbreviation{map}{MAP}{maximum-a-posteriori}
\newabbreviation{ssp}{SSP}{sparse stochastic processes}
\newabbreviation{iid}{i.i.d.}{independent and identically distributed}
\newabbreviation{dps}{DPS}{diffusion posterior sampling}
\newabbreviation{mala}{MALA}{Metropolis-adjusted Langevin algorithm}
\newabbreviation{glm}{GLM}{Gaussian latent machine}
\newabbreviation{ddpm}{DDPM}{denoising diffusion probabilistic model}
\newabbreviation{diffpir}{DiffPIR}{diffusion models for plug-and-play image restoration}
\newabbreviation{dpnp}{DPnP}{diffusion plug-and-play}
\newabbreviation{snr}{SNR}{signal-to-noise ratio}
\newabbreviation{cdps}{C-DPS}{Chung diffusion posterior sampling}
\newabbreviation{psnr}{PSNR}{peak signal-to-noise-ratio}
\newabbreviation{ssim}{SSIM}{structural similarity}
\glsdisablehyper
\definecolor{myblue}{rgb}{0.0627, 0.3333, 0.6039}
\definecolor{mygreen}{rgb}{0.0, 0.5, 0.0}
\definecolor{myred}{rgb}{0.8, 0.0, 0.0}
\newcommand{\plotposterior}[4]{
	\pgfplotsforeachungrouped \i in {1,2,3} {
		\edef\tmp{%
			\noexpand\nextgroupplot
			\noexpand\addplot table[x=x, y=y\i, col sep=comma] {\prefix/#1/measurements/#2/dps/cdps/#3/mean.csv};%
			\noexpand\ifthenelse{\i=3}{\noexpand\ifthenelse{#4=1}{\noexpand\addlegendentry{C-DPS}}{}}{}
			\noexpand\addplot table[x=x, y=y\i, col sep=comma] {\prefix/#1/measurements/#2/dps/diffpir/#3/mean.csv};%
			\noexpand\ifthenelse{\i=3}{\noexpand\ifthenelse{#4=1}{\noexpand\addlegendentry{DiffPIR}}{}}{}
			\noexpand\addplot table[x=x, y=y\i, col sep=comma] {\prefix/#1/measurements/#2/dps/dpnp/#3/mean.csv};%
			\noexpand\ifthenelse{\i=3}{\noexpand\ifthenelse{#4=1}{\noexpand\addlegendentry{DPnP}}{}}{}
			\noexpand\addplot[black, thin] table[x=x, y=y\i, col sep=comma] {\prefix/#1/measurements/#2/mmse/m.csv};%
			\noexpand\ifthenelse{\i=3}{\noexpand\ifthenelse{#4=1}{\noexpand\addlegendentry{Gibbs}}{}}{}
		}\tmp
	}
	\pgfplotsforeachungrouped \i in {1,2,3} {%
		\edef\tmp{%
			\noexpand\nextgroupplot[height=2.65cm, width=6.5cm, ymin=0]
			\noexpand\addplot table[x=x, y=y\i, col sep=comma] {\prefix/#1/measurements/#2/dps/cdps/#3/variance.csv};%
			\noexpand\addplot table[x=x, y=y\i, col sep=comma] {\prefix/#1/measurements/#2/dps/diffpir/#3/variance.csv};%
			\noexpand\addplot table[x=x, y=y\i, col sep=comma] {\prefix/#1/measurements/#2/dps/dpnp/#3/variance.csv};
			\noexpand\addplot table[x=x, y=y\i, col sep=comma] {\prefix/#1/measurements/#2/mmse/var_gibbs.csv};%
		}\tmp
	}
}
\newcommand{\plotgridsearchcurves}[2]{%
	\pgfplotsforeachungrouped \operator in {identity,convolution,sample,fourier} {%
		\edef\tmp{%
			\noexpand\nextgroupplot[xmode=log]
			\noexpand\addplot table[x=val, y=mse, col sep=comma] {\prefix/bernoulli-laplace/p=0.1_b=1.0/measurements/\operator/grid-search/#1/data.csv};%
			\noexpand\ifthenelse{\noexpand\equal{\operator}{sample}}{\noexpand\ifthenelse{#2=1}{\noexpand\addlegendentry{$\noexpand\BLDist(0.1, 1)$}}{}}{}
			\noexpand\addplot table[x=val, y=mse, col sep=comma] {\prefix/laplace/1.0/measurements/\operator/grid-search/#1/data.csv};%
			\noexpand\ifthenelse{\noexpand\equal{\operator}{sample}}{\noexpand\ifthenelse{#2=1}{\noexpand\addlegendentry{$\noexpand\LaplaceDist(1)$}}{}}{}
			\noexpand\addplot table[x=val, y=mse, col sep=comma] {\prefix/student/2.0/measurements/\operator/grid-search/#1/data.csv};%
			\noexpand\ifthenelse{\noexpand\equal{\operator}{sample}}{\noexpand\ifthenelse{#2=1}{\noexpand\addlegendentry{$\noexpand\tDist(2)$}}{}}{}
			\noexpand\addplot table[x=val, y=mse, col sep=comma] {\prefix/student/3.0/measurements/\operator/grid-search/#1/data.csv};%
			\noexpand\ifthenelse{\noexpand\equal{\operator}{sample}}{\noexpand\ifthenelse{#2=1}{\noexpand\addlegendentry{$\noexpand\tDist(3)$}}{}}{}
			\noexpand\addplot table[x=val, y=mse, col sep=comma] {\prefix/student/1.0/measurements/\operator/grid-search/#1/data.csv};%
			\noexpand\ifthenelse{\noexpand\equal{\operator}{sample}}{\noexpand\ifthenelse{#2=1}{\noexpand\addlegendentry{$\noexpand\tDist(1)$}}{}}{}
			\noexpand\addplot table[x=val, y=mse, col sep=comma] {\prefix/gauss/0.25/measurements/\operator/grid-search/#1/data.csv};%
			\noexpand\ifthenelse{\noexpand\equal{\operator}{sample}}{\noexpand\ifthenelse{#2=1}{\noexpand\addlegendentry{$\noexpand\GaussianDist(0, 0.25)$}}{}}{}
		}\tmp
	}
}
\newcommand{\plotunconditional}[1]{
	\nextgroupplot
	\addplot table[x=x, y=y1, col sep=comma] {./data/posterior_new/#1/unconditional/999.csv};%
	\addplot table[x=x, y=y2, col sep=comma] {./data/posterior_new/#1/unconditional/999.csv};%
	\addplot table[x=x, y=y3, col sep=comma] {./data/posterior_new/#1/unconditional/999.csv};%
	\nextgroupplot
	\addplot table[x=x, y=y1, col sep=comma] {./data/posterior_new/#1/unconditional/600.csv};%
	\addplot table[x=x, y=y2, col sep=comma] {./data/posterior_new/#1/unconditional/600.csv};%
	\addplot table[x=x, y=y3, col sep=comma] {./data/posterior_new/#1/unconditional/600.csv};%
	\nextgroupplot
	\addplot table[x=x, y=y1, col sep=comma] {./data/posterior_new/#1/unconditional/200.csv};%
	\addplot table[x=x, y=y2, col sep=comma] {./data/posterior_new/#1/unconditional/200.csv};%
	\addplot table[x=x, y=y3, col sep=comma] {./data/posterior_new/#1/unconditional/200.csv};%
	\nextgroupplot
	\addplot table[x=x, y=y1, col sep=comma] {./data/posterior_new/#1/unconditional/001.csv};%
	\addplot table[x=x, y=y2, col sep=comma] {./data/posterior_new/#1/unconditional/001.csv};%
	\addplot table[x=x, y=y3, col sep=comma] {./data/posterior_new/#1/unconditional/001.csv};%
}

\usepackage{siunitx}
\newcolumntype{Y}{S[round-mode=places,round-precision=2,table-format=1.2(2)]}
\title{A Statistical Benchmark for Diffusion Posterior Sampling Algorithms}
\author{%
    Martin Zach, Youssef Haouchat, Michael Unser%
    \thanks{M. Zach, Y. Haouchat, and M. Unser are with the Biomedical Imaging Group, École Polytechnique Fédérale de Lausanne, 1015 Lausanne, Switzerland. (e-mail: martin.zach@epfl.ch; youssef.haouchat@epfl.ch; michael.unser@epfl.ch)}%
    \thanks{M. Zach is with the Center for Biomedical Imaging, 1015 Lausanne, Switzerland.}
}
\newcommand{\dd}{\mathrm{d}}
\newcommand{\signal}{\mathbf{x}}
\newcommand{\increments}{\mathbf{u}}
\newcommand{\estimate}{\hat{\mathbf{x}}}
\newcommand{\mapest}{\estimate_{\mathrm{MAP}}}
\newcommand{\mmseest}{\estimate_{\mathrm{MMSE}}}
\newcommand{\Sigdim}{d}
\newcommand{\sigdim}{\mathbb{R}^{\Sigdim}}
\newcommand{\sigrv}{\mathbf{X}}
\newcommand{\sigdensity}{p_{\sigrv}}
\newcommand{\likelihood}{p_{\mathbf{Y} \mid \sigrv = \signal}}
\newcommand{\posterior}{p_{\sigrv \mid \mathbf{Y} = \meas}}
\newcommand{\posteriort}{p_{\sigrv_t \mid \mathbf{Y} = \meas}}
\newcommand{\meas}{\mathbf{y}}
\newcommand{\Measdim}{m}
\newcommand{\measdim}{\mathbb{R}^{\Measdim}}
\newcommand{\op}{\mathbf{A}}
\newcommand{\fdop}{\mathbf{D}}
\newcommand{\opdim}{\mathbb{R}^{\Measdim \times \Sigdim}}
\newcommand{\noise}{\mathbf{n}}
\newcommand{\noisevar}{\sigma_{\mathrm{n}}}
\newcommand{\R}{\mathbb{R}}
\newcommand{\argm}{\,\cdot\,}
\newcommand{\levy}{Lévy}
\newcommand{\statistic}{R}
\newcommand{\Ntrain}{N_{\mathrm{train}}}
\newcommand{\Nvalidation}{N_{\mathrm{val}}}
\newcommand{\Ntest}{N_{\mathrm{test}}}
\newcommand{\Nsamples}{N_{\mathrm{samples}}}

\DeclareMathOperator{\GaussianDist}{Gauss}
\DeclareMathOperator{\ExpDist}{Exp}
\DeclareMathOperator{\LaplaceDist}{Laplace}
\DeclareMathOperator{\tDist}{St}
\DeclareMathOperator{\GammaDist}{Gamma}
\DeclareMathOperator{\GIGDist}{GIG}
\DeclareMathOperator{\BLDist}{BL}

\DeclarePairedDelimiterX\norm[1]\lVert\rVert{
	\ifblank{#1}{\:\cdot\:}{#1}
}
\DeclarePairedDelimiterX\abs[1]{\lvert}{\rvert}{
	\ifblank{#1}{\:\cdot\:}{#1}
}

\DeclareMathOperator*{\argmin}{arg\,min}
\DeclareMathOperator*{\argmax}{arg\,max}

\theoremstyle{definition}
\newtheorem{definition}{Definition}[section]

\tikzset{line join=bevel}

\pgfplotsset{
	/tikz/line join=bevel,
	posteriorplot/.style={
			group/group size=3 by 8,
			group/group name=G,
			group/horizontal sep=0.7cm,
			group/vertical sep=0.35cm,
			group/xticklabels at=edge bottom,
			width=6.5cm,
			height=5cm,
			tick label style={font=\tiny},
			legend pos=north east,
			legend style={
				font=\scriptsize,
				legend cell align=left,
			},
			cycle list name=dashmark,
			/tikz/mark repeat={5},
			/tikz/line join=bevel,
	}
}
\pgfplotscreateplotcyclelist{dashmark}{%
  {myblue, solid, mark=*, line width=0.6pt, mark size=0.6pt},
  {myred, solid, mark=x, line width=0.6pt, mark size=1.2pt},
  {mygreen, solid, mark=triangle*, line width=0.6pt, mark size=0.7pt},
}

\definecolor{oiPurple} {HTML}{CC79A7}
\definecolor{oiSky}    {HTML}{56B4E9}
\definecolor{oiVerm}   {HTML}{D55E00}
\pgfplotscreateplotcyclelist{mutedColorMarkers}{%
  {myblue,   semithick, mark=*,        mark size=1.8pt, mark options={fill=white}},
  {myred, semithick, mark=square*,  mark size=1.9pt, mark options={fill=white}},
  {mygreen,  semithick, mark=triangle*,mark size=2.0pt, mark options={fill=white}},
  {oiPurple, semithick, mark=diamond*, mark size=2.0pt, mark options={fill=white}},
  {oiSky,    semithick, mark=+,        mark size=2.0pt},
  {oiVerm,   semithick, mark=x,        mark size=2.0pt},
  {black!55,semithick, mark=star,     mark size=1.9pt},
}
\pgfplotsset{every axis/.append style={
  cycle list name=mutedColorMarkers,
  mark options={solid, line width=0.4pt},
}}

\begin{document}
\maketitle
\begin{abstract}
    We propose a statistical benchmark for \gls{dps} algorithms for Bayesian linear inverse problems.
    The benchmark synthesizes signals from sparse \levy{}-process priors whose posteriors admit efficient Gibbs methods.
	These Gibbs methods can be used to obtain gold-standard posterior samples that can be compared to the samples obtained by the \gls{dps} algorithms.
    By using the Gibbs methods for the resolution of the denoising problems in the reverse diffusion, the framework also isolates the error that arises from the approximations to the likelihood score.
	We instantiate the benchmark with the minimum-mean-squared-error optimality gap and posterior coverage tests and provide numerical experiments for popular \gls{dps} algorithms on the inverse problems of denoising, deconvolution, imputation, and reconstruction from partial Fourier measurements.
	We release the benchmark code at \url{https://github.com/zacmar/dps-benchmark}.
	The repository exposes simple plug-in interfaces, reference scripts, and config-driven runs so that new algorithms can be added and evaluated with minimal effort.
	We invite researchers to contribute and report results.
\end{abstract}


\section{Introduction}%
\label{sec:introduction}
\IEEEPARstart{D}{iffusion} models are among the leading generative models in imaging~\cite{Rombach2022}, visual computing~\cite{Po2024}, finance and time-series analysis~\cite{huang2024generative,rasul2021autoregressive}, de novo protein and drug design~\cite{Watson2023,Alakhdar2024}, natural language processing~\cite{li2022diffusionlm,gong2023diffuseq}, and other domains.
Their ability to model complex distributions has motivated their use as priors in Bayesian inverse problems in those domains.
In fact, reconstruction methods that leverage diffusion models are competitive or state-of-the-art in, \textit{e.g.}, deconvolution~\cite{ren2023}, phase retrieval~\cite{xue2025}, \glsxtrlong{mri} and \glsxtrlong{ct} reconstruction~\cite{chung2022, liu2023}, weather-artifact removal~\cite{ozdenizci2023}, task-conditioned protein design~\cite{bogensperger2025variationalperspectivegenerativeprotein}, audio bandwidth extension and dereverberation~\cite{lemercier2024}, and denoising of financial time-series~\cite{wang2024}.

Despite this empirical success, diffusion models lack a natural strategy of conditioning on the measurements and active research explores how to incorporate the likelihood~\cite{ulugbek_cov,erbach2025solvinginverseproblemsflair}.
Currently, conditioning strategies are evaluated in one of two ways.
(i) With respect to downstream applications:
As an example, evaluations with respect to perceptual metrics such as the structural similarity~\cite{zhou_ssim}, the Fréchet inception distance~\cite{heusel_fid}, or the learned perceptual image patch similarity~\cite{zhang_lpips} are common in the imaging sciences.
However, as pointed out in~\cite{pierret_galerne_Gaussian_diffusion_exact_solutions_2025}, they are ill-suited for the statistical evaluation of the ability of algorithms to sample from the posterior distribution.
Such a statistical evaluation is critical in high-stakes applications such as medical imaging or remote sensing, where decisions based on reconstructions and their associated uncertainties may have significant consequences.
(ii) In overly simplistic settings:
A common fallback is to evaluate in synthetic settings with finite Gaussian mixture priors (often with only one component, like in the theoretical analysis in~\cite{pierret_galerne_Gaussian_diffusion_exact_solutions_2025}, or many components, like in the empirical evaluation in, \textit{e.g.},~\cite[Section E.3]{boys2024tweedie} or~\cite{crafts2025}).
Such finite Gaussian mixtures remain light-tailed with the tail decreasing exponentially like the widest component and, consequently, they cannot reproduce power-law-like extremes that are common in, \textit{e.g.}, asset returns~\cite{blattberg, Cont01022001} or the statistics of images~\cite{wainwright}.
Benchmarks built on such priors can therefore overstate posterior quality.

We propose a statistical evaluation framework for \gls{dps} algorithms for Bayesian linear inverse problems that addresses these issues.
We synthesize signals from discretized sparse \levy{}-process priors for which the resulting posteriors can be sampled from efficiently.
Indeed, they admit efficient Gibbs methods with exact conditionals that provide gold-standard posterior samples.
The framework supports general posterior-level comparisons---\textit{e.g.}, (sliced) Wasserstein or energy distances or calibration via coverage or posterior predictive checks---by furnishing matched samples obtained from the \gls{dps} algorithms and the gold-standard Gibbs methods.
In this paper, we instantiate the framework with two simple metrics: the \gls{mmse} optimality gap and highest posterior density coverage checks across the inverse problems of denoising, deconvolution, imputation, and reconstruction from partial Fourier measurements.
Moreover, the framework enables the isolation of the error that is attributable to the likelihood-score approximation by replacing the learned denoiser with an oracle \gls{mmse} denoiser computed from Gibbs samples at each reverse-diffusion step.

\subsection{Contributions}
We build on the statistical framework introduced in~\cite{bohra_20203_statistical} that benchmarked various neural network--based and model-based \gls{mmse} point estimators and provided optimality gaps by leveraging gold-standard estimators obtained by the Gibbs methods.
We generalize this benchmark from \gls{mmse} point estimators to posterior samplers and thereby enable the comparison of different methods on the distribution level instead of the point-estimation level.

To disentangle error sources within \gls{dps} algorithms, we replace any learned denoiser with an oracle \gls{mmse} denoiser that is computed via the Gibbs methods at each reverse-diffusion step.
This isolates the error due to the likelihood-score approximation from the error due to the prior-score surrogate.

To accommodate those \gls{dps} algorithms within our framework that do not follow a strict split between prior and likelihood---such as \gls{dpnp} \cite{xu2024provably}---we introduce a new template for \gls{dps} algorithms that utilize \emph{samples} from the denoising problem in their update steps, as opposed to only the \gls{mmse} point estimate.
This template arises naturally in our framework because Gibbs methods already provide these denoising-posterior samples and we show how several popular \gls{dps} algorithms can be re-expressed within this template.

Finally, we instantiate the benchmark with the \gls{mmse} optimality gap and posterior coverage tests.
We report results on denoising, deconvolution, imputation, and reconstruction from partial Fourier measurements, with and without learned components and provide an online repository that contains the open-source benchmarking code that is deliberately designed to facilitate the benchmarking of novel algorithms.

\section{Background}%
\label{sec:background}
\subsection{Bayesian Linear Inverse Problems}%
\label{ssec:bayesian linear inverse problems}
We seek to estimate a signal \( \signal \in \sigdim \) from the measurements
\begin{equation}
    \label{eq:inverse_problem}
    \meas = \op\signal + \noise,
\end{equation}
where the \emph{forward operator} \( \op \in \opdim \) models the noiseless linear measurement acquisition and \( \noise \in \measdim \) is additive noise.
In the Bayesian treatment of this problem, the signals are modeled as a random variable, denoted \( \sigrv \), with values in \( \sigdim \) and distribution \( \sigdensity \).
We refer to \( \sigdensity \) as the \emph{prior}.
Given any datum \( \meas \), the ultimate goal is to analyze the \emph{posterior} \( \posterior \) which is related to the \emph{likelihood} \( \likelihood \) and the prior \( \sigdensity \) via Bayes' rule, which states that
\begin{equation}
		\posterior(\signal) \propto \likelihood(\meas)\sigdensity(\signal).
\end{equation}

For a given signal \( \signal \), the likelihood \( \likelihood \) is determined by the distribution of the noise.
A common assumption on the noise is that it is a vector of \gls{iid} Gaussian random variables with mean zero and variance \( \noisevar^2 \).\footnote{%
	Our framework supports more general (possibly non-Gaussian) likelihoods, see \cref{ssec:efficient posterior sampling}.
}
In this case, the likelihood is given by
\begin{equation}
    \likelihood(\meas) \propto \exp\bigl( -\tfrac{1}{2\noisevar^2} \norm{\op\signal - \meas}^2 \bigr).
	\label{eq:likelihood}
\end{equation}

Thus, once the forward model and the noise law are specified, the remaining modeling choice is the prior distribution.
The ability of diffusion models to model complex distributions makes them a good candidate prior for the resolution of inverse problems.
However the lack of a natural conditioning mechanism prohibits a straightforward use, see~\cref{ssec:diffusion posterior sampling}.

\subsection{Bayes Estimators}
A benefit of the Bayesian approach over classical variational methods (see, e.g.,~\cite{Scherzer2008-dr}) is that different point estimates arise from a fixed prior.
For a given measurement \( \meas \), these point estimates summarize the posterior distribution \( \posterior \) with respect to a given loss \( \ell : \sigdim \times \sigdim \to \R \) via the optimization problem of finding the point \( \estimate_\ell(\meas) \) that minimizes the posterior risk:
\begin{equation}
	\estimate_\ell(\meas) = \argmin_{\estimate \in \sigdim} \biggl( \int_{\sigdim} \ell(\estimate, \signal)\ \posterior(\signal)\,\dd \signal \biggr).
    \label{eq:bayes estimator}
\end{equation}
In this paper, the Bayes estimator with respect to the \gls{mse} \( \ell = \tfrac{1}{d}\norm{\argm - \argm}^2 \) plays a key role due to its close relation to the prior \emph{score} in the reverse diffusion (see~\cref{ssec:diffusion models}) and because we later quantify the performance of \gls{dps} algorithms via an \gls{mmse} optimality gap.
With this choice of \( \ell \), \cref{eq:bayes estimator} can be written as
\begin{equation}
	\begin{aligned}
		\mmseest(\meas) &= \argmin_{\estimate \in \sigdim} \biggl(
				\int_{\sigdim}\tfrac{1}{d} \norm{\estimate - \signal}^2\ \posterior(\signal)\,\dd \signal
			\biggr)\\
			&= \int_{\sigdim} \signal \posterior(\signal)\,\dd \signal = \mathbb{E}[\sigrv \mid \mathbf{Y} = \meas],
	\end{aligned}
	\label{eq:mmse estimator}
\end{equation}
which is the expectation of the posterior \( \posterior \).

Another widely-used estimator arises through the choice
\begin{equation}
    \ell(\estimate, \signal) = -\chi_{\{ \estimate \}}(\signal)
\end{equation}
where
\begin{equation}
	\chi_{A}(\signal) \coloneqq \begin{cases} 
		1 & \text{if } \signal \in A, \\
		0 & \text{else},
	\end{cases}
\end{equation}
which leads to the \gls{map} estimator that seeks the mode of the posterior:\footnote{%
    This definition is informal but sufficient for the purposes of this paper.
    For continuous posteriors, the strict 0–1 loss Bayes' rule is ill-posed.
    A common formalization defines MAP as the limit of Bayes estimators under shrinking small-ball 0–1 losses;
    under additional regularity, this limit agrees with the posterior mode~\cite{Bassett2018,clason}.
    The \gls{map} estimator may also not be unique.
}
\begin{equation}
	\begin{aligned}
		\mapest(\meas) &= \argmin_{\estimate \in \sigdim} \biggl(
			\int_{\sigdim} -\chi_{\{ \estimate \}}(\signal)\ \posterior(\signal)\,\dd \signal
		\biggr)\\
				&= \argmax_{\signal \in \sigdim} \posterior(\signal).
	\end{aligned}
    \label{eq:map}
\end{equation}
Rewriting~\cref{eq:map} as
\begin{equation}
	\mapest = \argmin_{\signal \in \sigdim} \bigl(
		-\tfrac{1}{2\noisevar^2}\norm{\op\signal - \meas}^2  - \log \sigdensity(\signal)
	\bigr),
\end{equation}
reveals a close relation to classical variational approaches after identifying the regularizer with \( - \log \sigdensity \).

\subsection{Diffusion Models}%
\label{ssec:diffusion models}
As the likelihood is modeled explicitly in Bayesian inverse problems, the modeling burden lies with the prior.
Diffusion models are widely used to synthesize new signals that usefully generalize the training distribution and have proven effective across domains, which motivates their use as priors.
We first give a background on diffusion models and then describe the problem that arises when using these models in Bayesian inverse problems in~\cref{ssec:diffusion posterior sampling}.

Diffusion models were introduced by Song et al.\ in~\cite{song2020score} by unifying the discrete approaches presented in~\cite{song2019generative} and~\cite{ho2020} in a continuous theory based on \glspl{sde}.
In the sequel, we only introduce those concepts of \glspl{sde} that are relevant for the present work and refer to~\cite[Chapters 25 and 26]{Klenke2020} for more details.
We denote the (diffusion) \gls{sde} with \emph{drift coefficient} $\mathbf{f} : \sigdim \times \R_{\geq0} \to \sigdim$ and \emph{diffusion coefficient} $g : \R_{\geq 0} \to \R$ as
\begin{equation}
	\dd \sigrv_t = \mathbf{f}(\sigrv_t,t)\,\dd t + g(t)\,\dd \mathbf{W}_t,
	\label{eq:diffusion_SDE}
\end{equation}
where $\mathbf{W}_t$ is the standard Wiener process.
A solution to that \gls{sde} is a stochastic process \( (\sigrv_t)_t \) that satisfies the integral equation
\begin{equation}
    \sigrv_t = \sigrv_0 + \int_{0}^t\mathbf{f}(\sigrv_s,s)\,\dd s + \int_{0}^tg(s)\,\dd \mathbf{W}_s
    \label{eq:diffusion integral form}
\end{equation}
for some suitable initial condition \( \sigrv_0 \).
In our setup, the initial condition is the random variable that describes the signal, thus \( \sigrv_0 = \sigrv \).
Under suitable choices for \( \mathbf{f} \) and \( g \), the forward process admits a limiting marginal \( \sigrv_{\infty} \) as \( t \to \infty \).
Sampling from \( p_{\sigrv_0} \) can then proceed by simulating the \gls{sde} \cref{eq:diffusion_SDE} in reverse with initial condition \( \sigrv_\infty \).
By Anderson's theorem~\cite{anderson1982reverse}, the reverse \gls{sde} that reproduces the forward marginals satisfies
\begin{equation}
    \dd \sigrv_t = \bigl(\mathbf{f}(\sigrv_t,t) - g^2(t)\nabla\log p_{\sigrv_t}(\sigrv_t)\bigr)\,\dd t + g(t)\,\dd \mathbf{W}_t,
	\label{eq:BW_diff}
\end{equation}
where $p_{\sigrv_t}$ denotes the density of $\sigrv_t$ from \cref{eq:diffusion integral form}, and $\dd t$ is an infinitesimal negative time step.

The primary challenge in this approach lies in the computation of the \emph{scores} \( \nabla\log p_{\sigrv_t} \) for all \( t > 0 \).
We now derive an equality that relates \( \nabla\log p_{\sigrv_t} \) to \( \mathbb{E}[\sigrv_0 \mid \sigrv_t = \argm\,] \), \textit{i.e.}, the \gls{mmse} estimate of \( \sigrv_0 \) given that \( \sigrv_t \) takes on a certain value.
We restrict ourselves to the drift coefficient $\mathbf{f}(\signal,t) = -\frac{\beta(t)}{2}\signal$ and the diffusion coefficient $g(t) = \sqrt{\beta(t)}$, which correspond to the \emph{variance-preserving} formulation (see~\cite[Section 3.4]{song2020score}) whose limiting marginal is the standard normal (which is preferred for its numerical properties) and where \( \beta : \R_{\geq 0} \to \R_{\geq 0} \) controls how fast the process contracts towards zero and how much noise is injected.
Similar derivations can be found in, \textit{e.g.},~\cite{song2020score,chung2023diffusion,daras2024}, but we include it to underscore the relevance of the \gls{mmse} estimate in this paper and to facilitate the understanding of its relation to various objects.
Under this choice of the drift and diffusion coefficient, the diffusion \gls{sde}~\cref{eq:diffusion_SDE} simplifies to a time-inhomogeneous Ornstein--Uhlenbeck \gls{sde} (see~\cite[Example 26.5]{Klenke2020})
\begin{equation}
	\dd \sigrv_t = -\tfrac{\beta(t)}{2}\sigrv_t \,\dd t + \sqrt{\beta(t)}\,\dd \mathbf{W}_t,
\end{equation}
whose pathwise solution
\begin{equation}
	\sigrv_t = \alpha(t) \sigrv_0 + \int_{0}^t \frac{\alpha(t)}{\alpha(s)} \sqrt{\beta(t)}\,\dd\mathbf{W}_s
\end{equation}
where \( \alpha(t) = \exp\bigl( -\frac{1}{2} \int_{0}^t \beta(s)\,\dd s \bigr) \) can be computed with standard techniques, see, \textit{e.g.},~\cite[Section 4.4.4]{Gardiner1990-pk}.
In addition, since
\begin{equation}
	\begin{aligned}
		&\int_{0}^t \biggl(\frac{\alpha(t)}{\alpha(s)}\biggr)^2 \beta(t)\,\dd s \\
		&\quad = \int_{0}^t\beta(s)\exp\Bigl( -\int_{s}^t \beta(u)\,\dd u \Bigr)\,\dd s = 1 - \alpha^2(t)
	\end{aligned}
\end{equation}
we can write that
\begin{equation}
	\sigrv_t = \alpha(t) \sigrv_0 + \sigma(t) \mathbf{N}
	\label{eq:xt from x0 vp}
\end{equation}
in distribution, where \( \sigma^2(t) = 1 - \alpha^2(t) \).
Consequently, the density of \( \sigrv_t \) is given by the convolution of \( p_{\sigrv_0} \) with a Gaussian with variance \( \sigma^2(t) \) and appropriate scaling by \( \alpha(t) \):
\begin{equation}
	p_{\sigrv_t}(\signal) = \int_{\sigdim} g_{\mathbf{0}, \sigma(t)^2\mathbf{I}}(\signal - \alpha(t)\hat{\signal})p_{\sigrv_0}(\hat{\signal})\,\dd \hat{\signal},
\end{equation}
where \( g_{\bm{\mu}, \mathbf{\Sigma}}(\signal) = (2\pi)^{-\frac{d}{2}}|\mathbf{\Sigma}|^{-\frac{1}{2}}\exp\bigl( -\tfrac{1}{2}\norm{\signal - \bm{\mu}}_{\mathbf{\Sigma}^{-1}}^2 \bigr) \).
Finally, after taking the gradient we see that
\begin{equation}
	\begin{aligned}
		&\nabla p_{\sigrv_t}(\signal) \\
        &= \int_{\R^d} \nabla g_{\mathbf{0}, \sigma(t)^2\mathbf{I}}(\signal - \alpha(t)\hat{\signal})p_{\sigrv_0}(\hat{\signal})\,\dd \hat{\signal} \\
		&= \int_{\R^d} \bigl( -\tfrac{1}{\sigma^{2}(t)}(\signal - \alpha(t) \hat{\signal} \bigr)g_{\mathbf{0}, \sigma^2(t)\mathbf{I}}(\signal - \alpha(t) \hat{\signal})p_{\sigrv_0}(\hat{\signal})\,\dd\hat{\signal} \\
        &= -\tfrac{1}{\sigma^2(t)}\Bigl( \signal p_{\sigrv_t}(\signal) - \\&\qquad\quad\alpha(t)\int_{\R^d} \hat{\signal} g_{\mathbf{0}, \sigma^2(t)\mathbf{I}}(\signal - \alpha(t) \hat{\signal})p_{\sigrv_0}(\hat{\signal})\,\dd\hat{\signal}\Bigr) \\
		&= -\tfrac{1}{\sigma(t)^{2}}\bigl( \signal p_{\sigrv_t}(\signal) - \alpha(t) p_{\sigrv_t}(\signal)\mathbb{E}[\sigrv_0 \mid \sigrv_t = \signal] \bigr)
	\end{aligned}
\end{equation}
such that, after dividing by \( p_{\sigrv_t}(\signal) \) and since \( \frac{\nabla p_{\sigrv_t}(\signal)}{p_{\sigrv_t}(\signal)} = \nabla \log p_{\sigrv_t}(\signal) \), we find the celebrated Tweedie identity
\begin{equation}
	\nabla \log p_{\sigrv_t}(\signal) = - \sigma(t)^{-2} \bigl( \signal - \alpha(t) \mathbb{E}[\sigrv_0 \mid \sigrv_t = \signal] \bigr).
    \label{eq:tweedie}
\end{equation}
Thus, given any point \( \signal \) and time \( t > 0 \), this yields a practical way of computing \( \nabla \log p_{\sigrv_t}(\signal) \) through the resolution of the \gls{mmse} denoising problem of finding \( \mathbb{E}[\sigrv_0 \mid \sigrv_t = \signal] \).

In standard applications where the goal is the generation of new signals, this \gls{mmse} denoising problem is typically tackled through the off-line learning of the map \( (\signal, t) \mapsto \mathbb{E}[\sigrv_0 \mid \sigrv_t = \signal] \) via a neural network.\footnote{%
	This does not necessitate the off-line computation of any \gls{mmse} denoising result since it can be implemented by finding that \( f \) that minimizes \( \int_{t>0} \mathbb{E}_{(\sigrv_t,\sigrv_0)}[\norm{f(\sigrv_t, t) - \sigrv_0}^2] \), where a pair \( (\sigrv_t,\sigrv_0) \) from the joint distribution can be constructed through ancestral sampling (\textit{i.e.}, sampling \( \sigrv_0 \) and setting \( \sigrv_t \) according to~\cref{eq:xt from x0 vp})~\cite{vincent}.
}
In our benchmark, we instead obtain oracle \gls{mmse} denoisers via Gibbs methods and thereby eliminate approximation errors from a learned surrogate when isolating error sources in \gls{dps} algorithms.
\subsubsection{Discretization}
Implementing the reverse \gls{sde} for generation requires a time discretization, which is typically done by Euler--Maruyama techniques (see, e.g.,~\cite{higham2001}).
For the reverse Ornstein--Uhlenbeck \gls{sde}
\begin{equation}
    \dd \sigrv_t = 
    \bigl(
        -\tfrac{\beta(t)}{2}\sigrv_t
    - \beta(t)\nabla\log p_{\sigrv_t}(\sigrv_t)\bigr)\,\dd t 
    + \sqrt{\beta(t)}\,\dd\mathbf{W}_t,
	\label{eq:BW_vp}
\end{equation}
a first-order step from \( t \) to \( t - 1 \) (\(\dd t = -1 \)) gives the Euler--Maruyama update
\begin{equation}
    \sigrv_{t-1} = \bigl(1+\tfrac{\beta_t}{2}\bigr)\sigrv_{t} + \beta_t\nabla\log p_{\sigrv_t}(\sigrv_t) + \sqrt{\beta_t}\mathbf{Z}_t,
\end{equation}
where \( \beta_t \coloneqq \beta(t) \) and \( \mathbf{Z}_t \sim \GaussianDist(\mathbf{0}, \mathbf{I}) \).
\begin{figure*}
	\centering
	\def\prefix{./data/unconditional-sampling}
	\begin{tikzpicture}
		\begin{groupplot}[
			/tikz/line join=bevel,
			group style={
				group size=4 by 4,
				horizontal sep=0.7cm,
				vertical sep=0.35cm,
				group name=G,
				xticklabels at=edge bottom,
			},
			width=5.2cm,
			height=4.4cm,
			tick label style={font=\tiny},
			cycle list name=dashmark,
		]
			\plotunconditional{bernoulli-laplace/p=0.1_b=1.0}
			\plotunconditional{student/1.0}
			\plotunconditional{laplace/1.0}
			\plotunconditional{gauss/0.25}
		\end{groupplot}
		\node[anchor=south] at ($(G c1r1.north)+(0, 2pt)$) {\scriptsize \( t=999 \)};
		\node[anchor=south] at ($(G c2r1.north)+(0, 2pt)$) {\scriptsize \( t=600 \)};
		\node[anchor=south] at ($(G c3r1.north)+(0, 2pt)$) {\scriptsize \( t=200 \)};
		\node[anchor=south] at ($(G c4r1.north)+(0, 2pt)$) {\scriptsize \( t=0 \)};

		\node[rotate=90] at ($(G c1r1.west)+(-20pt, 0pt)$) {\scriptsize \( \BLDist(0.1, 1) \)};
		\node[rotate=90] at ($(G c1r2.west)+(-20pt, 0pt)$) {\scriptsize \( \tDist(1) \)};
		\node[rotate=90] at ($(G c1r3.west)+(-20pt, 0pt)$) {\scriptsize \( \LaplaceDist(1) \)};
		\node[rotate=90] at ($(G c1r4.west)+(-20pt, 0pt)$) {\scriptsize \( \GaussianDist(0.1, 1) \)};
    \end{tikzpicture}
	\caption{%
		Unconditional reverse-diffusion trajectories obtained by \glsxtrshort{ddpm} sampling with the oracle denoiser.
		Rows: Jump distributions.
		Columns: Diffusion times.
		Line styles: Different initializations and random states.
	}%
    \label{fig:ddpm}
\end{figure*}
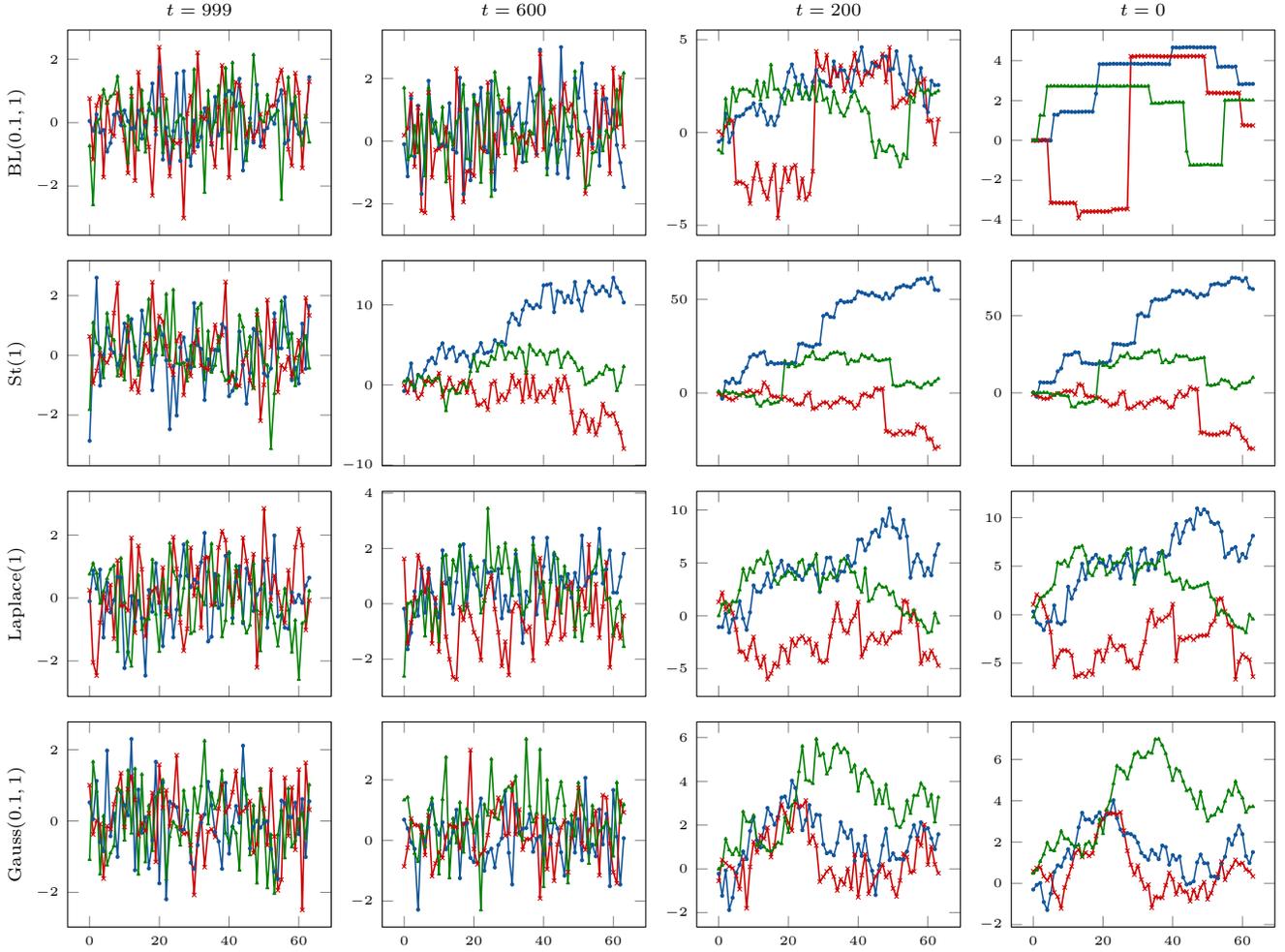
\begin{figure}
	\centering
	\def\prefix{./data/histogram-comparison}
	\begin{tikzpicture}
		\begin{axis}[
			width=9.0cm,
			height=5cm,
			xlabel={Jump values},
			ylabel={Histogram frequency},
			xmin=-15, xmax=15,
			ymin=0,
			tick label style={font=\scriptsize},
			label style={font=\scriptsize},
			title style={font=\scriptsize},
			legend style={
				font=\scriptsize,
				at={(0.97,0.97)},
				anchor=north east,
				draw=none
			},
			legend cell align={left},
			tick label style={/pgf/number format/fixed},
			scaled y ticks=false
			]
			\addplot+[ybar interval, fill=none, fill opacity=0.4, mark=none, draw=myred, draw opacity=0.6] table [x expr=\thisrow{x}-0.06, y=hist, col sep=comma] {\prefix/learned.csv};
			\addlegendentry{Learned}
			\addplot+[ybar interval, fill=none, fill opacity=0.4, mark=none, draw=myblue, draw opacity=0.6] table [x expr=\thisrow{x}-0.06, y=hist, col sep=comma] {\prefix/gibbs.csv};
			\addlegendentry{Oracle}
			\addplot[dashed, black] table [x=x, y=student_t_pdf, col sep=comma] {\prefix/reference.csv};
			\addlegendentry{Target}
		\end{axis}
	\end{tikzpicture}
	\caption{%
		Histogram of jumps of signals obtained by unconditional \glsxtrshort{ddpm} sampling with the oracle denoiser and the learned denoiser.%
	}%
	\label{fig:gibbs vs learned}
\end{figure}
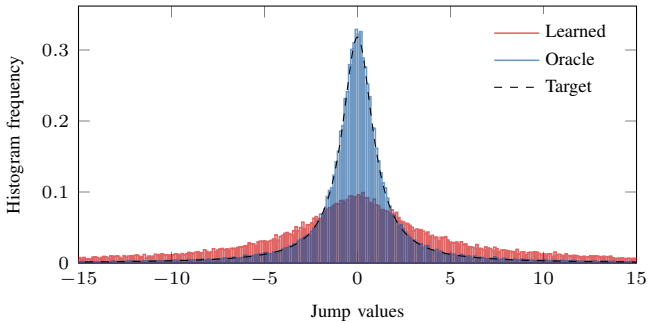

In practice, researchers typically use an alternative schedule from the discrete-time Markov chain that was initially proposed by Sohl-Dickstein et al.\ in~\cite{sohl-dickstein15} and revisited and popularized by Ho et al.\ in~\cite{ho2020}.
There, the discretization
\begin{equation}
    \sigrv_{t-1} = \tfrac{1}{\sqrt{1-\beta_t}} \left( \sigrv_{t} + \beta_t\nabla\log p_{\sigrv_t}(\sigrv_t) \right) + \sqrt{\beta_t}\mathbf{Z}_t,
\end{equation}
is used and typically called \gls{ddpm}-style sampling.
It can be related to the Euler--Maruyama discretization of the \gls{sde} via Taylor expansions; we provide a detailed derivation in~\cref{sec:reverse sde vs ddpm}.

We show a trajectory of signals generated by this discretization of the reverse \gls{sde} using the oracle \gls{mmse} denoiser in~\cref{fig:ddpm}.
The signals and the construction of the \gls{mmse} denoiser are described in~\cref{sec:proposed framework}.
To motivate our use of the oracle \gls{mmse} denoiser, we show in~~\cref{fig:gibbs vs learned} a histogram of jump distributions obtained by the learned denoiser versus the oracle denoiser for a \( \tDist(1) \) jump target (our notations of various distributions is summarized in~\cref{tab:all distributions}).
The signals generated by the oracle denoiser follow the jump distribution almost perfectly since the reverse diffusion is an exact sampler up to discretization error.
Example generated signals are shown in~\cref{fig:learned vs gibbs samples} in the appendix.

\subsection{Diffusion Posterior Sampling}%
\label{ssec:diffusion posterior sampling}
The reverse-diffusion sampler from the previous section can be adapted to sample from a posterior by replacing the prior score \( \nabla \log p_{\sigrv_t} \) with the posterior score \( \nabla \log \posteriort \) for some given measurement \( \meas \).
An inspection of \( \nabla \log \posteriort \) via Bayes' theorem reveals that
\begin{equation}
	\nabla \log \posteriort = \nabla \log p_{\sigrv_t} + \nabla \bigl( \signal \mapsto \log p_{\mathbf{Y} \mid \sigrv_t = \signal}(\meas) \bigr).
\end{equation}
Although the dependence between \( \mathbf{Y} \) and \( \sigrv_0 \) is known through \cref{eq:inverse_problem} and the likelihood is explicitly modeled via~\cref{eq:likelihood}, it is generally challenging to relate \( \mathbf{Y} \) and \( \sigrv_t \) for any \( t > 0 \) with an explicit likelihood.
Thus, the conditioning on the measurements (or, more generally, on various characteristics of the solution like a desired fitness value in de novo protein design) is usually done in one of two ways:
\begin{enumerate}
	\item A learned component directly models the posterior score and gets the measurements (or a quantity derived thereof) as input.
		This strategy is pursued in, \textit{e.g.},~\cite{liu2023,ozdenizci2023,bogensperger2025variationalperspectivegenerativeprotein,chitwan}, and is advantageous when the measurement process is unknown, difficult to model, or prohibitively expensive to evaluate.
		However, reconstructions obtained by this strategy typically degrade under shifts in measurement conditions, since the learned components cannot adapt to the new measurement conditions.
	\item The Bayesian separation that is described in~\cref{ssec:bayesian linear inverse problems} is pursued.
		This is done in, \textit{e.g.},~\cite{chung2022,xue2025} and the methods reviewed in~\cite{lemercier2024}, and is advantageous when the measurement process is known, relatively inexpensive to evaluate, and subject to change, but prior knowledge should be reused, which is frequently the case in, \textit{e.g.}, imaging or remote sensing applications.
		However, this requires approximations to \( \nabla (\signal \mapsto \log p_{\mathbf{Y} \mid \sigrv_t = \signal}(\meas)) \) for all \( t > 0 \).
\end{enumerate}
As an example, in~\cite{jalal2021} Jalal et al.\ proposed the approximation
\begin{equation}
	\nabla(\signal \mapsto \log p_{\mathbf{Y} \mid \sigrv_t = \signal}(\meas)) \approx \nabla(\signal \mapsto \log p_{\mathbf{Y} \mid \sigrv_0 = \signal}(\meas)).
\end{equation}
Another very popular approximation, known under the (unfortunately very generic) name \enquote{diffusion posterior sampling} was proposed by Chung et al.\ in~\cite{chung2023diffusion}.
There, the likelihood is approximated with
\begin{equation}
	\begin{aligned}
		&\nabla(\signal \mapsto \log p_{\mathbf{Y} \mid \sigrv_t = \signal}(\meas)) \\&\quad\approx \nabla(\signal \mapsto \log p_{\mathbf{Y} \mid \sigrv_0 = \mathbb{E}[\sigrv_0 \mid \sigrv_t = \signal]}(\meas)).
	\end{aligned}
    \label{eq:dps approx}
\end{equation}
To avoid confusion with the umbrella term of \gls{dps} algorithms, we abbreviate that specific method as \gls{cdps}.
The approximation that is made in \gls{diffpir}~\cite{zhu2023denoising} is to take the Moreau envelope of the potential of \( p_{\mathbf{Y} \mid \sigrv_0 = \argm}(\meas) \) and evaluating its gradient at \( \mathbb{E}[\sigrv_0 \mid \sigrv_t = \argm] \).

In principle, our benchmark can evaluate either strategy (and any other method that claims to sample from a posterior distribution).
The first approach, however, relies on black-box learning of the conditional posterior score and its performance heavily depends on various implementation details.
Thus, we primarily focus on the second approach that necessitates approximations of the likelihood score (and more general \gls{dps} algorithms with some kind of explicit conditioning, see our proposed generalization in~\cref{ssec:generalization}).
Our framework supplies reference objects---posterior samples and oracle denoisers via Gibbs methods---to isolate and quantify the impact of these approximations in \gls{dps} algorithms.

\subsection{A Generalization to Accommodate More Algorithms}%
\label{ssec:generalization}%
The description in the previous section---approximating the likelihood score inside the reverse diffusion---covers the family of algorithms named \enquote{explicit approximations for measurement matching} in the recent survey~\cite{daras2024surveydiffusionmodelsinverse}.
However, widely used methods such as \gls{dpnp}~\cite{xu2024provably}, fall outside this pattern.
We therefore introduce a simple template that accommodates a broader set of \gls{dps} algorithms and is natural in our setting, where the denoising posterior is \emph{sampled} via Gibbs methods.

We characterize \gls{dps} algorithms as an iteration rule that can be summarized into a two-stage process:
Given an iterate \( \signal_t \) with associated noise variance \( \sigma^2(t) \), the computation of the next iterate \( \signal_{t-1} \) is done by
\begin{enumerate}
	\item drawing \( S \) samples denoted \( \{ \bar{\signal}_{k} \}_{k=1}^S \) from the denoising posterior \( p_{\sigrv_0\mid\sigrv_t=\signal_t} \propto \exp\bigl( -\tfrac{1}{2\sigma_t^2} \norm{\argm - \signal_t}^2 \bigr)p_{\sigrv_0}(\argm) \), and
	\item computing the next iterate \( \signal_{t-1} \) through an update step \( \mathcal{S} \) that may utilize the current iterate \( \signal_t \), the samples \( \{ \bar{\signal}_{k} \}_{k=1}^S \), the measurements \( \meas \), the forward operator \( \op \), and any other possible algorithm-internal parameters such as a scalar that weights likelihood and prior terms or parameters that define the noise schedule.
\end{enumerate}
This template is summarized in \cref{alg:diffusion posterior sampling} and the specialized instances for the step \( \mathcal{S} \) that correspond to the three popular algorithms \gls{cdps}~\cite{chung2023diffusion}, \gls{diffpir}~\cite{zhu2023denoising}, and \gls{dpnp}~\cite{xu2024provably} are tabulated in \cref{tab:algorithms_Steps}.
We have absorbed the scaling by \( \alpha_t \) into the step \( \mathcal{S} \) since this template is not fundamentally limited to Ornstein--Uhlenbeck processes or even diffusion processes in general, but supports any (also not monotonically decreasing) variance schedules.
In addition, the noise variances \( \{ \sigma_t \}_{t=0}^T \) are usually derived from the algorithm-internal parameters \( \Theta \) that may include a noise-annealing schedule (\( \{ \alpha_t \}_{t=0}^T \) for \gls{cdps} and \gls{diffpir} and \( \{ \eta_t \}_{t=0}^T \) for \gls{dpnp}).

Through this construction, \gls{dps} algorithms can use any statistic \( \statistic \) of the samples \( \{ \bar{\signal}_k \}_{k=1}^S \) in their update steps.
Most methods use (possibly in addition to others) the mean \( \statistic(\bar{\signal}_1, \dotsc, \bar{\signal}_S) = \frac{1}{S}\sum_{k=1}^S \bar{\signal}_s \coloneqq \bar{\boldsymbol{\mu}} \), which is the Monte Carlo estimate of \( \mathbb{E}[\sigrv_0\mid\sigrv_t = \signal_t] \).
An example of a \gls{dps} algorithm that utilizes additional statistics is \gls{cdps} that requires the Jacobian of \( \signal_t \mapsto \mathbb{E}[\sigrv_0\mid\sigrv_t = \signal_t] \).
In settings where \( (\signal, t) \mapsto \mathbb{E}[\sigrv_0 \mid \sigrv_t = \signal] \) is learned, the Jacobian is typically obtained by automatic differentiation.
As we show in \cref{ssec:covariance in dps}, this Jacobian equals (up to the known VP scaling) the conditional covariance of \( \sigrv_0 \mid \sigrv_t = \signal_t \).
In our framework, an unbiased estimator of the covariance matrix can be obtained through the statistic \( \statistic(\bar{\signal}_1, \dotsc, \bar{\signal}_S) = \frac{1}{S-1} \sum_{k=1}^S (\bar{\signal}_k - \bar{\boldsymbol{\mu}}) (\bar{\signal}_s - \bar{\boldsymbol{\mu}})^T \).
An example of a \gls{dps} algorithm that utilizes an alternative statistic is the \gls{dpnp} algorithm that alternately samples a denoising problem and a data-proximal problem.
There, the statistic \( \statistic(\bar{\signal}_1, \dotsc, \bar{\signal}_S) = \bar{\signal}_1 \) is used to obtain one sample from the denoising problem.
\begin{algorithm}[t]
	\caption{Template for \gls{dps} algorithms.}%
    \label{alg:diffusion posterior sampling}
    \begin{algorithmic}[1]
		\Require Initial point \( \signal_{T} \), $\meas$, $\op$, $\Theta$, \( \{ \sigma_t \}_{t=0}^T \)
    \For{$t = T, \dotsc, 1$} \Comment{Diffusion process}
	\State Sample \( \{ \bar{\signal}_k \}_{k=1}^S \sim \exp\bigl( -\tfrac{\norm{\argm - \signal_t}^2}{2\sigma^2_t} \bigr)p_{\sigrv_0}(\argm) \)
	\State Update $\signal_{t-1} = \mathcal{S}(\signal_t, \{ \bar{\signal}_k \}_{k=1}^S, \meas, \op, \Theta, t)$
    \EndFor
    \State \Return \( \signal_0 \) \Comment{Posterior sample}
    \end{algorithmic}
\end{algorithm}
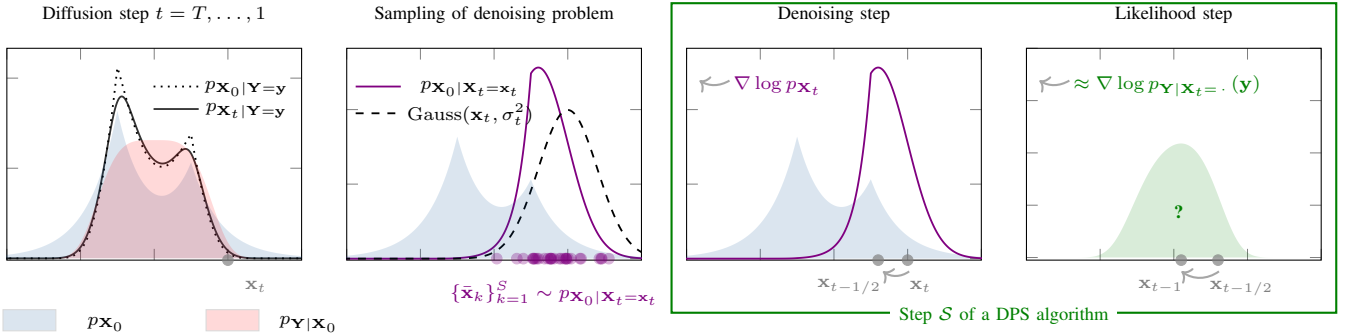
\begin{figure*}
	\centering
	\def\prefix{./data/teaser}
	\begin{tikzpicture}[font=\scriptsize]
		\begin{groupplot}[
			group style={
				group size=4 by 1,
				horizontal sep=0.6cm,
				vertical sep=1.5cm,
			},
			width=5.5cm,
			height=4.4cm,
			tick label style={font=\tiny},
			]
			\nextgroupplot[
			title={Diffusion step $t = T, \dotsc, 1$},
			xmin=-2., xmax=2,
			ymin=-0.01,
			yticklabels=\empty,
			xticklabels=\empty,
			legend style={
				font=\scriptsize,
				draw=none,    
				fill=none,
				at ={(1.,0.9)}
			}
			]
			\addplot[black, dotted, line width=0.7pt] table[x=x, y=pyx, col sep=comma] {\prefix/convolved_posterior_single.csv};
			\addlegendentry{$p_{\sigrv_0\mid \mathbf{Y} = \meas}$}
			\addplot[black, line width=0.7pt, opacity=0.8] table[x=x, y=convolved_posterior_single, col sep=comma] {\prefix/convolved_posterior_single.csv};
			\addlegendentry{$p_{\sigrv_t\mid \mathbf{Y} = \meas}$}
			\addplot[only marks, mark=*, mark size=2pt, gray, opacity=0.7] coordinates {(1, -0.01)};
			\addplot[draw=none, fill, color=myblue, opacity=0.15] table[x=x, y=px, col sep=comma] {\prefix/convolved_posterior_single.csv};
			\addplot[draw=none, fill, color=red, opacity=0.15] table[x=x, y=py, col sep=comma] {\prefix/convolved_posterior_single.csv};
			\addplot[draw=none] coordinates {(0,0)};

			\nextgroupplot[
			title={Sampling of denoising problem},
			xmin=-2., xmax=2,
			ymin=-0.01,
			yticklabels=\empty,
			xticklabels=\empty,
			legend style={
				font=\scriptsize,
				draw=none,    
				fill=none,
				at ={(0.68,0.9)}
			}
			]
			\addplot[violet, line width=0.7pt] table[x=x, y=denoising, col sep=comma] {\prefix/denoised_posterior.csv};
			\addlegendentry{$p_{\sigrv_0 \mid \sigrv_t = \signal_t}$};
			\addplot[dashed, line width=0.7pt] table[x=x, y=gaussian, col sep=comma] {\prefix/denoised_posterior.csv};
			\addlegendentry{$\GaussianDist(\signal_t, \sigma_t^2)$};
			\addplot[violet, only marks, , mark=*, mark size=2pt, violet, opacity=0.3] table[x=x, y=zero, col sep=comma] {\prefix/samples.csv};
			\addplot[draw=none, fill, color=myblue, opacity=0.15] table[x=x, y=px, col sep=comma] {\prefix/denoised_posterior.csv};
			\nextgroupplot[
			title={Denoising step},
			xmin=-2., xmax=2,
			ymin=-0.01,
			yticklabels=\empty,
			xticklabels=\empty,
			legend style={
				font=\scriptsize,
				draw=none,    
				fill=none,
				at ={(0.58,0.9)}
			}
			]
			\addplot[draw=none, fill, color=myblue, opacity=0.15] table[x=x, y=px, col sep=comma] {\prefix/denoised_posterior.csv};
			\addplot[only marks, mark=*, mark size=2pt, gray, opacity=0.7] coordinates {(1, -0.01)};
			\addplot[only marks, mark=*, mark size=2pt, gray, opacity=0.7] coordinates {(0.6, -0.01)};
			\addplot[violet, line width=0.7pt] table[x=x, y=denoising, col sep=comma] {\prefix/denoised_posterior.csv};
			\nextgroupplot[
			title={Likelihood step},
			xmin=-2., xmax=2,
			ymin=-0.01, ymax = 0.8,
			yticklabels=\empty,
			xticklabels=\empty,
			legend style={
				font=\scriptsize,
				draw=none,    
				fill=none,
				at ={(0.58,0.9)}
			}
			]
			\addplot[only marks, mark=*, mark size=2pt, gray, opacity=0.7] coordinates {(0.6, -0.01)};
			\addplot[only marks, mark=*, mark size=2pt, gray, opacity=0.7] coordinates {(0.1, -0.01)};
			\addplot[draw=none, fill, color=mygreen, opacity=0.15] table[x=x, y=intractable_posterior, col sep=comma] {\prefix/intractable_posterior.csv};
		\end{groupplot}
		\node[anchor=north, font=\scriptsize, color=gray] at ($(group c1r1.south) + (1.35,-0.15)$) {$\signal_t$};
		\node[anchor=north, font=\scriptsize, color=violet] at ($(group c2r1.south) + (0.8,-0.15)$) {$\{\bar{\signal}_k\}_{k=1}^S \sim p_{\sigrv_0\mid \sigrv_t = \signal_t}$};

		\coordinate (xt_label) at ($(group c3r1.south) + (1.15,-0.15)$);
		\node[anchor=north, font=\scriptsize, color=gray] at (xt_label) {$\signal_t$};
		\draw[->, thick, gray!70, bend left=35] 
			($(xt_label) + (-0.15, 0)$) to ($(xt_label) + (-0.5, 0)$);
		\node[anchor=east, font=\scriptsize, color=gray] 
			at ($(xt_label) + (-0.4, -0.2)$) {$\signal_{t -1/2}$};
		\coordinate (xthalf_label) at ($(group c4r1.south) + (0.75,-0.15)$);
		\node[anchor=north, font=\scriptsize, color=gray] at ($(xthalf_label) + (0.15, 0)$) {$\signal_{t-1/2}$};
		\draw[->, thick, gray!70, bend left=35] 
			($(xthalf_label) + (-0.15, 0)$) to ($(xthalf_label) + (-0.65, 0)$);
		\node[anchor=east, font=\scriptsize, color=gray] 
			at ($(xthalf_label) + (-0.5, -0.2)$) {$\signal_{t - 1}$};

		\draw[->, thick, gray!70, bend left=30] 
			([xshift=2pt, yshift=2pt] {$(group c3r1.north west) + (0.5,-0.5)$}) 
			to ([xshift=2pt, yshift=2pt] {$(group c3r1.north west) + (0.1,-0.5)$});
		\node[anchor=west, font=\scriptsize, color=violet] at ($(group c3r1.north west) + (0.5,-0.5)$) {$ \nabla \log p_{\sigrv_t}$};
		\draw[->, thick, gray!70, bend left=30] 
			([xshift=2pt, yshift=2pt] {$(group c4r1.north west) + (0.5,-0.5)$}) 
			to ([xshift=2pt, yshift=2pt] {$(group c4r1.north west) + (0.1,-0.5)$});
		\node[anchor=west, font=\scriptsize, color=mygreen] at ($(group c4r1.north west) + (0.5,-0.5)$) {$\approx \nabla \log p_{\mathbf{Y} \mid \sigrv_t = \argm}(\meas)$};
		\node[anchor=east, color=mygreen] at ($(xthalf_label) + (-0.5, 0.8)$) {\textbf{?}};
		\coordinate (rectSW) at ($(group c3r1.south west) + (-0.2, -.7)$);
		\coordinate (rectNE) at ($(group c4r1.north east) + (0.2, 0.6)$);
		\draw[thick, mygreen] (rectSW) rectangle (rectNE);
		\path (rectSW) -- (rectNE) coordinate[midway] (rectMidBottom);
		\coordinate (rectBottom) at ($(rectMidBottom) + (0, -2.1)$);
		\node[fill=white, text=mygreen] at (rectBottom) {Step $\mathcal{S}$ of a \gls{dps} algorithm};
		\node[fill=myblue, minimum width=20pt, minimum height=8pt, draw=black, opacity=0.15] at (0.3, -0.8) {};
		\node [right,font=\scriptsize] at (0.95, -.85) {$p_{\sigrv_0}$};
		\node[fill=red, minimum width=20pt, minimum height=8pt, draw=black, opacity=0.15] at (3, -0.8) {};
		\node [right, font=\scriptsize] at (3.5, -.85) {$p_{\mathbf{Y} \mid \sigrv_0}$};
	\end{tikzpicture}
	\caption{%
		Illustration of the proposed template for \gls{dps} algorithms.
		The benchmarked posterior sampler targets $\signal_0 \sim p_{\sigrv_0\mid \mathbf{Y} = \meas}$ via a diffusion process.
		At each diffusion time $t$, first the samples \( \{ \bar{\signal}_k \}_{k=1}^S \sim p_{\sigrv_0\mid \sigrv_t = \signal_t} \) are drawn from the denoising posterior.
		Then, the step $\mathcal{S}$ updates the iterate typically through a prior-guided update from the samples and a likelihood-guided update from the data.
		The likelihood guidance term is intractable and must be approximated, which constitutes the primary source of sampling error.
	}%
	\label{fig:algorithm illustration}
\end{figure*}

\section{Related Work}
For unconditional generation, many works derive theoretical bounds on various distances between a target distribution and the distribution produced by (approximations of) the reverse \gls{sde}~\cref{eq:BW_diff}.
These bounds relate the error to properties of the forward \gls{sde}, the target distribution, the score function, or the reverse-time discretization.
For example,~\cite{xuefeng_wasserstein} contains bounds on the Wasserstein-2 distance with respect to the discretization error under the assumption that the target distribution is smooth and log-concave.
These bounds translate directly to a bound on the number of reverse-diffusion steps that is needed to obtain a desired accuracy.
Under absolute continuity of the target with respect to a Gaussian,~\cite{strasman2025an} gives bounds on the Kullback--Leibler divergence that depend on the noise schedule \( \alpha \).
Additional results in total variation and other distances appear in the references cited therein.

A common simplification in analysis is to assume a Gaussian target.
In that case, many objects in the forward and reverse \gls{sde}---including the posterior score \( \nabla \log \posteriort \) for \( t \geq 0 \)---admit closed forms, which facilitates the computation of various bounds.
For example, the authors of~\cite{hurault2025scorematchingdiffusionfinegrained} analyze the effect of estimating the prior score (which is affine in this case) from finitely many samples of the target distribution and track the error propagation through the iterations of the reverse \gls{sde}.
The authors of~\cite{pierret_galerne_Gaussian_diffusion_exact_solutions_2025} derive explicit solutions to the \gls{sde} and use those to derive bounds on the Wasserstein-2 distance to the distributions that are obtained via the Euler--Maruyama discretizations.

The works that are closest to ours are~\cite{pierret2025exactevaluationaccuracydiffusion} and~\cite{crafts2025}.
The authors of~\cite{pierret2025exactevaluationaccuracydiffusion} study posterior sampling with Gaussian priors theoretically.
They derive expressions for the Wasserstein-2 distances between the conditional forward marginals (targets) and the distributions induced by specific likelihood approximations in the reverse \gls{sde}.
Because their setting assumes Gaussian priors and Gaussian conditional likelihoods, all involved distributions are Gaussian,\footnote{
    This also relies on the conditional likelihoods being Gaussian, which restricts their analysis to problems with Gaussian likelihoods and to those algorithms that adhere to the strict Bayesian structure described in \cref{ssec:diffusion posterior sampling}.
} and the expression for the Wasserstein-2 distances are given by standard formulas that involve the covariance of the target distribution and the forward operator.
Our framework is not restricted to Gaussian priors: we handle a broad family of priors that correspond to signals that are discretizations of \levy{} processes (detailed next).
Moreover, their framework targets algorithms that adhere to the strict likelihood-approximation structure that we detailed in \cref{ssec:diffusion posterior sampling}, which many algorithms, \textit{e.g.} DPnP, do not follow.
Finally, although their expressions are explicit, deriving them for new algorithms often requires a substantial amount of nontrivial mathematics.
In contrast, our benchmark and the accompanying code are deliberately designed such that novel algorithms can be easily benchmarked, and additionally allows the quantification of the error that is incurred by substituting the optimal denoisers with their learned counterparts.
The authors of~\cite{crafts2025} give a numerical evaluation of various \gls{dps} algorithms under the assumption of a (finite) Gaussian mixture prior.
Similar to the present work, the authors provide reference objects to the \gls{dps} algorithm for a fair evaluation.
However, their work is limited to the restrictive case of a Gaussian mixture prior which, as outlined in~\cref{sec:introduction}, cannot reproduce power-law-like extremes (heavy tails) and can overstate posterior quality.

Beyond diffusion-specific theory, the authors of~\cite{thong2024bayesianimagingmethodsreport} evaluate posterior calibration by checking coverage of credible regions produced by different Bayesian recovery strategies.
They find that recovery strategies that utilize diffusion models often under-report uncertainty or, in other words, are overly confident.
A shortcoming of their approach is that they use an empirical distribution of images as a surrogate for the prior distribution and, consequently, their approach is limited to the concrete application and dataset that is at hand.
Our framework, by contrast, relies on known priors from which we can generate infinitely many signals and corresponding measurements.
This isolates algorithmic error without resorting to surrogate priors and supports fair, repeatable comparisons across tasks and algorithms.

\section{Proposed Framework}%
\label{sec:proposed framework}
The prior distributions in our framework will be that of signals obtained by regularly spaced samples of processes with independent, stationary increments (\levy{} processes and their discrete-time counterparts).
We briefly recall the definition; see~\cite{Unser_Tafti_2014,Sato1999-sc} for background and the link to infinitely divisible laws.
\begin{definition}[\levy{} process]
	A stochastic process \( s = \{ s(t) : t \geq 0\} \) is a \levy{} process if
	\begin{enumerate}
		\item \(s(0)=0\) almost surely;
		\item (independent increments) for any \( N \in \mathbb{N} \setminus \{ 0, 1 \} \) and \( 0 \leq t_1 < t_2 < \cdots < t_N < \infty \), the increments \((s(t_2) - s(t_1)), (s(t_3) - s(t_2)), \dotsc, (s(t_N) - s(t_{N - 1}))\) are mutually independent;
		\item (stationary increments) for any given step \( h \), the increment process \( u_h = \{ s(t) - s(t - h) : t > h \} \) is stationary;
		\item (stochastic continuity) for any \( \varepsilon > 0\) and \( t \ge 0 \) \[ \lim_{h\to0}\Pr\bigl( |s(t + h) - s(t)| > \varepsilon \bigr)=0. \]
	\end{enumerate}
\end{definition}

We form discrete signals by sampling  \( s \) at integer times and stacking the values into \( \signal = (s(1), \dotsc, s(d)) \).
Let the unit-step increments be \( \increments_k = s(k) - s(k - 1) \) for \( k = 1, \dotsc, d \).
By independence and stationarity, the law of \( \increments_k \) does not depend on \( k \) and we denote it \( p_U \).\footnote{For our choices, it always has a density w.r.t.\ a suitable reference measure.}
We define the finite-difference matrix
\begin{equation}
	\fdop = \begin{bmatrix}
		1  & 0  & 0      & \cdots & 0 \\
		-1 & 1  & 0      & \cdots & 0 \\
		0  & -1 & 1      & \cdots & 0 \\
		\vdots &  \vdots  & \ddots & \ddots & 0 \\
		0  & 0  & \cdots & -1     & 1 \\
	\end{bmatrix}
\end{equation}
such that the increment vector satisfies
\begin{equation}
	\increments = \fdop \signal.
	\label{eq:increments and signal}
\end{equation}
Because \( s(0) = 0 \), \( \fdop \) is invertible and \( \fdop^{-1} \) is a lower-triangular matrix of ones, which also implies that for all \( k = 1, 2, \dotsc, \Sigdim \),
\begin{equation}
	\signal_k = \sum_{n=1}^k \increments_n
	\label{eq:cumsum}
\end{equation}
which is a convenient way to synthesize signals once \(\increments\) is drawn.
The combination of \cref{eq:increments and signal} with the independence of the increments implies that the density of the discrete signal is
\begin{equation}
	p_{\sigrv}(\signal) = \prod_{k=1}^d p_U\bigl( (\fdop\signal)_k \bigr).
\end{equation}

In this paper, we consider four increment families that are commonly used in sparse-process models: Gaussian, Laplace, Student-t, and Bernoulli--Laplace (spike-and-slab).
We provide precise definitions and our notation of these and other distributions that we use in this work in~\cref{tab:all distributions} in the appendix.
Such increment laws yield sparse or heavy-tailed signals according to the taxonomy in~\cite{Unser_Tafti_2014} and are relevant in signal and image processing, finance, and many other fields~\cite{Unser_Tafti_2014,Schoutens2003-zq}.

\subsection{Efficient Posterior Sampling}%
\label{ssec:efficient posterior sampling}
The Bayesian treatment of the inverse problem in~\cref{eq:inverse_problem} yields the posterior
\begin{equation}
    \begin{aligned}
        \posterior(\signal) 
        &\propto \exp\bigl( -\tfrac{1}{2\noisevar^2}\norm{\op\signal - \meas}^2 \bigr) \sigdensity(\signal)\\
        &= \exp\bigl( -\tfrac{1}{2\noisevar^2}\norm{\op\signal - \meas}^2 \bigr) \prod_{k=1}^d p_U \bigl((\fdop\signal)_k\bigr).
    \end{aligned}
    \label{eq:posterior}
\end{equation}
Unless \( p_U \) is a Gaussian (which is the simplified setting in~\cite{pierret_galerne_Gaussian_diffusion_exact_solutions_2025}), this posterior is not conjugate, so neither closed-form sampling nor direct evaluation of moments is available.
Nevertheless, for the increment laws used in this paper, the posterior distributions admit efficient Gibbs methods via standard latent-variable augmentations.
Under these augmentations, all conditionals are either Gaussian or many independent one-dimensional distributions, both of which can be sampled efficiently, which results in rapidly mixing Gibbs methods.
Such methods were recently shown to be significantly faster than other standard sampling routines that are commonly used in such settings in~\cite{kuric2025gaussianlatentmachineefficient}.
They report sampling efficiencies of close to \num{1}, while alternatives, such as the Metropolis-adjusted Langevin algorithm, achieve sampling efficiencies of around \num{1e-3}.\footnote{
    Sampling efficiency refers to effective samples per iteration; an efficiency of \( \rho \) means roughly \( 1 / \rho \) iterations per \enquote{effective sample} \cite[Section 11.5]{Gelman2013}.
}
In addition, Gibbs methods require no step-size or acceptance-rate tuning and introduce no discretization bias.
These properties motivate our use of Gibbs methods for the fast and robust posterior sampling throughout this work.

\subsubsection{Gibbs Methods}%
\label{ssec:gibbs}
Gibbs methods are \gls{mcmc} methods to sample from a joint distribution \( p_{\mathbf{X}, \mathbf{Z}_1, \mathbf{Z}_2, \dotsc, \mathbf{Z}_n} \) of \( (n + 1) \) variables that are advantageous when the direct sampling is computationally difficult but sampling from the conditional distributions \( p_{\mathbf{X} | \mathbf{Z}_1, \mathbf{Z}_2, \dotsc, \mathbf{Z}_n}, p_{\mathbf{Z}_1 \mid \mathbf{X}, \mathbf{Z}_2, \dotsc, \mathbf{Z}_n}, \dotsc \) is easy.
Gibbs methods cycle through the conditional distributions with repeated draws, which maintains the joint distribution invariant~\cite{Casella1992}.
The naming of the variables \( \mathbf{X}, \mathbf{Z}_1, \mathbf{Z}_2, \dotsc, \mathbf{Z}_n \) is deliberately chosen to emphasize that we use \emph{latent-variable} Gibbs methods that rely on auxiliary variables that are introduced solely to make the conditionals simple.
The steps of a general latent variable Gibbs sampler are shown in~\cref{alg:gibbs}, where the iteration counter in the sampling of the latent variables is omitted since they need not be stored and previous iterations can immediately be overwritten.
\begin{algorithm}[t]
	\caption{Latent-variable Gibbs sampling of \( p_{\mathbf{X}, \mathbf{Z}_1, \dotsc, \mathbf{Z}_N} \).}%
	\label{alg:gibbs}
	\begin{algorithmic}[1]
		\Require{}Burn-in period \( B \in \mathbb{N} \), number of samples \( S \in \mathbb{N} \), initial point \( (\mathbf{x}_0, \mathbf{z}_1, \dotsc\, \mathbf{z}_N) \).
		\For{$s=1, 2, \dots, B + S$}
			\State{}\( \mathbf{x}_{s} \sim p_{\mathbf{X} \mid \mathbf{Z}_1 = \mathbf{z}_{1}, \dotsc, \mathbf{Z}_N = \mathbf{z}_{N} }\)
			\State{}\( \mathbf{z}_{1} \sim p_{\mathbf{Z}_1 \mid \mathbf{X} = \mathbf{x}_{k}, \dotsc, \mathbf{Z}_N = \mathbf{z}_{N} }\)
			\State{}\( \vdots \)
		\EndFor{}
		\State{}\Return{}\( \{ \mathbf{x}_{B+s} \}_{s=1}^{S} \)
	\end{algorithmic}
\end{algorithm}

Like all \gls{mcmc} methods, in practice Gibbs methods benefit from discarding some number of initial samples, the \emph{burn-in period}, when the initial point is located in low-density regions.
After the burn-in period, it is crucial to tune the number of samples such that the Monte Carlo estimates of various quantities, such as the \gls{mmse} estimate in~\cref{eq:mmse estimator}, are sufficiently accurate.
We discuss our choice of the burn-in period and the number of samples for the various problems in~\cref{ssec:implementation details}.

The Gaussian, Laplace, and Student-t distributions admit latent representations as infinite-component Gaussian mixture models, which makes them suitable for the \gls{glm} framework that was recently introduced in~\cite{kuric2025gaussianlatentmachineefficient}.
The \gls{glm} framework is generally applicable to distributions of the form
\begin{equation}
    p_{\sigrv} (\signal) \propto \prod_{k=1}^{n}\phi_k\bigl( (\mathbf{K}\signal)_k \bigr)
    \label{eq:glm}
\end{equation}
where \( \mathbf{K} \in \R^{n \times d} \) and all distributions \( \phi_1, \phi_2, \dotsc, \phi_n : \R \to \R \) have a latent representation
\begin{equation}
    \phi_k(t) = \int_{\R}g_{\mu_k(z), \sigma_k^2(z)}(t)f_k(z)\,\dd z,
\end{equation}
where the \emph{latent distribution} \( f_i \) and the \emph{latent maps} \( \mu_i \) and \( \sigma_i^2 \) depend on the distribution \( \phi_i \) and are tabulated in~\cref{tab:distributions} for the distributions that are relevant in this paper.

The introduction of an appropriate \( n \)-dimensional random variable \( \mathbf{Z} \) with non-trivial distribution (see the details in~\cite{kuric2025gaussianlatentmachineefficient}) enables the efficient sampling from the conditionals:
Sampling \( \sigrv \mid \mathbf{Z} = \mathbf{z} \) amounts to sampling a Gaussian with covariance
\begin{equation}
    \mathbf{\Sigma}(\mathbf{z}) = (\mathbf{K}^T\mathbf{\Sigma}_0(\mathbf{z})^{-1}\mathbf{K})^{-1}
    \label{eq:glm covariance}
\end{equation}
and mean
\begin{equation}
    \bm{\mu}(\mathbf{z}) = \mathbf{\Sigma}(\mathbf{z}) \mathbf{K}^T \mathbf{\Sigma}_0(\mathbf{z})^{-1} \bm{\mu}_0(\mathbf{z})
    \label{eq:glm mean}
\end{equation}
where \( \mathbf{\Sigma}_0(\mathbf{z}) = \operatorname{diag}\bigl(\sigma_1^2(\mathbf{z}_1), \sigma_2^2(\mathbf{z}_2), \dotsc, \sigma_n^2(\mathbf{z}_n)\bigr) \) and \( \bm{\mu}_0(\mathbf{z}) = \bigl(\mu_1(\mathbf{z}_1), \mu_2(\mathbf{z}_2), \dotsc, \mu_n(\mathbf{z}_n)\bigr) \).
Sampling \( \mathbf{Z} \mid \sigrv = \signal \) amounts to sampling \( n \) independent one-dimensional \emph{conditional latent distributions} \( p_{Z_i\mid X = (\mathbf{K}\signal)_i} \) that depend on the distributions \( \phi_1, \dotsc, \phi_n \) and are also given in~\cref{tab:distributions}.

The posterior distribution in~\cref{eq:posterior} can be cast into this framework by rewriting it as
\begin{equation}
	\begin{aligned}
		&p_{\sigrv|\mathbf{Y}=\meas}(\signal) \\
        &\propto\exp\biggl(-\tfrac{1}{2\noisevar^2}\sum_{k=1}^m\bigl((\op \signal)_k - \meas_k\bigr)^2 \biggr) \prod_{k=1}^d p_U((\fdop\signal)_k)\\
		&\propto\prod_{k=1}^m g_{\meas_k, \noisevar^2}\bigl((\op\signal)_k\bigr)\prod_{k=1}^d p_U\bigl((\fdop\signal)_k\bigr)\\
		&=\prod_{k=1}^{m+d}\phi_k\bigl( (\mathbf{K}\signal)_k \bigr)
	\end{aligned}
\end{equation}
by setting
\begin{equation}
	\mathbf{K} = \begin{bmatrix} \op \\ \fdop \end{bmatrix}
\end{equation}
and
\begin{equation}
	\phi_k = \begin{cases}
		g_{\meas_k, \noisevar^2} & \text{for } k = 1,\dotsc, m, \\
		p_U & \text{for } k = m+1, \dotsc, m+d.
	\end{cases}
\end{equation}
We summarize the \gls{glm} sampling in~\cref{alg:glm}.
It is easily adapted non-Gaussian likelihoods by substituting the respective first \( m \) distributions.
\begin{algorithm}[t]
	\caption{\gls{glm} Gibbs sampler.}%
    \label{alg:glm}
    \begin{algorithmic}[1]
		\Require{}Initial point $\mathbf{x}_0 \in \mathbb{R}^d$, operator $\mathbf{K} \in \mathbb{R}^{n\times d}$, distributions $\{ \phi_i \}_{i=1}^n$ with corresponding conditional latent distributions $\{ p_{Z_i\mid X} \}_{i=1}^n$ and latent maps $\{ \mu_i, \sigma_i^2 \}_{i=1}^n$
		\For{$s = 1, \dotsc, B + S$}
			\State{}Draw \( \mathbf{z}_i \sim p_{Z_i\mid X = (\mathbf{K}\signal_{s-1})_{i}} \) \Comment{par. \( i = 1, \dotsc, n \)}
			\State{}Draw \( \mathbf{x}_s \sim \GaussianDist(\bm{\mu}(\mathbf{z}), \mathbf{\Sigma}(\mathbf{z})) \) \Comment{see \cref{eq:glm covariance,eq:glm mean}}
		\EndFor
		\State \Return \( \{ \mathbf{x}_{B+s} \}_{s=1}^S \)
    \end{algorithmic}
\end{algorithm}

The Bernoulli--Laplace distribution requires the introduction of two latent variables to effectively handle the distribution of jumps due to the Bernoulli being \enquote{on} and the corresponding jump height that is distributed according to a Laplace distribution, and we use the algorithm that was proposed in~\cite{bohra_20203_statistical}.
We start by noting that the Bernoulli--Laplace density
\begin{equation}
	p_U(u) = \lambda \delta(u) + (1 - \lambda)\tfrac{b}{2}\exp(-b|u|)
\end{equation}
with Bernoulli parameter \( \lambda \) and scale parameter \( b \) admits the representation
\begin{equation}
	p_U(u) = \int_{\R}\biggl( \sum_{v=0}^1 p_{U\mid V=v, W=w}(u)p_V(v) \biggr)p_W(w)\,\dd w,
\end{equation}
where
\begin{equation}
	p_V(v) = \lambda^{1-v}(1-\lambda)^v
\end{equation}
for \( v \in \{ 0, 1\} \) is a Bernoulli distribution,
\begin{equation}
	p_W(w) = \frac{b^2}{2}\exp\biggl( -\frac{b^2w}{2} \biggr)\chi_{\R_{\geq 0}}(w)
\end{equation}
is an exponential distribution, and
\begin{equation}
	p_{U \mid V=v, W=w}(u) = \begin{cases}
		\delta(u) & \text{if}\ v = 0, \\
		\GaussianDist(0, w) & \text{if}\ v = 1.
	\end{cases}
	\label{eq:u given v is zero}
\end{equation}
The algorithm relies on the introduction of two latent vectors \( \mathbf{v}, \mathbf{w} \in \sigdim \) that satisfy
\begin{equation}
	p_{\mathbf{U} \mid \mathbf{V}=\mathbf{v},\mathbf{W}=\mathbf{w}}(\increments) = \prod_{k=1}^d p_{U\mid V=\mathbf{v}_k, W=\mathbf{w}_k}(\mathbf{u}_k)
\end{equation}
such that, as a result, the distribution conditioned on the measurements can be written as
\begin{equation}
	\begin{aligned}
		&p_{\mathbf{U}, \mathbf{V}, \mathbf{W} \mid \mathbf{Y}=\meas}(\mathbf{u}, \mathbf{v}, \mathbf{w}) \\
        &\propto\exp\bigl( -\tfrac{1}{2\noisevar^2}\norm{\mathbf{H}\increments - \meas}^2 \bigr) \prod_{k=1}^d p_{U\mid V=\mathbf{v}_k,W=\mathbf{w}_k}(\mathbf{u}_k)\\
		& \quad\times \prod_{k=1}^d \lambda^{1-\mathbf{v}_k}(1-\lambda)^{\mathbf{v}_k}\prod_{k=1}^d \frac{b^2}{2}\exp\biggl( -\frac{b^2\mathbf{w}_k}{2} \biggr),
	\end{aligned}
	\label{eq:joint}
\end{equation}
where \( \mathbf{H} = \op\fdop^{-1} \).
\Cref{eq:u given v is zero,eq:joint} imply that any sample from \( p_{\mathbf{U} \mid \mathbf{V}=\mathbf{v},\mathbf{W}=\mathbf{w}, \mathbf{Y}=\meas} \) takes the value zero at those indices where \( \mathbf{v} \) is zero, and values from a multivariate Gaussian distribution with covariance \( \mathbf{C} = \bigl( \noisevar^2 \mathbf{H}\mathbf{H}^T + \operatorname{diag}(\mathbf{w}) \bigr)^{-1}\) and mean \( \noisevar^{-2}\mathbf{C}\mathbf{H}^{T}\meas \) otherwise.
Sampling \( \mathbf{W} \mid \mathbf{U}=\mathbf{u}, \mathbf{V}=\mathbf{v}, \mathbf{Y}= \meas \) amounts to the independent sampling of \( d \) one-dimensional distributions, which are  \( \ExpDist(2 / b^2) \) at those indices where \( \mathbf{v} \) is zero and \( \GIGDist( b^2, \mathbf{u}_k^2, 0.5) \) those indices \( k \) where \( \mathbf{v} \) is one.
The conditional distribution of the binary support vector is
\begin{equation}
	\begin{aligned}
		p_{\mathbf{V} \mid \mathbf{W}=\mathbf{w}, \mathbf{Y}=\meas}(\mathbf{v}) \propto\ &|\mathbf{B}(\mathbf{v},\mathbf{w})|^{-\frac{1}{2}} \exp\bigl( -\tfrac{1}{2}\meas^T\mathbf{B}(\mathbf{v},\mathbf{w})^{-1}\meas \bigr)\\
																							 &\times\prod_{k=1}^d\lambda^{1-\mathbf{v}_k}(1-\lambda)^{\mathbf{v}_k},
	\end{aligned}
\end{equation}
where \( \mathbf{B}(\mathbf{v}, \mathbf{w}) = \noisevar^2 \mathbf{I} + \mathbf{H}\operatorname{diag}(\mathbf{v} \odot \mathbf{w}) \mathbf{H}^T \).\footnote{%
	This is a different but equivalent formulation to what is presented in~\cite{bohra_20203_statistical}, where the authors explicitly \enquote{slice} the matrices \( \mathbf{H} \) and \( \operatorname{diag}(\mathbf{w}) \) with the indices where \( \mathbf{v} \) is one.
	We stick to this formulation since it requires less notation and emphasizes that implementations need not build variable-sized matrices, which is crucial for an efficient implementation on modern compute units that utilize highly parallelized computations.
}
The standard way to sample from this distribution is to use a coordinate-wise Gibbs sampler that updates \( \mathbf{v}_k \sim \mathrm{Bernoulli}(p_k) \) with
\begin{equation}
	p_k = (1 + \exp(-\Delta_k))^{-1}
    \label{eq:bernoulli probability}
\end{equation}
where the log-odds increment 
\begin{equation}
	\begin{aligned}
		\Delta_k =\ &\log\tfrac{1-\lambda}{\lambda} - \tfrac{1}{2}\bigl( \log |\mathbf{B}(\mathbf{v}_{k=1}, \mathbf{w})| - \log|\mathbf{B}(\mathbf{v}_{k=0}, \mathbf{w})| \bigr) \\
					&- \tfrac{1}{2}\bigl( \meas^T\mathbf{B}(\mathbf{v}_{k=1}, \mathbf{w})^{-1}\meas - \meas^T\mathbf{B}(\mathbf{v}_{k=0}, \mathbf{w})^{-1}) \meas \bigr),
	\end{aligned}
\end{equation}
where \( \mathbf{v}_{k=\argm} \coloneqq (\mathbf{v}_1, \dotsc, \mathbf{v}_{k-1}, \argm, \mathbf{v}_{k+1}, \dotsc, \mathbf{v}_d) \) is the difference between the log-posterior when the bit is on and when it is off.
This resulting algorithm, given in~\cref{alg:bl}, can be interpreted as \( (d + 2) \)-variable (\textit{i.e.}, dimension-dependent) Gibbs method\footnote{%
	This is not strictly correct since the density violates the classical positivity conditions that are needed for Gibbs methods.
	It is a \emph{partially collapsed} Gibbs method, see~\cite{bohra_20203_statistical,vandyck}.
} and an efficient implementation is crucial.
We detail our implementation, which utilizes incremental updates of \( \mathbf{B} \) based on the Woodbury--Sherman--Morrison identities, in~\cref{ssec:implementation details}.
\begin{algorithm}[t]
	\caption{Bernoulli--Laplace Gibbs sampler.}%
	\label{alg:bl}
	\begin{algorithmic}[1]
		\Require{}Initial jumps $\mathbf{u}_0 \in \mathbb{R}^n$
		\For{$s = 1, \dotsc, B + S$}
			\State{}Draw \( \mathbf{w} \sim p_{\mathbf{W} \mid \mathbf{U} = \mathbf{u}_{s - 1}, \mathbf{V} = \mathbf{v}, \mathbf{Y} = \meas} \) \Comment{see \cref{eq:joint}}
			\For{\( k = 1, \dotsc, d \)}
				\State{}Draw \( \mathbf{v}_k \sim \operatorname{Bernoulli}(p_k) \) \Comment{incremental, see \cref{eq:bernoulli probability}}
			\EndFor
			\State{}Draw \( \mathbf{u}_s \sim p_{\mathbf{U} \mid \mathbf{V=\mathbf{v},\mathbf{W}=\mathbf{w}, \mathbf{Y}=\meas}} \) \Comment{see \cref{eq:joint}}
		\EndFor
		\State \Return \( \{\fdop^{-1} \mathbf{u}_{B+s} \}_{s=1}^S \)
	\end{algorithmic}
\end{algorithm}

The Gibbs methods that we just described are suitable for the generation of the gold-standard samples of the posterior that corresponds to the initial inverse problem~\cref{eq:inverse_problem} as well as the generation of samples from the denoising problems that arise in the various \gls{dps} algorithms.
For the latter, the forward operator \( \op \) is the identity, the measurements are the noisy intermediate reconstructions \( \signal_t \), and \( \noisevar = \sigma_t \) is the noise schedule at timestep \( t \).

\subsection{Benchmark Implementation Details}%
\label{ssec:implementation details}
The benchmarking pipeline starts with the generation of \( \Ntest \) test signals denoted \( \{ \signal^{\mathrm{test}}_k \}_{k=1}^{\Ntest} \) per jump distribution, each of which is independently synthesized by first drawing \gls{iid} increments from the respective jump distribution and forming the signals via~\cref{eq:cumsum}.
It then proceeds to synthesize the \( \Ntest \) measurements (\textit{i.e.} we use one noise instance per signal) denoted \( \{ \meas^{\mathrm{test}}_k \}_{k=1}^{\Ntest} \) according to~\cref{eq:inverse_problem} and, for each of the measurements, computes the gold-standard posterior samples of the various inverse problems via the Gibbs methods described in \cref{ssec:efficient posterior sampling}.
This stage is off-line (no reverse-diffusion loop) and trivially parallel across the measurements, which allows us to run long chains with burn-in periods of \num{1e5} iterations and obtain \num{2e5} draws from the posterior distribution.
This far exceeds any values reported in~\cite{kuric2025gaussianlatentmachineefficient} or~\cite{bohra_20203_statistical} and results in precise \gls{mmse} estimates.

The dataset generation stage also involves the generation of \( \Ntrain \) training signals \( \{ \signal^{\mathrm{train}}_k \}_{k=1}^{\Ntrain} \) and \( \Nvalidation \) validation signals (mutually disjoint from the test signals) \( \{ \signal^{\mathrm{val}}_k \}_{k=1}^{\Nvalidation} \) and the corresponding validation measurements \( \{ \meas^{\mathrm{val}}_k \}_{k=1}^{\Nvalidation} \).
The training signals are used for the learning of a neural score function like those that are used for the resolution of inverse problems when the prior is unknown or too expensive to evaluate.
The validation signals are used to monitor the performance of the neural score function on unseen signals during the training stage and to tune the regularization parameters for the model-based approaches as well as the parameters of the \gls{dps} algorithms, see~\cref{sssec:model based methods} and~\cref{sssec:dps algorithms}.

Unlike for the computation of the gold-standard \gls{mmse} estimate of the initial inverse problem, the denoising posteriors are sampled \( T \) times per trajectory (we use \( T = \num{1000} \)).
To ensure acceptable runtimes in this setting, we therefore pick the smallest burn-in period and sample count that still yield accurate estimates of the required statistics.
We determine these settings with a rigorous protocol that is detailed in~\cref{sec:burn in and samples protocol}.
Ultimately, this protocol resulted in the choice of a burn-in period of \num{100} iterations and a sample count of \num{300}.

\subsubsection{Practical Gibbs Implementations}
Sampling \( \sigrv \mid \mathbf{Z} \) in the \gls{glm} and \( \mathbf{U} \mid \mathbf{V}, \mathbf{W}, \mathbf{Y} \) for the Bernoulli--Laplace case reduces to drawing from a high-dimensional Gaussian, which is a well-studied problem.
For settings that necessitate a matrix-free implementation such as those that are commonly encountered in imaging applications, the authors of~\cite{kuric2025gaussianlatentmachineefficient} advocate a Perturb-and-MAP sampler with preconditioned conjugate gradient solvers.
For our moderate-dimensional problems with \( \Sigdim = \num{64} \), a standard implementation based on the Cholesky factorization of the covariance matrix offered significantly faster (approximately one order of magnitude) sampling.
The sampling of the different latent variables necessitates the sampling of the one-dimensional conditional latent distributions.
All the conditional latent distributions that are relevant in this paper admit efficient samplers that are readily available in standard scientific computing packages or can be implemented with little effort.
We reuse the CUDA implementation of the generalized inverse Gaussian sampler from~\cite{kuric2025gaussianlatentmachineefficient} that implements the method proposed in~\cite{Devroye2012} and rely on \texttt{pytorch}~\cite{paszke2017automatic} for all others.
Wherever possible, latent updates are parallelized.

In the Gibbs methods for the Bernoulli--Laplace jumps, the sequential drawing of the binary support vector \( \mathbf{V} \) is embedded in the outer Gibbs loop, which, in turn, may be embedded in the reverse-diffusion loop.
This makes it crucial to minimize the use of heavy linear algebra operations to achieve acceptable runtimes.
Writing  \( \mathbf{B}(\mathbf{v}, \mathbf{w}) = \noisevar^2 \mathbf{I} + \mathbf{H}\operatorname{diag}(\mathbf{v} \odot \mathbf{w}) \mathbf{H}^T \), we recognize that flipping the \( k \)th bit of \( \mathbf{v} \) adds or removes a rank-one term \( \mathbf{w}_k \mathbf{H}_k\mathbf{H}_k^T \), where \( \mathbf{H}_k \) is the \( k \)th column of \( \mathbf{H} \). 
Using the matrix determinant lemma and Woodbury--Sherman--Morrison, we update
\begin{equation}
	\log|\mathbf{B}(\mathbf{v}_{k=1}, \mathbf{w})| = \log|\mathbf{B}(\mathbf{v}_{k=0}, \mathbf{w})| + \log(1 + \mathbf{w}_k \tau_k)
\end{equation}
and
\begin{equation}
	\begin{aligned}
		\meas^T\mathbf{B}(\mathbf{v}_{k=1}, \mathbf{w})^{-1}\meas =\ &\meas^T\mathbf{B}(\mathbf{v}_{k=0}, \mathbf{w})^{-1}\meas \\
																	 &- \frac{\mathbf{w}_k (\mathbf{H}_k^T \mathbf{B}(\mathbf{v}_{k=0}, \mathbf{w})^{-1} \meas)^2}{1 + \mathbf{w}_k\tau_k},
	\end{aligned}
\end{equation}
where \( \tau_k = \mathbf{H}_k^T\mathbf{B}(\mathbf{v}_{k=0}, \mathbf{w})^{-1}\mathbf{H}_k \).
Thus, an efficient implementation factors \( \mathbf{B}(\mathbf{v}, \mathbf{w}) \) once per latent state, obtains the needed scalars via triangular solves, and performs rank-one updates as bits flip.

\begin{table*}[t]
    \centering
	\scalebox{0.93}{%
	\begin{tabular}{lll}
		\toprule
		\gls{cdps} & \gls{diffpir} & \gls{dpnp} \\
		\midrule
		\( \Theta = \{(1-\bar{\alpha}_t)/\bar{\alpha}_t\}_{t=1}^T, \zeta \) & \( \Theta = \{(1-\bar{\alpha}_t)/\bar{\alpha}_t\}_{t=1}^T,\lambda, \zeta \) & \( \Theta = \{\eta_t\}_{t=1}^T \)\\
		\midrule
		\(
		\begin{aligned}
		& \hat{\signal}_{0} = \dfrac{1}{S} \sum_{k=1}^S \bar{\signal}_k \\
		& \mathbf{C} = \dfrac{1}{S} \sum_{k=1}^S (\bar{\signal}_k - \hat{\signal}_{0})(\bar{\signal}_k - \hat{\signal}_{0})^T \\
		& \signal'_{t-1} = \dfrac{\sqrt{\bar{\alpha}_t} (1-\bar{\alpha}_{t-1})}{1-\bar{\alpha}_t} \signal_{t} + \dfrac{\sqrt{\bar{\alpha}_{t-1}} \beta_t}{1-\bar{\alpha}_t} \hat{\signal}_{0} + \sigma_t z \\
		& \signal_{t-1} = \signal'_{t-1} - \zeta_t \dfrac{\sqrt{\bar{\alpha}_t}}{1-\bar{\alpha}_t} \mathbf{C}^T \op^T (\op \hat{\signal}_{0} - \meas)\\
		& \signal_{t-1} = \signal_{t - 1}/\sqrt{\bar{\alpha}_{t-1}} \\
		\end{aligned}
		\) &
		\(
		\begin{aligned}
		& \hat{\signal}_{0} = \dfrac{1}{S} \sum_{k=1}^S \bar{\signal}_k \\
		& \rho_t = \lambda \noisevar^2 / \sigma_t^2 \\
		& \bar{\signal}_{0} = \argmin_{\signal} \bigl( \tfrac{1}{2} \| \op\signal - \meas \|^2 + \tfrac{\rho_t}{2} \| \signal - \hat{\signal}_{0} \|^2 \bigr)\\
		& \hat{\epsilon} = \dfrac{1}{\sqrt{1-\bar{\alpha}_t}} (\signal_{t} - \sqrt{\bar{\alpha}_t} \bar{\signal}_{0}) \\
		& \signal_{t-1}^\prime = \sqrt{\bar{\alpha}_{t-1}} \bar{\signal}_{0} + \sqrt{1 - \bar{\alpha}_{t-1}} (\sqrt{1 - \zeta} \hat{\epsilon} + \sqrt{\zeta} z)\\
		& \signal_{t-1} = \signal_{t-1}^\prime/\sqrt{\bar{\alpha}_{t-1}} \\
		\end{aligned}
		\) &
		\(
		\begin{aligned}
		& \signal_{0} = \bar{\signal}_{1} \\
		& \signal_{t - 1} \sim \exp\Bigl(-\tfrac{1}{2}\|\op\cdot - \meas\|^2 - \tfrac{1}{2 \eta_t^2} \| \cdot - \signal_{0} \|^2 \Bigr)
		\end{aligned}
		\) \\
		\bottomrule
\end{tabular}}
	\caption{%
		Instantiations of the update step \( \mathcal{S}(\signal_t, \{ \bar{\signal}_k \}_{k=1}^S, \meas, \op, \Theta, t) \) and description of the algorithm parameters \( \Theta \) for \gls{cdps}, \gls{diffpir}, and \gls{dpnp}.
		Each \( z \) is a \( d \)-dimensional random vector with \gls{iid} standard Gaussian entries.
	}%
    \label{tab:algorithms_Steps}
\end{table*}

\section{Numerical Experiments}
\subsection{Forward Operators}
We consider four forward operators that are frequently encountered in various estimation tasks throughout the natural sciences.
First, the identity \( \op = \mathbf{I} \in \R^{d \times d} \).
This choice is motivated by the fundamental role that denoising algorithms currently play in many restoration algorithms and even labeling problems such as edge detection~\cite{le2025}.
Second, a convolution operator \( \op \in \R^{d\times d} \) that implements the convolution with a kernel that consists of the \num{13} central samples of a truncated Gaussian with variance \num{2} and is normalized to unit sum.
We adopt circular boundary conditions to enable a fast computation of the proximal map that arises in the update step of \gls{diffpir} (see~\cref{tab:algorithms_Steps}) via the fast Fourier transform.
Deconvolution is a relevant problem with applications in, \textit{e.g.}, microscopy and astronomy.
Third, a sampling operator \( \op \in \R^{m \times d} \) that returns \( m < d \) entries of its argument unchanged.
This operator is also relevant in many fields such as image reconstruction and time-series forecasting.
In particular, a forecasting or prediction problem can be modeled by returning the first \( m \) known entries recovering the remaining \( (d - m) \) entries through the resolution of the inverse problem.
In our experiments, each entry has an independent chance of \qty{40}{\percent} being kept.
Fourth, an operator \( \op = \mathbf{MF} \in \R^{m \times d} \) where \( \mathbf{F} \in \R^{2(\lfloor d/2 \rfloor + 1)\times d} \) is the matrix representation of the \enquote{real} one-dimensional discrete Fourier transform with separated real and imaginary components, and \( \mathbf{M} \in \R^{m \times 2(\lfloor d/2 \rfloor + 1)} \) is a sampling operator.
Such operators are relevant in, \textit{e.g.}, medical imaging and astronomy.
The sampling operator is constructed such that the \num{5} lowest frequencies (including the DC term) are acquired, and the remaining frequencies independently have a \qty{40}{\percent} chance of being kept.

For all operators, the noise variance \( \noisevar^2 \) is chosen such that the median measurement \gls{snr} is around \qty{25}{\decibel}.
We set \( \Ntrain = \num{1e6} \), \( \Nvalidation = \num{1e3} \), and \( \Ntest = \num{1e3} \).

\subsection{Reconstruction Algorithms}
We now describe the reconstruction algorithms that we compare and how we obtain suitable parameters.
For the sake of simplicity, the parameter estimation procedures are given for a general jump distribution and forward operator.
The parameters are estimated independently for each jump distribution and forward operator.

\subsubsection{Model-Based Methods}%
\label{sssec:model based methods}
As baseline reconstruction algorithms we consider the model-based methods
\begin{equation}
    \estimate^{\ell_2}(\meas, \lambda) = \argmin_{\signal \in \sigdim}\, \bigl( \tfrac{1}{2} \norm{\op\signal - \meas}^2 + \lambda \norm{\fdop \signal}^2 \bigr),
\end{equation}
and
\begin{equation}
    \estimate^{\ell_1}(\meas, \lambda) = \argmin_{\signal \in \sigdim}\, \bigl( \tfrac{1}{2} \norm{\op\signal - \meas}^2 + \lambda \norm{\fdop \signal}_1 \bigr),
\end{equation}
which coincide with the \gls{map} estimators of \levy{} processes associated with Laplace Gaussian and Laplace distributions, respectively.
The adjustable regularization parameter for the method \( \mathrm{est} = \ell_2, \ell_1 \) was found by
\begin{equation}
	\lambda^{\mathrm{est},\star} = \argmin_{\lambda \in \Lambda} \frac{1}{\Nvalidation} \sum_{k=1}^{\Nvalidation} \tfrac{1}{d}\norm{\estimate^{\mathrm{est}}(\meas_{k}^{\mathrm{val}}, \lambda) - \signal_k^{\mathrm{val}}}^2,
\end{equation}
where \( \Lambda \) is a finite set of real numbers.
\( \Lambda \) is detailed~\cref{ssec:parameter identification} and the obtained \gls{mse} curves are shown in~\cref{fig:grid search}.

\subsubsection{Diffusion Posterior Sampling Algorithms}%
\label{sssec:dps algorithms}
We consider three \gls{dps} algorithms that are popular in the literature.
First, the \gls{cdps} algorithm due to Chung et al.\ \cite{chung2023diffusion}, which was one of the first algorithms that was proposed for the resolution of general noisy inverse problems with diffusion priors.
Second, the \gls{diffpir} algorithm due to Zhu et al.\ \cite{zhu2023denoising} that can be regarded as an extension of the \gls{cdps} algorithm and typically reports superior results in standard perception-based evaluations.
Third, the \gls{dpnp} algorithm due to Xu et al.\ \cite{xu2024provably} that alternates between sampling the denoising subproblem and a data-proximal subproblem.
We include the \gls{dpnp} algorithm to showcase the broad applicability of our framework to nonstandard setups that utilize various statistics of the denoising problem.

For each \gls{dps} algorithm, we benchmark two variants: One where the sampling of the denoising problem is done with the gold-standard Gibbs methods (\enquote{oracle} denoiser) and statistics are computed from those samples, and one where the sampling (or the direct estimation of any point estimate) is done with learned components.
For the second case, a noise-conditional neural network with UNet architecture (\num{305761} learnable parameters) is trained in an off-line step on the \( \Ntrain \) training signals in a standard setup (Adam optimizer with learning rate \num{1e-4} with exponential decay with factor \num{0.9999}, \num{100000} parameter updates, batch size \num{10000}).
The noise schedule in \gls{cdps} and \gls{diffpir} is defined by the two endpoints \( \beta_0 = \num{1e-4} \) and \( \beta_T = \num{2e-2} \) with linear equidistant samples in-between.
The learned variant of \gls{dpnp} is the \enquote{DDS-DDPM} variant~\cite[Algorithms 1 and 3]{xu2024provably} that contains an inner denoising-sampling loop.
The oracle variant does not require an inner loop at all (except for the burn-in period), which makes the oracle variant the faster one for this case.

Before advancing, we introduce some notation.
For any given measurements \( \meas \), any \gls{dps} algorithm \( \mathrm{alg} \) that depends on any parameters \( \bm{\theta} \) produces samples denoted \( \{ \estimate^{\mathrm{alg}}_k (\meas, \bm{\theta}) \}_{k=1}^{N_{\mathrm{samples}}} \).
We denote \( \estimate^{\mathrm{alg}}_{\mathrm{MMSE}}(\meas, \bm{\theta}) \coloneqq \frac{1}{\Nsamples} \sum_{k=1}^{\Nsamples}\estimate^{\mathrm{alg}}_k (\meas, \bm{\theta}) \).
We tune the parameters of the methods by finding
\begin{equation}
	\bm{\theta}^{\mathrm{alg},\star} = \argmin_{\bm{\theta} \in \bm{\Theta}^{\mathrm{alg}}} \sum_{k=1}^{\Nvalidation} \norm{\estimate^{\mathrm{alg}}_{\mathrm{MMSE}}(\meas_{k}^{\mathrm{val}}, \bm{\theta}) - \signal_k^{\mathrm{val}}}^2
\end{equation}
where the grid \( \bm{\Theta}^{\mathrm{alg}} \) is method-dependent.
Note that this tuning is specifically tailored towards the evaluation with respect to the \gls{mmse} optimality gap.
Due to resource constraints, the parameters are tuned for the learned denoiser.
The parameter grids for the different methods is detailed in~\cref{ssec:parameter identification} and the corresponding \gls{mse} curves are shown in~\cref{fig:grid search}.
We use \( N_{\mathrm{samples}} = 10 \) for the grid search on the validation set and \( N_{\mathrm{samples}} = 50 \) for the experiments on the test set.

\subsubsection{Gold-Standard Gibbs Methods}
The Gibbs methods are used to obtain gold-standard samples from the posterior.
As described in~\cref{ssec:efficient posterior sampling}, the Gibbs methods are parameter-free and efficient and, consequently, well-suited for this purpose.
Chain lengths, diagnostics, and implementation details are given in \cref{ssec:implementation details}; we reuse the same settings across operators and increment families.

\subsection{Results}
For a method \( \mathrm{est} \) and test data \( \meas_k^{\mathrm{test}} \) with corresponding data-generating signal \( \signal_k^\mathrm{test} \) we measure the \gls{mmse} optimality gap (in decibel) defined by
\begin{equation}
	10\log_{10}\biggl( \frac{\norm{\estimate^{\mathrm{est}}(\meas_k^\mathrm{test}) - \signal_k^\mathrm{test}}^2}{\norm{\estimate^{\mathrm{Gibbs}}_{\mathrm{MMSE}}(\meas_k^{\mathrm{test}}) - \signal_k^\mathrm{test}}^2} \biggr)
	\label{eq:optimality gap}
\end{equation}
where \( \estimate^{\mathrm{est}}(\meas) = \estimate^{\mathrm{est}}(\meas, \lambda^{\mathrm{est},\star}) \) for model-based methods and \( \estimate^{\mathrm{est}}_{\mathrm{MMSE}}(\meas, \bm{\theta}^{\mathrm{est},\star}) \) for \gls{dps} algorithms.
The \gls{mmse} optimality gap shows the degree of optimality of a method at a glance without the need to compare to a reference:
A gap of \num{0} indicates a perfect recovery of the gold-standard \gls{mmse} estimate and the positive nonzero values show the orders of magnitude of the error relative to the reference error.

We report the mean and standard deviation of the \gls{mmse} optimality gap defined in~\cref{eq:optimality gap} for the learned denoiser over the entire test set in~\cref{tab:optimality gap}.
The Gaussian jump distribution validates the implementation:
Since the \gls{mmse} and the \gls{map} point estimates coincide, the model-based \( \ell_2 \) estimator matches the Gibbs reference up to the error due to the finite grid resolution.
When the posterior mean is smooth (\textit{e.g.}, imputation and some deconvolution cases), \( \ell_2 \) is the best model-based choice and frequently beats the \gls{dps} algorithms.
When the posterior mean is (close to) piecewise constant (typical in denoising of signals with sparse jumps), the \( \ell_1 \) estimator is preferred.
Among \gls{dps} algorithms, \gls{diffpir} is typically the top performer and exceeds \( \ell_2 \) and \( \ell_1 \) baselines in deconvolution, imputation, and reconstruction from partial Fourier measurements.
For spike-and-slab settings (Bernoulli--Laplace), \gls{dps} algorithms substantially outperform the model-based baselines across operators.
In deconvolution, and reconstruction from partial Fourier measurements, \gls{dps} algorithms frequently match or surpass the best model-based estimator, with \gls{dpnp} and \gls{diffpir} alternating as the strongest.
\begin{table*}
	\centering
	\begin{tabular}{ll%
			*{2}{S[table-format=1.2 +- 1.2,separate-uncertainty]}
			*{2}{S[table-format=2.2 +- 1.2,separate-uncertainty]}
			*{2}{S[table-format=1.2 +- 1.2,separate-uncertainty]}
		}
		\toprule
		& & {\( \GaussianDist(0.25) \)} & {\( \LaplaceDist(1) \)} & {\( \BLDist(0.1, 1) \)} & {\( \tDist(1) \)} & {\( \tDist(2) \)} & {\( \tDist(3) \)} \\
    \midrule
    \multirow{5}{*}{Denoising} & C-DPS & 0.12 \pm 0.18 & 0.12 \pm 0.20 & 2.22 \pm 2.26 & 3.26 \pm 1.01 & 0.28 \pm 0.30 & 0.10 \pm 0.18 \\
    & DiffPIR & 0.16 \pm 0.21 & 0.09 \pm 0.16 & 0.72 \pm 1.10 & 0.93 \pm 1.06 & 0.07 \pm 0.14 & 0.15 \pm 0.21 \\
    & DPnP & 0.24 \pm 0.25 & 0.11 \pm 0.17 & 1.33 \pm 2.12 & 1.19 \pm 1.38 & 0.10 \pm 0.17 & 0.10 \pm 0.17 \\
    & L1 & 0.15 \pm 0.21 & 0.06 \pm 0.12 & 3.44 \pm 2.38 & 0.38 \pm 0.43 & 0.14 \pm 0.19 & 0.11 \pm 0.18 \\
    & L2 & 0.00 \pm 0.01 & 0.16 \pm 0.21 & 8.61 \pm 3.10 & 3.25 \pm 0.99 & 0.74 \pm 0.83 & 0.25 \pm 0.33 \\
    \midrule
    \multirow{5}{*}{Deconvolution} & C-DPS & 0.12 \pm 0.20 & 0.12 \pm 0.23 & 4.30 \pm 3.87 & 18.30 \pm 5.28 & 0.46 \pm 1.40 & 0.17 \pm 0.53 \\
    & DiffPIR & 0.07 \pm 0.17 & 0.07 \pm 0.19 & 1.09 \pm 2.22 & 10.45 \pm 6.10 & 0.09 \pm 0.57 & 0.08 \pm 0.26 \\
    & DPnP & 0.10 \pm 0.18 & 0.13 \pm 0.22 & 1.71 \pm 2.49 & 7.84 \pm 5.66 & 0.35 \pm 1.39 & 0.14 \pm 0.41 \\
    & L1 & 1.65 \pm 0.84 & 1.38 \pm 0.86 & 1.86 \pm 3.14 & 1.87 \pm 4.01 & 1.10 \pm 1.19 & 1.28 \pm 0.94 \\
    & L2 & 0.00 \pm 0.01 & 0.07 \pm 0.23 & 6.11 \pm 4.49 & 21.50 \pm 4.46 & 1.44 \pm 2.85 & 0.36 \pm 1.09 \\
    \midrule
    \multirow{5}{*}{Imputation} & C-DPS & 0.15 \pm 0.29 & 0.18 \pm 0.39 & 2.99 \pm 2.82 & 23.33 \pm 8.69 & 0.50 \pm 1.09 & 0.14 \pm 0.57 \\
    & DiffPIR & 0.09 \pm 0.23 & 0.08 \pm 0.24 & 0.24 \pm 1.14 & 0.88 \pm 3.50 & 0.11 \pm 0.62 & 0.08 \pm 0.42 \\
    & DPnP & 0.14 \pm 0.32 & 0.17 \pm 0.36 & 0.50 \pm 1.28 & 10.89 \pm 5.92 & 0.25 \pm 0.82 & 0.27 \pm 0.58 \\
    & L1 & 1.74 \pm 1.12 & 1.77 \pm 1.35 & 1.25 \pm 2.78 & 13.32 \pm 5.32 & 1.37 \pm 2.56 & 1.55 \pm 1.58 \\
    & L2 & -0.00 \pm 0.01 & 0.01 \pm 0.05 & 1.10 \pm 1.88 & 0.42 \pm 0.95 & 0.06 \pm 0.34 & 0.02 \pm 0.28 \\
    \midrule
    \multirow{5}{*}{Fourier} & C-DPS & 0.15 \pm 0.36 & 0.26 \pm 0.65 & 5.90 \pm 4.41 & 4.29 \pm 5.78 & 0.53 \pm 0.83 & 0.35 \pm 0.77 \\
    & DiffPIR & 0.11 \pm 0.29 & 0.08 \pm 0.31 & 0.83 \pm 1.44 & 3.19 \pm 4.37 & 0.11 \pm 0.39 & 0.12 \pm 0.37 \\
    & DPnP & 0.11 \pm 0.35 & 0.20 \pm 0.51 & 1.88 \pm 2.47 & 2.45 \pm 4.83 & 0.39 \pm 0.89 & 0.24 \pm 0.64 \\
    & L1 & 1.50 \pm 1.59 & 0.73 \pm 0.94 & 3.57 \pm 2.82 & 1.07 \pm 2.98 & 0.71 \pm 0.99 & 0.78 \pm 0.97 \\
    & L2 & -0.00 \pm 0.02 & 0.36 \pm 0.73 & 12.22 \pm 4.53 & 9.47 \pm 8.34 & 2.66 \pm 3.57 & 1.03 \pm 1.79 \\
    \bottomrule
	\end{tabular}
	\caption{%
		\gls{mmse} optimality gap in decibel (mean \( \pm \) standard deviation; lower is better; \num{0} is a perfect reconstruction) of various estimation methods over the test set.
	}%
	\label{tab:optimality gap}
\end{table*}

In addition to the reconstruction performance obtained with the learned denoisers---for which the parameters of the algorithms wast tuned---we inspect the robustness of the algorithms when substituting the learned denoiser with the oracle denoiser by reporting the difference in the \gls{mmse} optimality gap between the two in~\cref{tab:transfer}.
\Gls{dpnp} is the most robust to swapping the learned denoiser with the oracle denoiser.
Indeed, \gls{dpnp} significantly benefits from the oracle denoiser in the most challenging cases of the spike-and-slab prior and the extremely heavy-tailed t-distribution with degree of freedom parameter \( \nu = 1 \).
By contrast, \gls{cdps} and \gls{diffpir} can require retuning when the denoiser changes.
This sensitivity is most visible under heavy-tailed jumps (\textit{e.g.}, \( \tDist(1) \) jumps):
reusing hyperparameters tuned on the learned denoiser can degrade scores for the oracle denoiser (see, \textit{e.g.}, deconvolution and imputation for \( \BLDist(0.1, 1) \) and \( \tDist(1) \)), whereas a brief hand-tuning on the validation set improves \gls{diffpir} way beyond the learned denoiser (\textit{e.g.}, \( \zeta = 0.7 \) and \( \rho = 4\) decreased the optimality gap by almost \qty{10}{\decibel}).
We did not exhaustively retune all methods for the oracle denoiser due to resource limits; the framework supports this, and we view full oracle-side tuning as a community task.
\Cref{tab:transfer} also reports for which cases the oracle denoiser reports significantly better results than the learned denoiser according to a Wilcoxon signed-rank test (\( p = 0.05 \), \( \Ntest \) pairs, two-sided test with the winner determined by the median of differences).
The differences between the algorithms are generally greater than the differences between the learned/oracle variants except for the heavy-tailed cases, which confirms the findings in~\cite{bohra_20203_statistical} and indicates that the research of efficient and robust \gls{dps} algorithms is still crucial.
\begin{table*}
	\centering
	\begin{tabular}{ll%
			*{3}{S[table-format=-1.2 +- 1.2\textsuperscript{*},separate-uncertainty]}
			*{1}{S[table-format=-2.2 +- 1.2\textsuperscript{*},separate-uncertainty]}
			*{2}{S[table-format=-1.2 +- 1.2\textsuperscript{*},separate-uncertainty]}
		}
		\toprule
		& & {\( \GaussianDist(0.25) \)} & {\( \LaplaceDist(1) \)} & {\( \BLDist(0.1, 1) \)} & {\( \tDist(1) \)} & {\( \tDist(2) \)} & {\( \tDist(3) \)} \\
    \midrule
    \multirow{3}{*}{Denoising} & C-DPS & -0.00 \pm 0.11 & 0.00 \pm 0.16 & -0.46 \pm 1.16\textsuperscript{*} & -0.00 \pm 0.01 & 0.02 \pm 0.79\textsuperscript{*} & -0.01 \pm 0.14 \\
    & DiffPIR & 0.00 \pm 0.13 & 0.00 \pm 0.17 & -0.05 \pm 0.78\textsuperscript{*} & -0.41 \pm 0.80\textsuperscript{*} & -0.00 \pm 0.20 & 0.00 \pm 0.15 \\
    & DPnP & 0.04 \pm 0.27\textsuperscript{*} & -0.01 \pm 0.22 & -0.55 \pm 1.31\textsuperscript{*} & -0.77 \pm 1.31\textsuperscript{*} & -0.00 \pm 0.24 & -0.00 \pm 0.23 \\
    \midrule
    \multirow{3}{*}{Deconvolution} & C-DPS & -0.01 \pm 0.24 & -0.00 \pm 0.26 & 0.09 \pm 0.97\textsuperscript{*} & 6.64 \pm 3.21\textsuperscript{*} & -0.12 \pm 1.11\textsuperscript{*} & -0.03 \pm 0.43 \\
    & DiffPIR & -0.01 \pm 0.23 & -0.00 \pm 0.23 & 0.04 \pm 1.12 & 13.56 \pm 9.90\textsuperscript{*} & -0.01 \pm 0.47 & 0.00 \pm 0.31 \\
    & DPnP & -0.00 \pm 0.25 & -0.01 \pm 0.27\textsuperscript{*} & -0.02 \pm 1.20 & -4.98 \pm 3.86\textsuperscript{*} & 0.06 \pm 0.77 & -0.02 \pm 0.34 \\
    \midrule
    \multirow{3}{*}{Imputation} & C-DPS & 0.00 \pm 0.30 & 0.01 \pm 0.35 & 0.41 \pm 1.51\textsuperscript{*} & 3.41 \pm 4.99\textsuperscript{*} & -0.12 \pm 1.01\textsuperscript{*} & -0.01 \pm 0.57 \\
    & DiffPIR & 0.00 \pm 0.29 & -0.00 \pm 0.33 & 0.03 \pm 1.05 & -0.20 \pm 3.05\textsuperscript{*} & 0.03 \pm 0.71 & 0.00 \pm 0.47 \\
    & DPnP & 0.00 \pm 0.35 & -0.02 \pm 0.38 & -0.02 \pm 1.02 & -10.46 \pm 5.70\textsuperscript{*} & 0.02 \pm 0.67 & -0.01 \pm 0.48 \\
    \midrule
    \multirow{3}{*}{Fourier} & C-DPS & -0.02 \pm 0.43 & -0.01 \pm 0.49 & 0.80 \pm 1.43\textsuperscript{*} & 0.09 \pm 5.63\textsuperscript{*} & -0.03 \pm 0.79\textsuperscript{*} & 0.01 \pm 0.49 \\
    & DiffPIR & -0.01 \pm 0.39 & -0.00 \pm 0.40 & 0.12 \pm 0.83\textsuperscript{*} & -0.64 \pm 1.70\textsuperscript{*} & -0.03 \pm 0.42\textsuperscript{*} & -0.02 \pm 0.38 \\
    & DPnP & -0.01 \pm 0.43 & -0.00 \pm 0.45 & -0.33 \pm 1.13\textsuperscript{*} & -1.32 \pm 3.18\textsuperscript{*} & -0.00 \pm 0.54 & 0.01 \pm 0.46 \\
    \bottomrule
	\end{tabular}
	\caption{%
		Change in \gls{mmse} optimality gap (mean \( \pm \) standard deviation) after substituting the learned denoiser with the oracle denoiser.
		An asterisk indicates a significant changes according to a Wilcoxon signed-rank test (\( p = 0.05 \)):
		Negative number with asterisk: oracle denoiser is significantly better;
		positive number with asterisk: learned denoiser is significantly better.
	}%
	\label{tab:transfer}
\end{table*}

\begin{figure*}
	\centering
	\def\prefix{./data/posterior_new}
	\begin{tikzpicture}
		\begin{groupplot}[posteriorplot]
			\plotposterior{bernoulli-laplace/p=0.1_b=1.0}{convolution}{gibbs}{1}
			\plotposterior{student/1.0}{convolution}{gibbs}{0}
			\plotposterior{student/2.0}{convolution}{gibbs}{0}
			\plotposterior{laplace/1.0}{convolution}{gibbs}{0}
		\end{groupplot}

		\node[rotate=90] at ($(G c1r1.west)+(-20pt, 0pt)$) {\scriptsize \( \BLDist(0.1, 1) \)};
		\node[rotate=90] at ($(G c1r3.west)+(-20pt, 0pt)$) {\scriptsize \( \tDist(1) \)};
		\node[rotate=90] at ($(G c1r5.west)+(-20pt, 0pt)$) {\scriptsize \( \tDist(2) \)};
		\node[rotate=90] at ($(G c1r7.west)+(-20pt, 0pt)$) {\scriptsize \( \LaplaceDist(1) \)};
	\end{tikzpicture}
	\caption{%
		Qualitative results for deconvolution using the oracle denoising sampler.
		Rows: Jump distributions.
		For each jump distribution, the \gls{mmse} estimates obtained by the different \gls{dps} algorithms and the gold-standard Gibbs methods are shown on top of the corresponding index-wise marginal variances.
		Columns: Different measurements.
	}%
	\label{fig:results deconvolution gibbs}
\end{figure*}
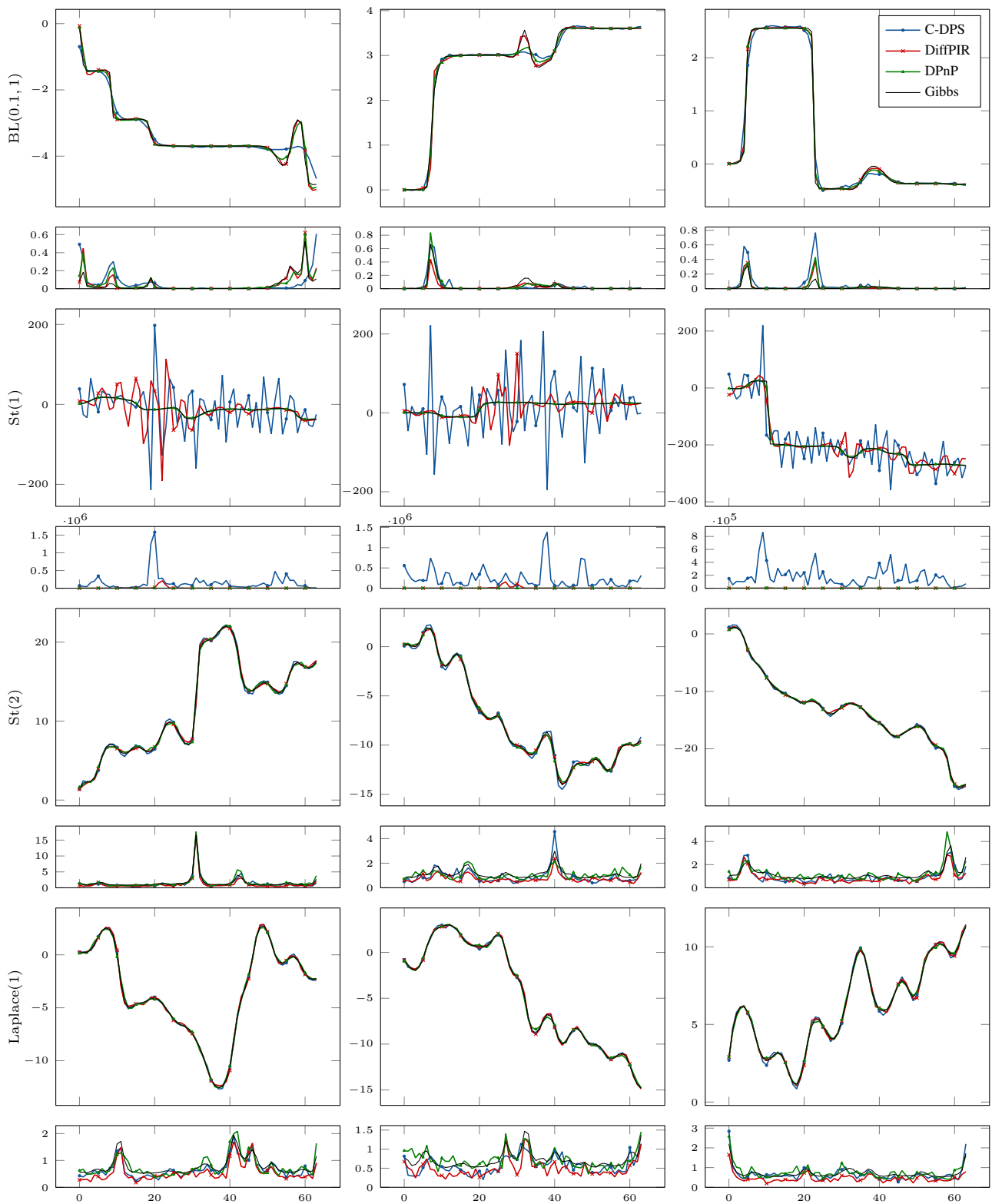

Uncurated examples of the \gls{mmse} estimates and the marginal variances obtained by the \gls{dps} algorithms and the gold-standard Gibbs methods for deconvolution with the oracle denoiser for \( \BLDist(0.1, 1) \), \( \tDist(1) \), \( \tDist(2) \), and \( \LaplaceDist(1) \) jump distributions are shown in~\cref{fig:results deconvolution gibbs}.
The learned counterpart and the results for denoising, imputation, and reconstruction from partial Fourier measurements are provided in~\cref{fig:results deconvolution learned,fig:results denoising gibbs,fig:results denoising learned,fig:results imputation gibbs,fig:results imputation learned,fig:results fourier gibbs,fig:results fourier learned}.
In all of those figures, we deliberately omit the underlying signal to emphasize that the targets are the \gls{mmse} estimate and the marginal variance that are obtained by the gold-standard Gibbs methods.

\Cref{fig:example reconstruction with trajectory student gibbs} shows a typical conditional reverse-diffusion trajectory (here obtained by \gls{diffpir} for deconvolution and a \( \BLDist(0.1, 1) \) jump distribution) that highlights a key distinction:
Posterior \emph{samples} often preserve high-frequency structure and reflect prior variability, whereas the \emph{\gls{mmse} point estimate}---obtained by averaging all samples---is much smoother.
This explains why \gls{dps} methods tend to score higher on preception-oriented metrics, while regressors trained with an \gls{mse} loss target the \gls{mmse} and therefore excel on metrics distortion metrics (\textit{e.g.}, \gls{psnr}, \gls{mse}).
Consistent with this distinction,~\cite{chitwan} fairly compare a sampling-based method to an \gls{mmse} regressor and find the expected trade-off: higher \gls{psnr} and \gls{ssim} for the regressor and better perceptual scores for the sampler.
We therefore recommend making the Bayesian target explicit---point estimate versus sample quality---and using evaluation protocols that are aligned to that target.
Our framework supports this by offering gold-standard posterior samples and oracle denoisers.

In addition to the evaluation of the \gls{mmse} optimality gap, which is on the point-estimator level, we analyze the high-posterior density coverage of the algorithms.
Specifically, for any datum \( \meas \) and any \( k = 1, \dotsc, \Nsamples \), denote \( l_k(\meas) \coloneqq \log \posterior(\estimate^{\mathrm{alg}}_{P(k)} (\meas, \bm{\theta}^{\mathrm{alg},\star})) \)\footnote{%
	With slight abuse of notation, \( \log \posterior \) is the unnormalized ground-truth log-posterior~\cref{eq:posterior};
	the additive constant is the same across all methods so ranking is valid.
} where \( P \) is the permutation that ensures that  \( l_1(\meas) \geq l_2(\meas) \geq \cdots \geq l_{\Nsamples}(\meas) \) and define the empirical highest-posterior-density threshold at \( \alpha \in [0, 1] \) as \( l_{\lceil \alpha \Nsamples \rceil}(\meas) \).
We declare the data-generating signal \( \signal \) covered if \( \log \posterior(\signal) \geq  l_{\lceil \alpha \Nsamples \rceil}(\meas) \) and define the coverage of a method as
\begin{equation}
	\frac{1}{\Ntest} \sum_{k=1}^{\Ntest} \chi_{\R_{\geq l_{\lceil \alpha \Nsamples \rceil}(\meas^{\mathrm{test}}_k)}}\bigl(\log p_{\sigrv\mid\mathbf{Y}=\meas_k^{\mathrm{test}}}(\signal_{k}^{\mathrm{test}})\bigr).
\end{equation}
The coverage of a sampling method that really draws samples from the posterior will be \( \alpha \) up to Monte Carlo error.
A coverage result that is significantly less than \( \alpha \) indicates that the samples obtained by the method concentrate too heavily around the mode; a coverage result that is greater than \( \alpha \) indicates that the samples are too spread out.
The coverage results for \( \alpha = 0.9 \) are presented in~\cref{tab:coverage}.
The Gibbs row again validates the implementation; for all forward operators, they achieve coverages that are very close to \num{0.9}.
In contrast, the coverage values obtained by the \gls{dps} algorithms are generally much smaller than \num{0.9}.
For \gls{cdps} and \gls{diffpir}, the reported coverage values are almost always \num{0} except for \( \BLDist(0.1, 1) \) and \( \tDist(1) \) jumps, where the coverages are usually (close to) \num{1} for \gls{cdps} and inconsistent for \gls{diffpir}.
For almost all jump distributions and forward operators, \gls{dpnp} reports coverage values that are closest to but typically smaller than \num{0.9}.\footnote{Note that a coverage of \num{1} can be considered the worst case even at a target of \num{0.9}. For instance, it is achieved by setting all samples to a constant vector with extremely large (i.e.\ \enquote{unlikely}) entries.}
\begin{table*}
    \centering
	\setlength{\tabcolsep}{5pt}
\begin{tabular}{ll*{12}{S[round-mode=places,round-precision=2,table-format=1.2]}}
    \toprule
	 & & \multicolumn{2}{c}{\( \GaussianDist(0, 0.25) \)} & \multicolumn{2}{c}{\( \LaplaceDist(1) \)} & \multicolumn{2}{c}{\( \BLDist(0.1, 1) \)} &  \multicolumn{2}{c}{\( \tDist(1) \)} &  \multicolumn{2}{c}{\( \tDist(2) \)} &  \multicolumn{2}{c}{\( \tDist(3) \)} \\
     \cmidrule(lr){3-4}\cmidrule(lr){5-6}\cmidrule(lr){7-8}\cmidrule(lr){9-10}\cmidrule(lr){11-12}\cmidrule(lr){13-14}
     & & {Learned} & {Oracle} & {Learned} & {Oracle} & {Learned} & {Oracle} & {Learned} & {Oracle}& {Learned} & {Oracle}& {Learned} & {Oracle} \\
    \midrule
	\multirow{4}{*}{Denoising} & Gibbs & {---} & 0.90 & {---} & 0.91 & {---} & 0.91 & {---} & 0.89 & {---} & 0.91 & {---} & 0.89 \\
    & C-DPS & 0.00 & 0.00 & 0.00 & 0.00 & 1.00 & 1.00 & 1.00 & 1.00 & 0.00 & 0.00 & 0.00 & 0.00 \\
    & DiffPIR & 0.00 & 0.00 & 0.00 & 0.00 & 1.00 & 1.00 & 0.28 & 0.02 & 0.00 & 0.00 & 0.00 & 0.00 \\
    & DPnP & 0.58 & 0.67 & 0.11 & 0.11 & 1.00 & 0.41 & 0.53 & 0.08 & 0.09 & 0.09 & 0.09 & 0.10 \\
    \midrule
	\multirow{4}{*}{Deconvolution} & Gibbs & {---} & 0.89 & {---} & 0.90 & {---} & 0.90 & {---} & 0.91 & {---} & 0.91 & {---} & 0.91 \\
    & C-DPS & 0.00 & 0.00 & 0.01 & 0.00 & 1.00 & 1.00 & 1.00 & 0.83 & 0.01 & 0.00 & 0.00 & 0.00 \\
    & DiffPIR & 0.00 & 0.00 & 0.00 & 0.00 & 1.00 & 1.00 & 0.97 & 0.92 & 0.00 & 0.00 & 0.00 & 0.00 \\
    & DPnP & 0.12 & 0.12 & 0.06 & 0.07 & 1.00 & 0.31 & 0.50 & 0.06 & 0.06 & 0.06 & 0.07 & 0.06 \\
    \midrule
	\multirow{4}{*}{Imputation} & Gibbs & {---} & 0.89 & {---} & 0.90 & {---} & 0.86 & {---} & 0.91 & {---} & 0.91 & {---} & 0.91 \\
    & C-DPS & 0.00 & 0.00 & 0.00 & 0.00 & 1.00 & 1.00 & 0.94 & 0.78 & 0.15 & 0.15 & 0.00 & 0.00 \\
    & DiffPIR & 0.00 & 0.00 & 0.00 & 0.00 & 1.00 & 1.00 & 0.72 & 0.32 & 0.00 & 0.00 & 0.00 & 0.00 \\
    & DPnP & 0.28 & 0.31 & 0.09 & 0.08 & 1.00 & 0.41 & 0.56 & 0.07 & 0.14 & 0.13 & 0.12 & 0.13 \\
    \midrule
	\multirow{4}{*}{Fourier} & Gibbs & {---} & 0.91 & {---} & 0.90 & {---} & 0.90 & {---} & 0.91 & {---} & 0.92 & {---} & 0.91 \\
    & C-DPS & 0.00 & 0.00 & 0.00 & 0.00 & 1.00 & 1.00 & 0.96 & 0.74 & 0.01 & 0.01 & 0.00 & 0.00 \\
    & DiffPIR & 0.00 & 0.00 & 0.00 & 0.00 & 1.00 & 1.00 & 0.92 & 0.65 & 0.00 & 0.01 & 0.00 & 0.00 \\
    & DPnP & 0.19 & 0.19 & 0.08 & 0.06 & 1.00 & 0.32 & 0.50 & 0.06 & 0.07 & 0.07 & 0.07 & 0.06 \\
    \bottomrule
\end{tabular}
	\caption{%
		Posterior coverage of various estimation methods at \( \alpha = 0.9 \).
	}%
    \label{tab:coverage}
\end{table*}

\section{Conclusion}
We introduced a framework for the objective benchmarking of diffusion posterior sampling algorithms.
The framework relies on the construction of signals with known distribution, the simulation of the measurement process, and the subsequent generation of samples from the posterior distribution that arises through the combination of the known prior and the known likelihood.
Gold-standard samples from that distribution are then acquired by efficient Gibbs methods and these samples can be compared to those obtained by the diffusion posterior sampling algorithms.
In addition, the Gibbs methods can serve as oracle \gls{mmse} denoisers in the denoising problems that are encountered in each iteration of the reverse \gls{sde}.
Consequently, the framework also enables the quantification of the additional errors that are incurred through any suboptimal learned components.

We provided numerical results for three common posterior sampling algorithms applied to four common inverse problems and invite other researcher to benchmark their algorithms on our open implementation that is deliberately designed such that novel \gls{dps} algorithms can be benchmarked in a plug-and-play manner.
A common theme among all tested algorithms is that the samples they produce are not calibrated, which demonstrates that research for algorithms that perform better in this respect is still crucial.
\printbibliography%
\appendix%

\subsection{Connection between the Reverse SDE and DDPM Sampling}%
\label{sec:reverse sde vs ddpm}
The \gls{ddpm} model has been introduced in~\cite{sohl-dickstein15} as a discrete-time Markov chain of length $T$ with Gaussian transitions
\begin{equation}
    p_{\sigrv_t \mid \sigrv_{t-1} = \signal_{t-1}} = \GaussianDist(\sqrt{1-\beta_t}\signal_{t-1}, \beta_t \mathbf{I})
\end{equation}
such that the transitions from $\sigrv_0$ to $\sigrv_t$ are also tractable as
\begin{align}
    \sigrv_t = \sqrt{\bar{\alpha}_t}\sigrv_0  + \sqrt{1-\bar{\alpha}_t} \mathbf{Z}_t
	\label{ddpm_full}
\end{align}
where $\alpha_t = 1 - \beta_t$, $\bar{\alpha}_t = \prod_{s = 0}^t \alpha_s$ and \( \mathbf{Z}_t \sim \GaussianDist(\mathbf{0}, \mathbf{I}) \).
By definition,
\begin{align}
	\sigrv_{t} = \sqrt{1-\beta_t}\sigrv_{t-1} + \sqrt{\beta_t}\mathbf{Z}_{t - 1},
	\label{eq:ddpm_fwd}
\end{align}
which, by a straightforward application of Tweedie's formula
\begin{equation}
	\mathbb{E}[\sigrv_{t-1}|\sigrv_t] = \dfrac{1}{\sqrt{\alpha}_t} \left( \sigrv_t + (1-\alpha_t) \nabla\log p_{\sigrv_t}(\sigrv_t) \right)
\end{equation}
leads to the backward transitions
\begin{equation}
    \sigrv_{t-1} = \dfrac{1}{\sqrt{1-\beta_t}} \left( \sigrv_{t} + \beta_t\nabla\log p_{\sigrv_t}(\sigrv_t) \right) + \sqrt{\beta_t}\mathbf{Z}_t.
	\label{eq:ddpm_bw}
\end{equation}
The Euler--Maruyama discretization of the backward Ornstein--Uhlenbeck \gls{sde} given in~\cref{eq:BW_vp} and repeated here as
\begin{equation}
    \sigrv_{t-1} = \bigl(1+\tfrac{\beta_t}{2}\bigr)\sigrv_{t} + \beta_t\nabla\log p_{\sigrv_t}(\sigrv_t) + \sqrt{\beta_t}\mathbf{Z}_t
\end{equation}
can be related to the \gls{ddpm} updates via Taylor expansions since
\begin{equation}
	\dfrac{1}{\sqrt{1-\beta_t}} = 1+\frac{\beta_t}{2} + O(\beta_t^2)
\end{equation}
and
\begin{equation}
	\dfrac{\beta_t}{\sqrt{1-\beta_t}} = \beta_t + O(\beta_t^2).
\end{equation}


\subsection{Covariance in DPS}%
\label{ssec:covariance in dps}
\gls{cdps} \cite{chung2023diffusion} uses the approximation of the likelihood
\begin{equation}
	p_{\mathbf{Y} \mid \sigrv_t = \signal}(\meas) \approx p_{\mathbf{Y} \mid \sigrv_0 = \mathbb{E}[\sigrv_0 \mid \sigrv_t = \signal]}(\meas).
\end{equation}
When the noise in the inverse problem is Gaussian, the computation of \( \nabla \log \bigl(\signal \mapsto p_{\mathbf{Y} \mid \sigrv_0 = \mathbb{E}[\sigrv_0 \mid \sigrv_t = \signal]}(\meas) \bigr) \) necessitates the computation of
\begin{equation}
	\nabla \left( \signal \mapsto \tfrac{1}{2}\| \op \mathbb{E}[\sigrv_0 \mid \sigrv_t = \signal] - \meas \|^2 \right),
	\label{grad_dps}
\end{equation}
which is
\begin{equation}
    \mathbf{J}\left ( {\signal} \mapsto \mathbb{E}[\sigrv_0 \mid \sigrv_t = \signal] \right)(\argm) \op^T (\op\mathbb{E}[\sigrv_0 \mid \sigrv_t = \argm] - \meas)
\end{equation}
after an application of the chain rule.
The Jacobian \( \mathbf{J} \left( \signal \mapsto \mathbb{E}[\sigrv_0 \mid \sigrv_t = \signal] \right) \) is typically computed with automatic differentiation when \( (\signal, t) \mapsto \mathbb{E}[\sigrv_0 \mid \sigrv_t = \signal] \) is approximated with a neural network.
In our framework, we use the connection with the covariance matrix \( \operatorname{Cov}[\sigrv_0 \mid \sigrv_t = \argm] \).
Indeed, as also shown in, \textit{e.g.},~\cite{rissanen2025free}, if $\sigrv_0$ and $\sigrv_t$ verify \cref{ddpm_full}, then
\begin{equation}
    \tfrac{1}{1-\bar{\alpha}_t}\text{Cov}[\sigrv_0\mid \sigrv_t = \signal] = \tfrac{1}{\bar{\alpha}_t} \left(\mathbf{I} + (1 - \bar{\alpha}_t)^2 \nabla^2\log p_{\sigrv_t}(\signal)\right).
\end{equation}
This identity combined with the derivative of \cref{eq:tweedie} yields
\begin{equation}
	\mathbf{J}\bigl( \signal \mapsto \mathbb{E}[\sigrv_0 \mid \sigrv_t = \signal] \bigr)(\signal_t) = \dfrac{\sqrt{\bar{\alpha}_t}}{1-\bar{\alpha}_t}\text{Cov}[\sigrv_0\mid \sigrv_t = \signal_t].
\end{equation}
\subsection{A Protocol to Determine the Burn-In Period and the Number of Samples}%
\label{sec:burn in and samples protocol}
As discussed in \cref{ssec:implementation details}, the burn-in period and the number of samples of the Gibbs samplers needs to be chosen appropriately to ensure an acceptable runtime of the benchmark when they serve as the gold-standard samplers of the denoising problems that are encountered in the \gls{dps} algorithms.
We determine the burn-in period and the number of samples through the following protocol that is run in an off-line stage prior to running the benchmark.
We take one datum \( \signal_t = \signal_0 + \noisevar \noise \) where \( \signal_0 \) is constructed via \cref{eq:cumsum} where \( p_U \) is a \( \tDist(1) \) distribution and \( \noise \) is some unknown but fixed vector of standard Gaussian noise.
We then launch \( C = \num{1000} \) parallel Gibbs chains on the corresponding denoising problem and run those chains for \( N_{\mathrm{sufficient}} \) iterations, where \( N_{\mathrm{sufficient}} \) is a sufficiently large natural number that guarantees that the chains are stationary for at least \( N_{\mathrm{avg}} \) (which is also relatively large) iterations and that, consequently, we can compute precise estimates of various statistics of the posterior distribution from the iterates from the last \( N_{\mathrm{avg}} \) iterations across all \( C \) chains.

\begin{figure}
	\centering
	\begin{tikzpicture}
		\begin{axis}[
			/tikz/line join=bevel,
			width=7.5cm,
			height=5.5cm,
			xlabel style={font=\small},
			ylabel style={font=\small},
			tick label style={font=\tiny},
			legend style={
				font=\scriptsize,
				draw=none,    
				fill=none,
				at ={(1.,0.9)},
				legend cell align=left,
			},
			grid=none,
			xlabel={Gibbs iterations},
			ylabel={Wasserstein-1 distance},
			ymode=log,
			legend pos=north east,
			/tikz/mark repeat={10},
		]
			\addplot+ table[x=x, y=sigma_6.174, col sep=comma] {./data/burnin-and-sample-protocol/wasserstein_curves_student.csv};
			\addlegendentry{$\sigma=6.17$}
			\addplot+ table[x=x, y=sigma_25.736, col sep=comma] {./data/burnin-and-sample-protocol/wasserstein_curves_student.csv};
			\addlegendentry{$\sigma=25.7$}
			\addplot+ table[x=x, y=sigma_123.865, col sep=comma] {./data/burnin-and-sample-protocol/wasserstein_curves_student.csv};
			\addlegendentry{$\sigma=123$}
		\end{axis}
	\end{tikzpicture}
	\caption{Wasserstein-1 distance of the intermediate distribution of \( \sigrv_{33} - \sigrv_{32} \) during the iterations to that of the final sample.}%
	\label{fig:wasserstein burn in}
\end{figure}
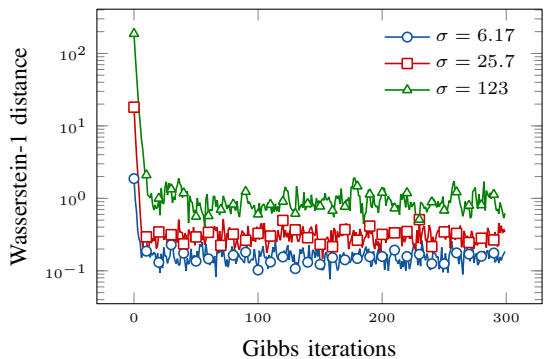

To determine the burn-in period, we then proceed to calculate a statistic that we can monitor throughout the iterations and that we can compare against the reference statistic.
Specifically, denoting with \( \sigrv \) the random variable of the Gibbs sampler, we compute the empirical distribution of the jumps at index \num{32}, that is, \( \sigrv_{33} - \sigrv_{32} \).
The distribution of differences that is obtained by taking the last \( N_{\mathrm{avg}} \) iterations across all \( C \) chains is considered the reference distribution.
Then, we compute the Wasserstein-1 distance of that distribution to the one obtained by taking the average across \( N_{\mathrm{avg}} \) iterations and all \( C \) in a sliding-window starting from the first Gibbs iterations.
This allows us to gauge the burn-in period through a visual inspection of the Wasserstein-1 distance through the Gibbs iterations.
In particular, we expect the Wasserstein-1 distance to be large for a number of initial samples where the Gibbs sampler is not stationary and then to oscillate around a small but nonzero value.
The value will be nonzero due to the finite sample size.
The Wasserstein-1 distance between the reference distribution and the one obtained through the Gibbs iterations is shown in \cref{fig:wasserstein burn in}.
We observe that the empirical distribution of jumps converges rapidly to the reference one.
The Wasserstein-1 distance reaches the noise level after a single-digit number of iterations, which is in line with the analysis provided in \cite{kuric2025gaussianlatentmachineefficient}.
Based on these findings, we chose the burn-in period as \( B = \num{100} \) iterations for all our experiments, which is more than sufficient to reach stationarity and has acceptable runtime.

\begin{figure}
	\centering
	\begin{tikzpicture}
		\begin{axis}[
			width=7.5cm,
			height=5.5cm,
			xlabel style={font=\small},
			ylabel style={font=\small},
			tick label style={font=\tiny},
			legend style={
				font=\scriptsize,
				draw=none,
				fill=none,
				at ={(1.,0.9)},
				legend cell align=left,
			},
			grid=none,
			xlabel={Averaging window length},
			ylabel={MSE},
			ymode=log,
			legend pos=north east,
			xtick={0, 100, 200, 300, 400, 500, 600, 700},
			xticklabels={ 0, 100, 200, {\textcolor{myblue}{\textbf{300}}}, 400, 500, 600, 700 },
			/tikz/mark size=0.6pt,
			/tikz/mark repeat={10},
		]
			\addplot+ table[x=x, y=sigma_6.174, col sep=comma] {./data/burnin-and-sample-protocol/mmse_curves_student.csv};
			\addlegendentry{$\sigma=6.17$}
			\addplot+ table[x=x, y=sigma_25.736, col sep=comma] {./data/burnin-and-sample-protocol/mmse_curves_student.csv};
			\addlegendentry{$\sigma=25.7$}
			\addplot+ table[x=x, y=sigma_123.865, col sep=comma] {./data/burnin-and-sample-protocol/mmse_curves_student.csv};
			\addlegendentry{$\sigma=123$}
			\addplot[dotted, black, thick, domain=0:750] {0.01};
			\addlegendentry{Tolerance $10^{-2}$}
			\draw[dotted, thick, black] (axis cs:300, 0.0001) -- (axis cs:300, 0.01);
			\node[anchor=north, red, font=\scriptsize] at (axis cs:100, \pgfkeysvalueof{/pgfplots/ymin}) {100};
		\end{axis}
	\end{tikzpicture}
	\caption{\gls{mse} gap to the long-run \gls{mmse} normalized by $\sigma^2$.}%
	\label{fig:mmse convergence}
\end{figure}
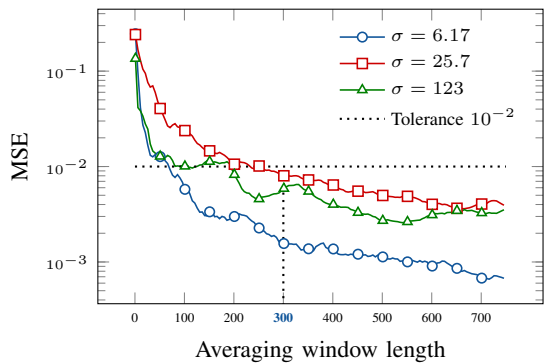

To determine the number of samples that are needed for a sufficiently accurate computation of various statistics that any \gls{dps} algorithm may utilize in their update steps, we compute a precise estimation of the \gls{mmse} estimate of the denoising problem by averaging the last \( N_{\mathrm{avg}} \) iterations across all \( C \) chains.
Then, we pick an arbitrary chain and grow a window from iteration \(  (N_{\mathrm{avg}} - 1) \) to the left, average the samples in that window, and compute the \gls{mse} from the \gls{mmse} estimates obtained in the one-chain window to the precise estimate obtained by averaging the \( C \) chains and the last  \( N_{\mathrm{avg}} \) iterations.
We pick a tolerance of \num{1e-2} and monitor at which window length the \gls{mse} falls below that tolerance.
The results in \cref{fig:mmse convergence} show that this tolerance is consistently reached when the averaging window is \num{300} samples long, which motivates our choice of using \( S = \num{300} \) samples for all our experiments.

\subsection{Algorithm Parameter Identification}%
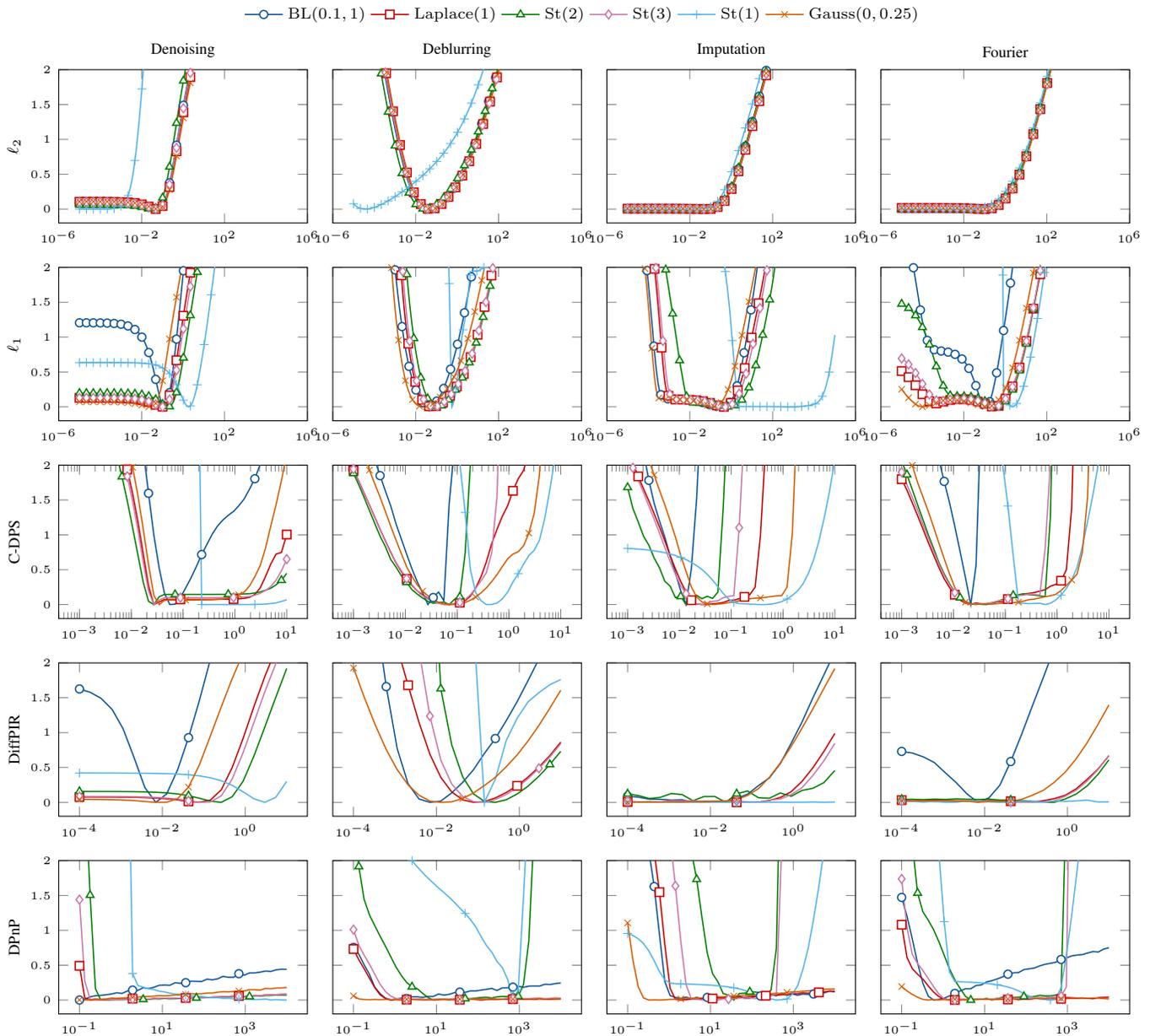
\begin{figure*}
	\def\prefix{./data/posterior_new}
	\centering
	\begin{tikzpicture}
		\begin{groupplot}[
			group style={
				group size=4 by 5,
				horizontal sep=0.4cm,
				vertical sep=.7cm,
				yticklabels at=edge left,
				group name=G,
			},
				width=5.5cm,
				height=4cm,
				ymax=2,
				clip limits=false,
				tick label style={font=\tiny},
				legend pos=south east,
				legend columns=-1,
				legend style={
					font=\scriptsize,
					at={(1.3, 1.25)},
					draw=none,
				},
				/tikz/mark size=0.5pt,
				/tikz/mark repeat={10},
			]
			\plotgridsearchcurves{l2}{1}
			\plotgridsearchcurves{l1}{0}
			\plotgridsearchcurves{cdps}{0}
			\plotgridsearchcurves{diffpir}{0}
			\plotgridsearchcurves{dpnp}{0}
		\end{groupplot}
		\node[anchor=south] at ($(G c1r1.north)+(0, 2pt)$) {\scriptsize Denoising};
		\node[anchor=south] at ($(G c2r1.north)+(0, 2pt)$) {\scriptsize Deblurring};
		\node[anchor=south] at ($(G c3r1.north)+(0, 2pt)$) {\scriptsize Imputation};
		\node[anchor=south] at ($(G c4r1.north)+(0, 2pt)$) {\scriptsize Fourier};

		\node[rotate=90] at ($(G c1r1.west)+(-20pt, 0pt)$) {\scriptsize \( \ell_2 \)};
		\node[rotate=90] at ($(G c1r2.west)+(-20pt, 0pt)$) {\scriptsize \( \ell_1 \)};
		\node[rotate=90] at ($(G c1r3.west)+(-20pt, 0pt)$) {\scriptsize C-DPS};
		\node[rotate=90] at ($(G c1r4.west)+(-20pt, 0pt)$) {\scriptsize DiffPIR};
		\node[rotate=90] at ($(G c1r5.west)+(-20pt, 0pt)$) {\scriptsize DPnP};
	\end{tikzpicture}
	\caption{
		Grid search diagnostics (logarithm of the \gls{mse} over the validation data set) for the \gls{dps} algorithms.
		Rows: \( \ell_2 \); \( \ell_1 \); \gls{cdps}; \gls{diffpir}; \gls{dpnp}.
		Columns: Denoising; deconvolution; imputation; reconstruction from partial Fourier measurements.
		For better visualization, each curve has had its minimum subtracted.
		To avoid clutter, marks are placed only at every 10th grid point.
	}%
	\label{fig:grid search}
\end{figure*}

\label{ssec:parameter identification}
The grid for the model-based methods was a loglinear grid \( \Lambda = \{ \lambda_1, \lambda_2, \dotsc, \lambda_{N_{\mathrm{mb}}} \} \) where
\begin{equation}
	\lambda_i = 10^{a + (i - 1)\frac{(b - a)}{N_{\mathrm{mb}} - 1}} 
\end{equation}
with \( a = -5 \) and \( b = 5 \).
Since the model-based methods are very fast, we can use the relatively high \( N_{\mathrm{mb}} = \num{1000} \).
The \gls{mse} over the validation data set of the \( \ell_2 \) and \( \ell_1 \) estimators are shown in~\cref{fig:grid search}.

For the \gls{dps} algorithms we define a modest number of \( N_{\mathrm{dps}} = \num{40} \) grid-points and found the extreme points of the grid (i.e. values of the parameters that clearly lead to worse results) by hand.
For \gls{cdps} and \gls{diffpir}, we fix the diffusion schedule to standard choices (\( \beta_0 = \num{1e-4}, \beta_T = \num{0.02} \)).
In addition to the diffusion schedule, \gls{cdps} has one tunable parameter \( \zeta \) that we tune on \num{40} loglinear grid points (\( i = 1, \dotsc, N_{\mathrm{dps}} \))
\begin{equation}
	10^{a + (i - 1)\frac{(b - a)}{N_{\mathrm{dps}} - 1}},
\end{equation}
where \( a = -3 \) and \( b = 1 \).
\Gls{diffpir} has two tunable parameters \( \zeta \) and \( \rho \), although \( \zeta \) is typically considered not so critical.
Thus, we split the \num{40} grid points into a two-dimensional grid \( \Theta^{\mathrm{DiffPIR}} = \{ 0.3, 0.7 \} \times \Theta^{\rho} \), \textit{i.e.}, \num{2} points for \( \zeta \) and \num{20} points for \( \rho \) given by \( \Theta^\rho = \{\Theta^\rho_1, \dotsc, \Theta^\rho_{N_{\mathrm{dps}}/ 2} \} \) where
\begin{equation}
	\Theta^\rho_1 = 10^{a + (i - 1)\frac{(b - a)}{(N_{\mathrm{dps}}/2) - 1}}
\end{equation}
with \( a = -4 \) and \( b = 1 \).
The \gls{dpnp} algorithm only has the schedule \( \{ \eta_t \}_{t=1}^T \) to tune.
In this case, since \gls{dpnp} is asymptotically correct, the schedule is a practical vehicle that enables to trade off between speed and accuracy.
Therefore, we use a schedule that is similar to the one that was proposed in the original publication~\cite{xu2024provably}:
We fix a small \( \eta_{\mathrm{final}} = 0.15 \), and linearly decrease eta from some \( \eta_{\mathrm{initial}} \) to \( \eta_{\mathrm{final}} \) after \( K / 5 \) initial iterations with \( \eta_{\mathrm{initial}} \):
\begin{equation}
	\eta_i = \begin{cases}
		\eta_{\mathrm{initial}} & \text{if} i = 1, \dotsc, K/5\\
		\frac{\eta_{\mathrm{final}}}{\eta_{\mathrm{initial}}}^\frac{i - K/5}{K - K/5}\eta_{\mathrm{initial}} & \text{if} i = K/5+1, \dotsc, K\\
	\end{cases}
\end{equation}
We treat \( \eta_{\mathrm{initial}} \) as a tunable parameter and search over \( \Theta^{\mathrm{DPnP}} = \{ \eta_{1}, \eta_{2}, \dotsc, \eta_{40} \} \) where for \( i = 1, \dotsc, 40 \),
\begin{equation}
	\eta_{i} = 10^{a + (i - 1)\frac{(b - a)}{40 - 1}}
\end{equation}
with \( a = -1 \) and \( b = 4 \).
Like in the original publication, we use the comparatively small \( K = 40 \).

The \gls{mse} over the validation data set of \gls{cdps}, \gls{diffpir}, and \gls{dpnp} for the various parameter choices is shown in~\cref{fig:grid search}.
Since the \( \zeta \) parameter of \gls{diffpir} is considered not so critical, we only show the values of the \gls{mse} for various choices of \( \rho \) where \( \zeta \) is set to the value of the optimal \( (\zeta, \rho) \) pair.

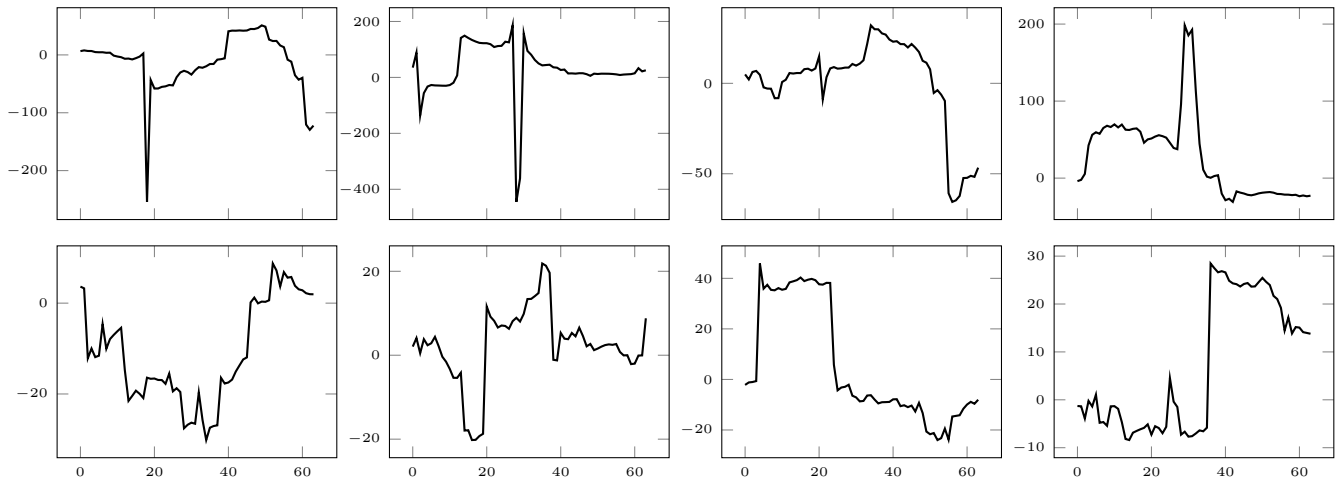
\begin{figure*}
    \centering
	\begin{tikzpicture}
		\begin{groupplot}[
			group style={
				group size=4 by 2,
				horizontal sep=0.7cm,
				vertical sep=0.35cm,
				xticklabels at=edge bottom,
			},
			width=5.3cm,
			height=4.4cm,
			tick label style={font=\tiny},
			title style={font=\scriptsize},
			]
			\nextgroupplot
			\addplot[black, thick] table[x=x, y=y1, col sep=comma] {./data/posterior_new/student/1.0/unconditional/learned-vs-gibbs/learned.csv};%
			\nextgroupplot
			\addplot[black, thick] table[x=x, y=y2, col sep=comma] {./data/posterior_new/student/1.0/unconditional/learned-vs-gibbs/learned.csv};%
			\nextgroupplot
			\addplot[black, thick] table[x=x, y=y3, col sep=comma] {./data/posterior_new/student/1.0/unconditional/learned-vs-gibbs/learned.csv};%
			\nextgroupplot
			\addplot[black, thick] table[x=x, y=y4, col sep=comma] {./data/posterior_new/student/1.0/unconditional/learned-vs-gibbs/learned.csv};%
			\nextgroupplot
			\addplot[black, thick] table[x=x, y=y1, col sep=comma] {./data/posterior_new/student/1.0/unconditional/learned-vs-gibbs/gibbs.csv};%
			\nextgroupplot
			\addplot[black, thick] table[x=x, y=y2, col sep=comma] {./data/posterior_new/student/1.0/unconditional/learned-vs-gibbs/gibbs.csv};%
			\nextgroupplot
			\addplot[black, thick] table[x=x, y=y3, col sep=comma] {./data/posterior_new/student/1.0/unconditional/learned-vs-gibbs/gibbs.csv};%
			\nextgroupplot
			\addplot[black, thick] table[x=x, y=y4, col sep=comma] {./data/posterior_new/student/1.0/unconditional/learned-vs-gibbs/gibbs.csv};%
		\end{groupplot}
	\end{tikzpicture}
    \caption{%
		Unconditional samples obtained by \gls{ddpm} with the learned denoiser (top) and the oracle denoiser (bottom).
		Columns: Different initializations and random states.
	}%
    \label{fig:learned vs gibbs samples}
\end{figure*}
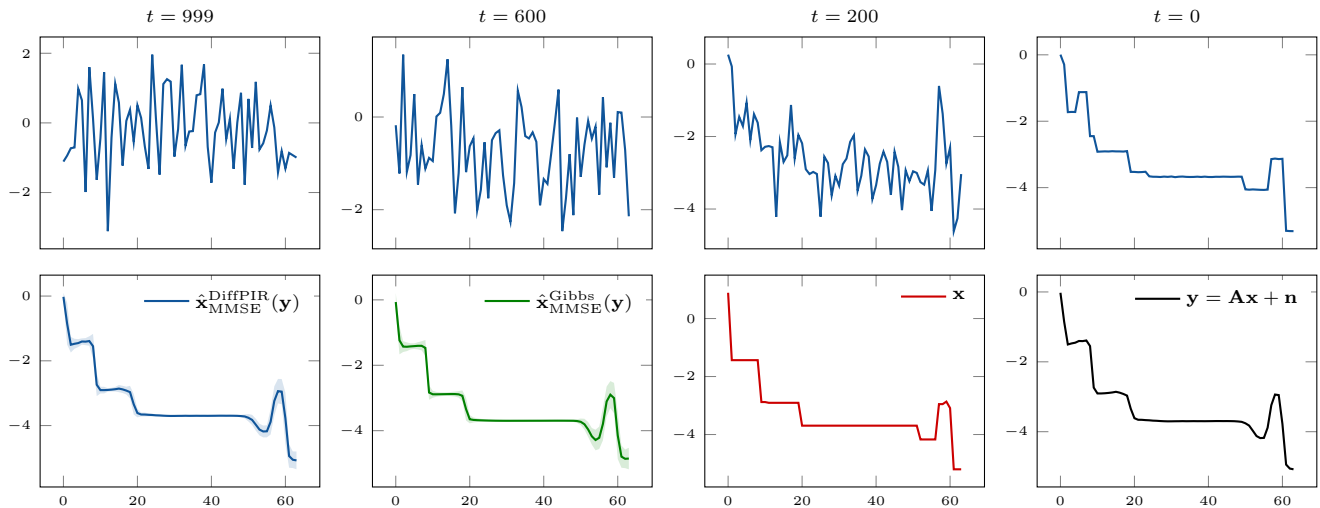
\begin{figure*}
	\centering
	\begin{tikzpicture}
	\pgfplotstableread[col sep=comma]{./data/posterior_new/conditional/mmse_diffpir.csv}\meantablediffpir
	\pgfplotstableread[col sep=comma]{./data/posterior_new/conditional/std_diffpir.csv}\stdtablediffpir
	\pgfplotstablecreatecol[copy column from table={\stdtablediffpir}{y1}]{sigma}\meantablediffpir
	\pgfplotstablecreatecol[create col/expr={\thisrow{y1} + \thisrow{sigma}}]{upper}\meantablediffpir
	\pgfplotstablecreatecol[create col/expr={\thisrow{y1} - \thisrow{sigma}}]{lower}\meantablediffpir
	\pgfplotstableread[col sep=comma]{./data/posterior_new/conditional/mmse_gibbs.csv}\meantablegibbs
	\pgfplotstableread[col sep=comma]{./data/posterior_new/conditional/std_gibbs.csv}\stdtablegibbs
	\pgfplotstablecreatecol[copy column from table={\stdtablegibbs}{y1}]{sigma}\meantablegibbs
	\pgfplotstablecreatecol[create col/expr={\thisrow{y1} + \thisrow{sigma}}]{upper}\meantablegibbs
	\pgfplotstablecreatecol[create col/expr={\thisrow{y1} - \thisrow{sigma}}]{lower}\meantablegibbs
		\begin{groupplot}[
			group style={
				group size=4 by 2,
				group name=G,
				horizontal sep=0.7cm,
				vertical sep=0.35cm,
				xticklabels at=edge bottom,
			},
			legend style={font=\scriptsize, draw=none},
			width=5.3cm,
			height=4.4cm,
			tick label style={font=\tiny},
		]
			\nextgroupplot
			\addplot [myblue, thick] table[x=x, y=y1, col sep=comma] {./data/posterior_new/conditional/999.csv};%
			\nextgroupplot
			\addplot [myblue, thick] table[x=x, y=y1, col sep=comma] {./data/posterior_new/conditional/600.csv};%
			\nextgroupplot
			\addplot [myblue, thick] table[x=x, y=y1, col sep=comma] {./data/posterior_new/conditional/200.csv};%
			\nextgroupplot
			\addplot [myblue, thick] table[x=x, y=y1, col sep=comma] {./data/posterior_new/conditional/001.csv};%
			\nextgroupplot
			\addplot[myblue, thick] table[x=x, y=y1]{\meantablediffpir};
			\addlegendentry{$\estimate_{\mathrm{MMSE}}^{\mathrm{DiffPIR}}(\meas)$}
			\addplot[name path=dupper, draw=none] table[x=x, y=upper]{\meantablediffpir};
			\addplot[name path=dlower, draw=none] table[x=x, y=lower]{\meantablediffpir};
			\addplot[fill=myblue, fill opacity=0.16, draw=none] fill between[of=dupper and dlower];
			\nextgroupplot
			\addplot[mygreen, thick] table[x=x, y=y1]{\meantablegibbs};
			\addlegendentry{$\estimate_{\mathrm{MMSE}}^{\mathrm{Gibbs}}(\meas)$}
			\addplot[name path=gupper, draw=none] table[x=x, y=upper]{\meantablegibbs};
			\addplot[name path=glower, draw=none] table[x=x, y=lower]{\meantablegibbs};
			\addplot[fill=mygreen, fill opacity=0.15, draw=none] fill between[of=gupper and glower];
			\nextgroupplot
			\addplot [myred, thick] table[x=x, y=y1, col sep=comma] {./data/posterior_new/conditional/signal.csv};%
			\addlegendentry{$\signal$}
			\nextgroupplot
			\addplot [black, thick] table[x=x, y=y1, col sep=comma] {./data/posterior_new/conditional/mmse_diffpir.csv};%
			\addlegendentry{$\meas = \op\signal + \noise$}
		\end{groupplot}

		\node[anchor=south] at ($(G c1r1.north)+(0, 2pt)$) {\scriptsize \( t=999 \)};
		\node[anchor=south] at ($(G c2r1.north)+(0, 2pt)$) {\scriptsize \( t=600 \)};
		\node[anchor=south] at ($(G c3r1.north)+(0, 2pt)$) {\scriptsize \( t=200 \)};
		\node[anchor=south] at ($(G c4r1.north)+(0, 2pt)$) {\scriptsize \( t=0 \)};
	\end{tikzpicture}
	\caption{%
		Conditional generation for deconvolution of a signal with \( \BLDist(0.1, 1) \) jumps with \gls{diffpir}.
		Top: Prototypical sampling trajectory at times \( t = 999, 600, 200, 0 \).
		Bottom: From left to right: \gls{mmse} estimate obtained by averaging all \gls{diffpir} samples; gold-standard \gls{mmse} estimate obtained by the Gibbs method; the data-generating signal; the data.
	}%
    \label{fig:example reconstruction with trajectory student gibbs}
\end{figure*}

\subsection{Additional Results}
\begin{figure*}
	\centering
	\def\prefix{./data/posterior_new}
	\begin{tikzpicture}
		\begin{groupplot}[posteriorplot]
			\plotposterior{bernoulli-laplace/p=0.1_b=1.0}{convolution}{learned}{1}
			\plotposterior{student/1.0}{convolution}{learned}{0}
			\plotposterior{student/2.0}{convolution}{learned}{0}
			\plotposterior{laplace/1.0}{convolution}{learned}{0}
		\end{groupplot}
		\node[rotate=90] at ($(G c1r1.west)+(-20pt, 0pt)$) {\scriptsize \( \BLDist(0.1, 1) \)};
		\node[rotate=90] at ($(G c1r3.west)+(-20pt, 0pt)$) {\scriptsize \( \tDist(1) \)};
		\node[rotate=90] at ($(G c1r5.west)+(-20pt, 0pt)$) {\scriptsize \( \tDist(2) \)};
		\node[rotate=90] at ($(G c1r7.west)+(-20pt, 0pt)$) {\scriptsize \( \LaplaceDist(1) \)};
	\end{tikzpicture}
	\caption{%
		Qualitative results for deconvolution using the learned denoiser.
		Rows: Jump distributions.
		For each jump distribution, the \gls{mmse} estimates obtained by the different \gls{dps} algorithms and the gold-standard Gibbs methods are shown on top of the corresponding index-wise marginal variances.
		Columns: Different measurements.
	}%
	\label{fig:results deconvolution learned}
\end{figure*}
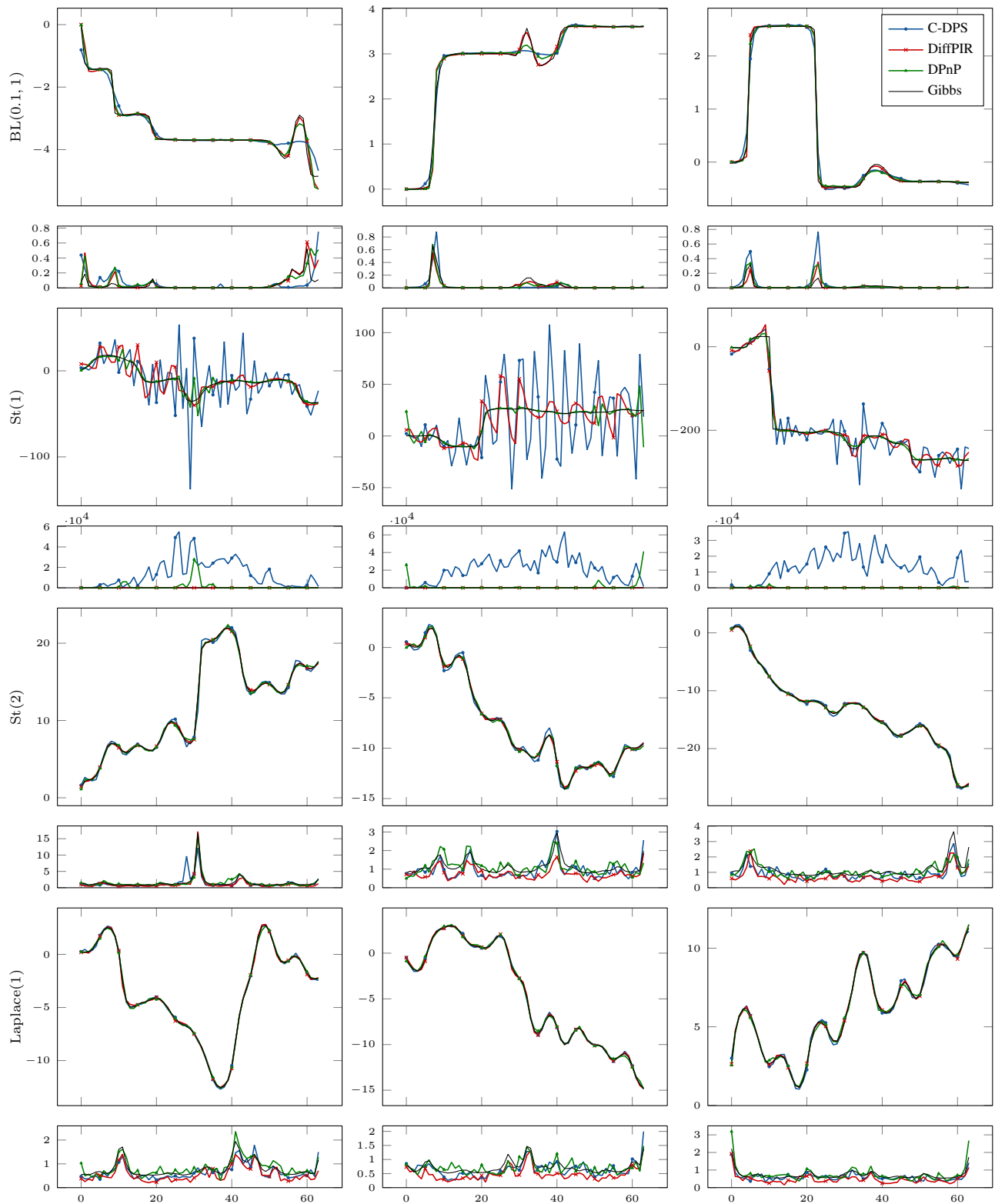

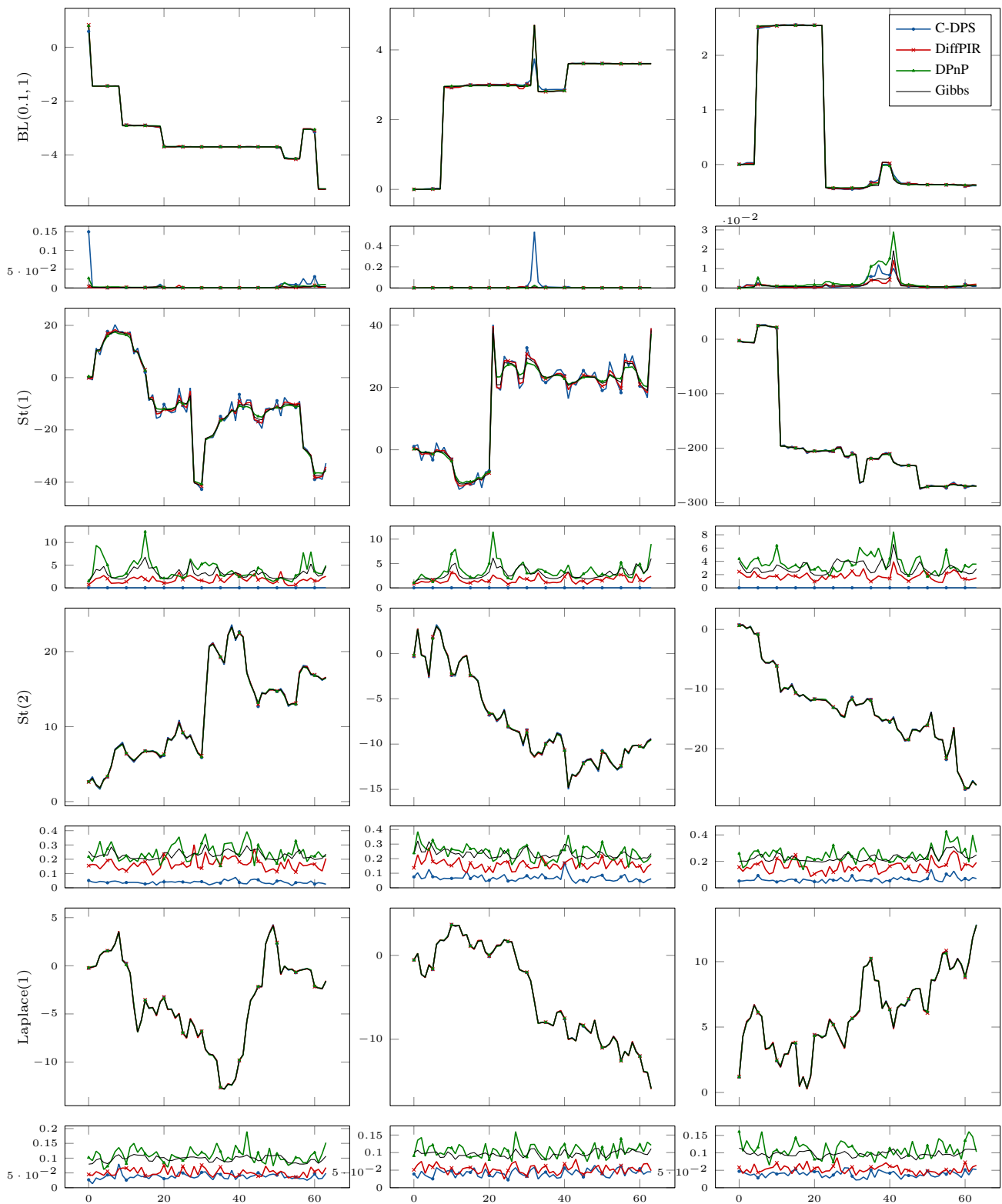
\begin{figure*}
	\centering
	\def\prefix{./data/posterior_new}
	\begin{tikzpicture}
		\begin{groupplot}[posteriorplot]
			\plotposterior{bernoulli-laplace/p=0.1_b=1.0}{identity}{gibbs}{1}
			\plotposterior{student/1.0}{identity}{gibbs}{0}
			\plotposterior{student/2.0}{identity}{gibbs}{0}
			\plotposterior{laplace/1.0}{identity}{gibbs}{0}
		\end{groupplot}
		\node[rotate=90] at ($(G c1r1.west)+(-20pt, 0pt)$) {\scriptsize \( \BLDist(0.1, 1) \)};
		\node[rotate=90] at ($(G c1r3.west)+(-20pt, 0pt)$) {\scriptsize \( \tDist(1) \)};
		\node[rotate=90] at ($(G c1r5.west)+(-20pt, 0pt)$) {\scriptsize \( \tDist(2) \)};
		\node[rotate=90] at ($(G c1r7.west)+(-20pt, 0pt)$) {\scriptsize \( \LaplaceDist(1) \)};
	\end{tikzpicture}
	\caption{%
		Qualitative results for denoising using the oracle \gls{mmse} denoiser.
		Rows: Jump distributions.
		For each jump distribution, the \gls{mmse} estimates obtained by the different \gls{dps} algorithms and the gold-standard Gibbs methods are shown on top of the corresponding index-wise marginal variances.
		Columns: Different measurements.
	}%
	\label{fig:results denoising gibbs}
\end{figure*}

\begin{figure*}
	\centering
	\def\prefix{./data/posterior_new}
	\begin{tikzpicture}
		\begin{groupplot}[posteriorplot]
			\plotposterior{bernoulli-laplace/p=0.1_b=1.0}{fourier}{gibbs}{1}
			\plotposterior{student/1.0}{fourier}{gibbs}{0}
			\plotposterior{student/2.0}{fourier}{gibbs}{0}
			\plotposterior{laplace/1.0}{fourier}{gibbs}{0}
		\end{groupplot}
		\node[rotate=90] at ($(G c1r1.west)+(-20pt, 0pt)$) {\scriptsize \( \BLDist(0.1, 1) \)};
		\node[rotate=90] at ($(G c1r3.west)+(-20pt, 0pt)$) {\scriptsize \( \tDist(1) \)};
		\node[rotate=90] at ($(G c1r5.west)+(-20pt, 0pt)$) {\scriptsize \( \tDist(2) \)};
		\node[rotate=90] at ($(G c1r7.west)+(-20pt, 0pt)$) {\scriptsize \( \LaplaceDist(1) \)};
	\end{tikzpicture}
	\caption{%
		Qualitative results for reconstruction from partial Fourier measurements using the oracle denoiser.
		Rows: Jump distributions.
		For each jump distribution, the \gls{mmse} estimates obtained by the different \gls{dps} algorithms and the gold-standard Gibbs methods are shown on top of the corresponding index-wise marginal variances.
		Columns: Different measurements.
	}%
	\label{fig:results fourier gibbs}
\end{figure*}
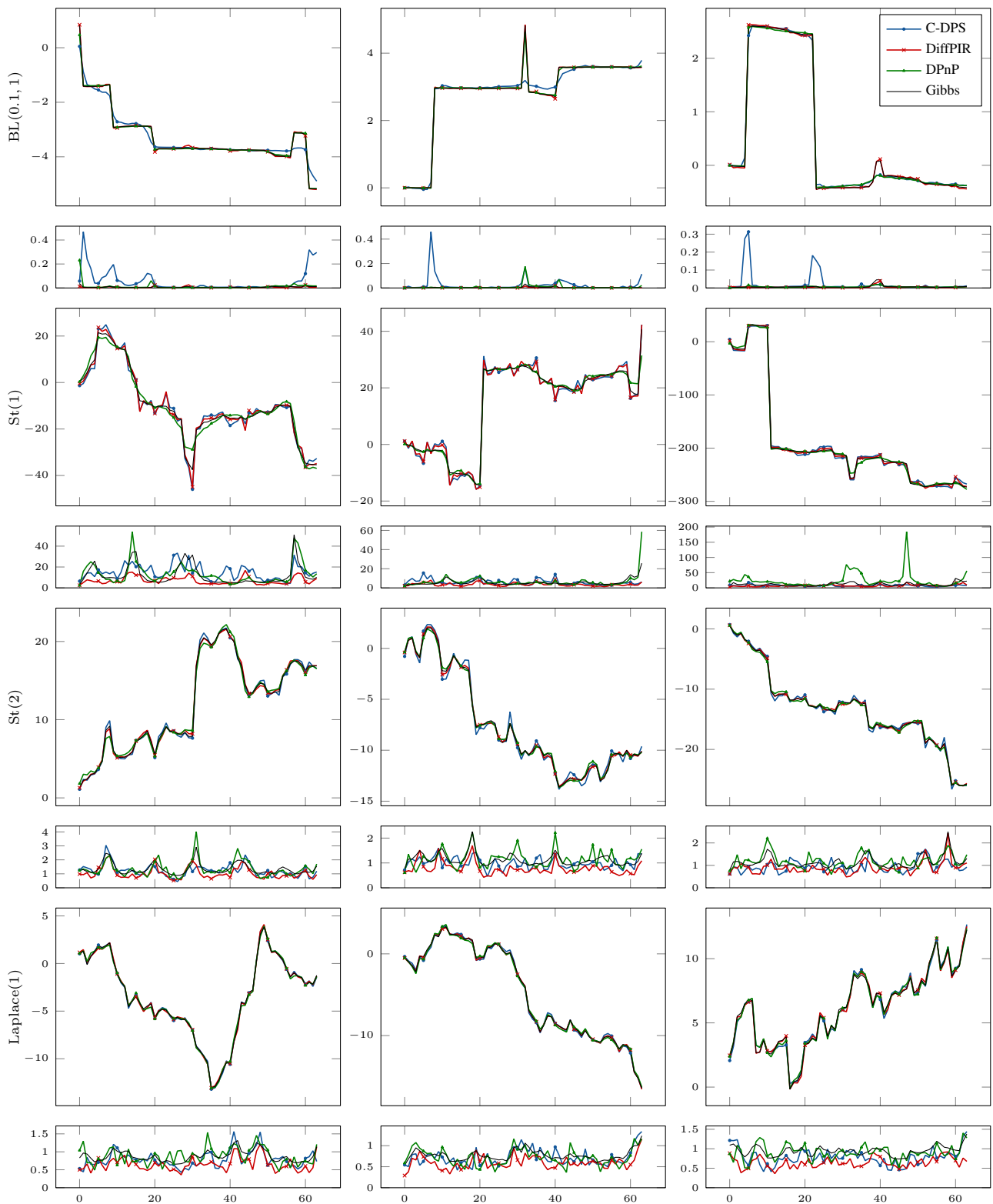

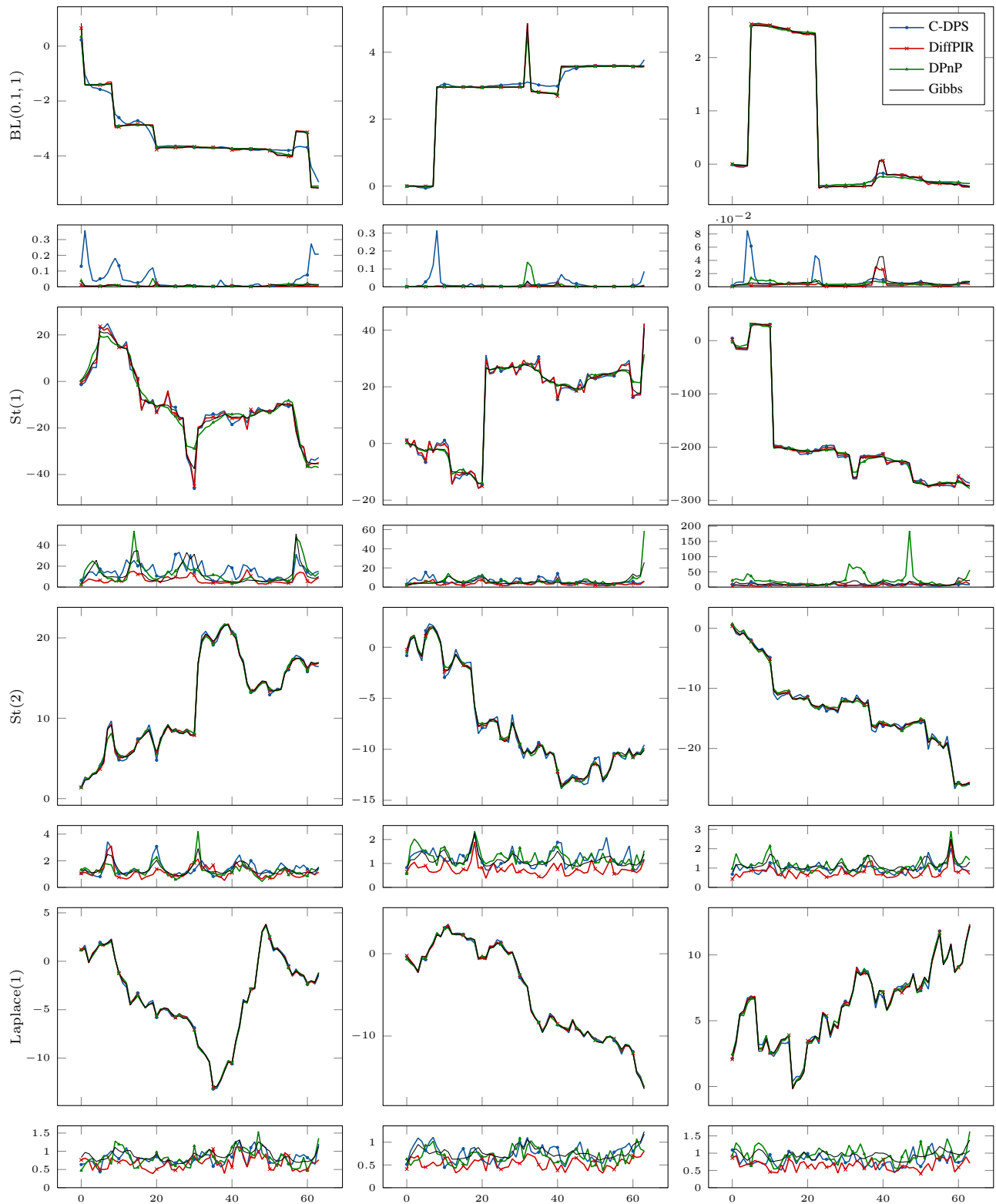
\begin{figure*}
	\centering
	\def\prefix{./data/posterior_new}
	\begin{tikzpicture}
		\begin{groupplot}[posteriorplot]
			\plotposterior{bernoulli-laplace/p=0.1_b=1.0}{fourier}{learned}{1}
			\plotposterior{student/1.0}{fourier}{gibbs}{0}
			\plotposterior{student/2.0}{fourier}{learned}{0}
			\plotposterior{laplace/1.0}{fourier}{learned}{0}
		\end{groupplot}
		\node[rotate=90] at ($(G c1r1.west)+(-20pt, 0pt)$) {\scriptsize \( \BLDist(0.1, 1) \)};
		\node[rotate=90] at ($(G c1r3.west)+(-20pt, 0pt)$) {\scriptsize \( \tDist(1) \)};
		\node[rotate=90] at ($(G c1r5.west)+(-20pt, 0pt)$) {\scriptsize \( \tDist(2) \)};
		\node[rotate=90] at ($(G c1r7.west)+(-20pt, 0pt)$) {\scriptsize \( \LaplaceDist(1) \)};
	\end{tikzpicture}
	\caption{%
		Qualitative results for reconstruction from partial Fourier measurements using the learned denoiser.
		Rows: Jump distributions.
		For each jump distribution, the \gls{mmse} estimates obtained by the different \gls{dps} algorithms and the gold-standard Gibbs methods are shown on top of the corresponding index-wise marginal variances.
		Columns: Different measurements.
	}%
	\label{fig:results fourier learned}
\end{figure*}

\Cref{fig:results deconvolution gibbs} in the main body of the manuscript showed the results for deconvolution using the oracle denoiser.
\Cref{fig:results deconvolution learned} shows the results for deconvolution using the learned denoiser.
\Cref{fig:results denoising gibbs,fig:results denoising learned,fig:results imputation gibbs,fig:results imputation learned,fig:results fourier gibbs,fig:results fourier learned} show the results for denoising and imputation using the oracle and learned denoisers.

\begin{figure*}
	\centering
	\def\prefix{./data/posterior_new}
	\begin{tikzpicture}
		\begin{groupplot}[posteriorplot]
			\plotposterior{bernoulli-laplace/p=0.1_b=1.0}{identity}{learned}{1}
			\plotposterior{student/1.0}{identity}{learned}{0}
			\plotposterior{student/2.0}{identity}{learned}{0}
			\plotposterior{laplace/1.0}{identity}{learned}{0}
		\end{groupplot}
		\node[rotate=90] at ($(G c1r1.west)+(-20pt, 0pt)$) {\scriptsize \( \BLDist(0.1, 1) \)};
		\node[rotate=90] at ($(G c1r3.west)+(-20pt, 0pt)$) {\scriptsize \( \tDist(1) \)};
		\node[rotate=90] at ($(G c1r5.west)+(-20pt, 0pt)$) {\scriptsize \( \tDist(2) \)};
		\node[rotate=90] at ($(G c1r7.west)+(-20pt, 0pt)$) {\scriptsize \( \LaplaceDist(1) \)};
	\end{tikzpicture}
	\caption{%
		Qualitative results for denoising using the learned denoiser.
		Rows: Jump distributions.
		For each jump distribution, the \gls{mmse} estimates obtained by the different \gls{dps} algorithms and the gold-standard Gibbs methods are shown on top of the corresponding index-wise marginal variances.
		Columns: Different measurements.
	}%
	\label{fig:results denoising learned}
\end{figure*}
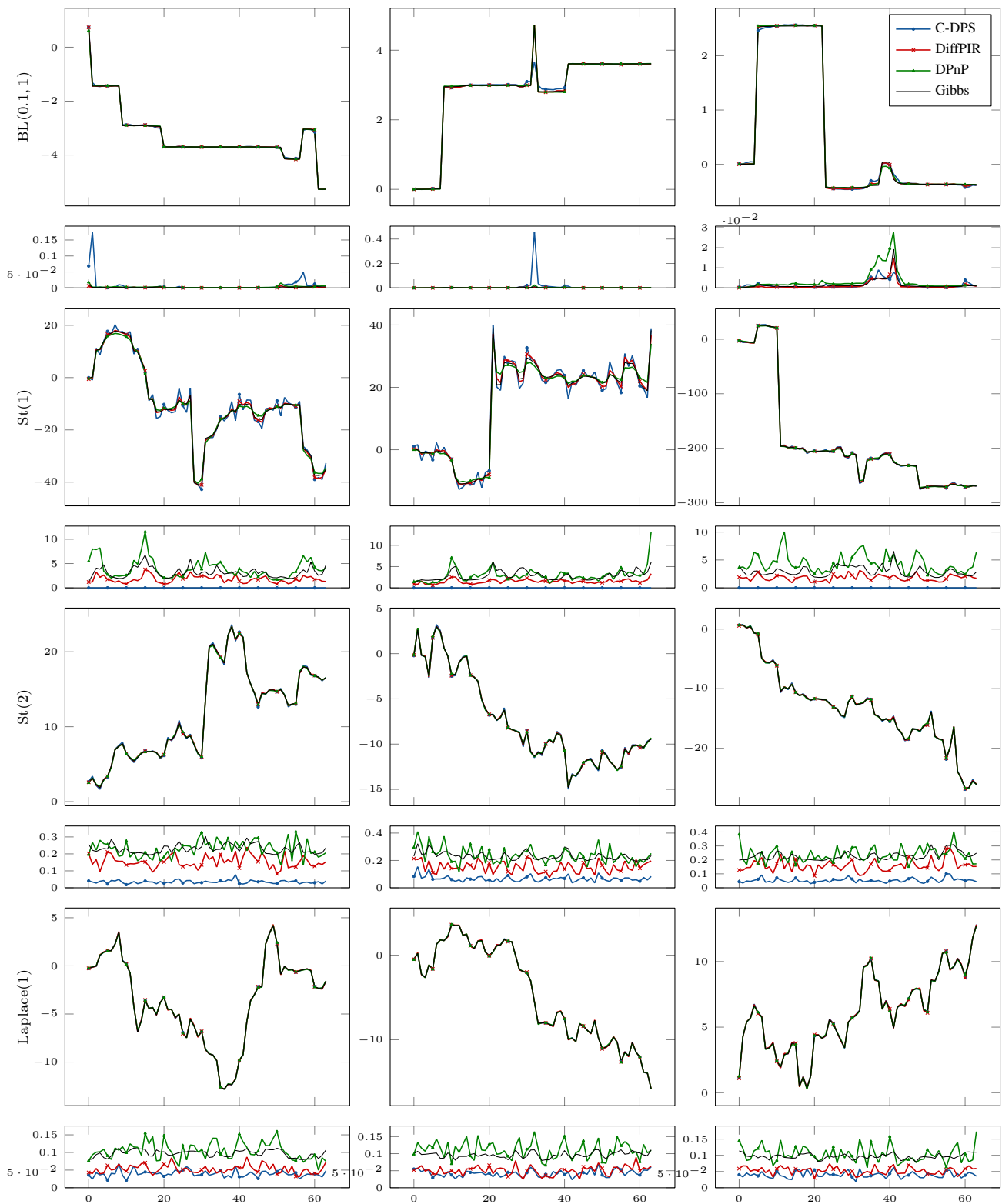

\begin{figure*}
	\centering
	\def\prefix{./data/posterior_new}
	\begin{tikzpicture}
		\begin{groupplot}[posteriorplot]
			\plotposterior{bernoulli-laplace/p=0.1_b=1.0}{sample}{gibbs}{1}
			\plotposterior{student/1.0}{sample}{gibbs}{0}
			\plotposterior{student/2.0}{sample}{gibbs}{0}
			\plotposterior{laplace/1.0}{sample}{gibbs}{0}
		\end{groupplot}
		\node[rotate=90] at ($(G c1r1.west)+(-20pt, 0pt)$) {\scriptsize \( \BLDist(0.1, 1) \)};
		\node[rotate=90] at ($(G c1r3.west)+(-20pt, 0pt)$) {\scriptsize \( \tDist(1) \)};
		\node[rotate=90] at ($(G c1r5.west)+(-20pt, 0pt)$) {\scriptsize \( \tDist(2) \)};
		\node[rotate=90] at ($(G c1r7.west)+(-20pt, 0pt)$) {\scriptsize \( \LaplaceDist(1) \)};
	\end{tikzpicture}
	\caption{%
		Qualitative results for imputation using the oracle sampler.
		Rows: Jump distributions.
		For each jump distribution, the \gls{mmse} estimates obtained by the different \gls{dps} algorithms and the gold-standard Gibbs methods are shown on top of the corresponding index-wise marginal variances.
		Columns: Different measurements.
	}%
	\label{fig:results imputation gibbs}
\end{figure*}
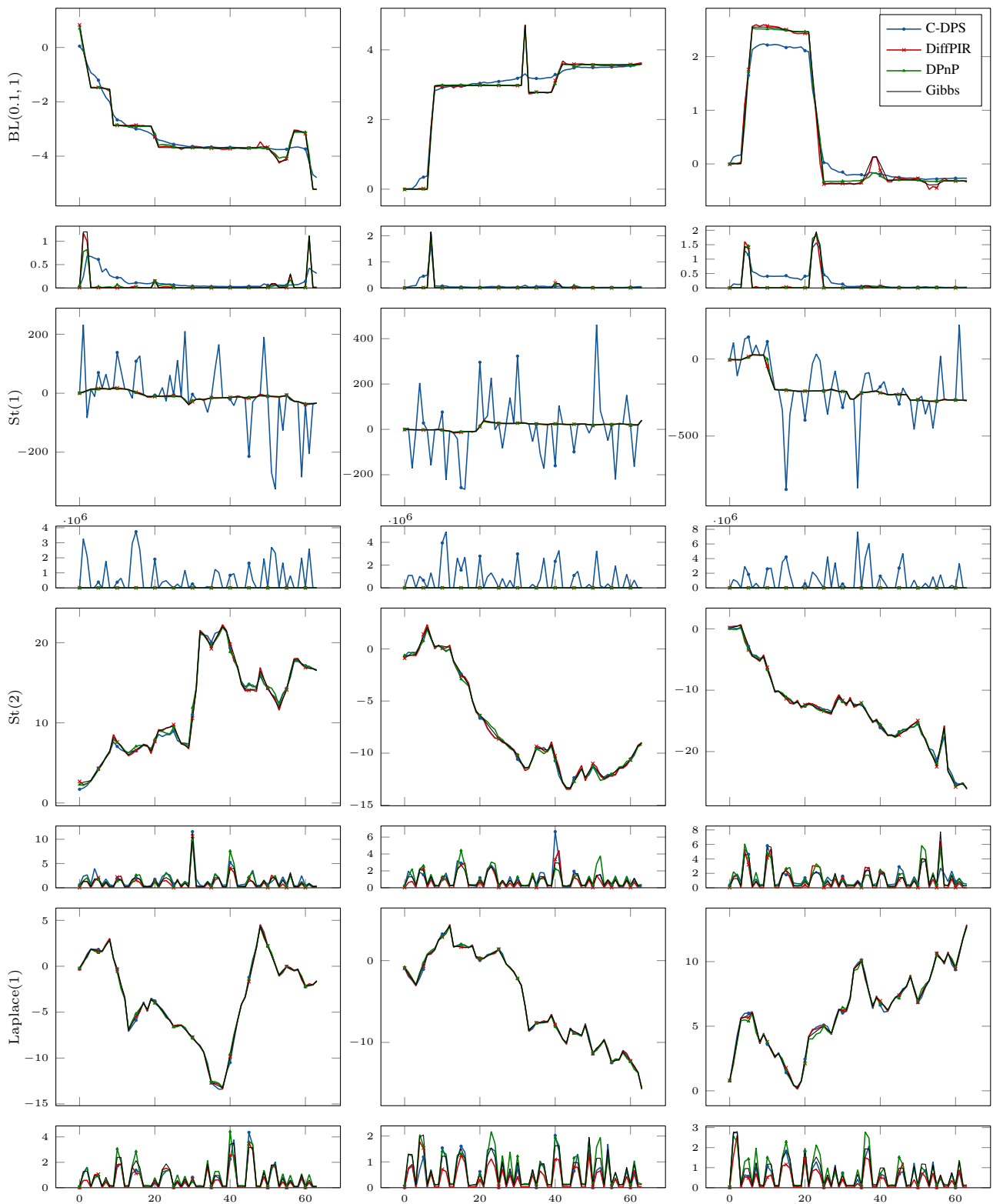

\begin{figure*}
	\centering
	\def\prefix{./data/posterior_new}
	\begin{tikzpicture}
		\begin{groupplot}[posteriorplot]
			\plotposterior{bernoulli-laplace/p=0.1_b=1.0}{sample}{learned}{1}
			\plotposterior{student/1.0}{sample}{learned}{0}
			\plotposterior{student/2.0}{sample}{learned}{0}
			\plotposterior{laplace/1.0}{sample}{learned}{0}
		\end{groupplot}
		\node[rotate=90] at ($(G c1r1.west)+(-20pt, 0pt)$) {\scriptsize \( \BLDist(0.1, 1) \)};
		\node[rotate=90] at ($(G c1r3.west)+(-20pt, 0pt)$) {\scriptsize \( \tDist(1) \)};
		\node[rotate=90] at ($(G c1r5.west)+(-20pt, 0pt)$) {\scriptsize \( \tDist(2) \)};
		\node[rotate=90] at ($(G c1r7.west)+(-20pt, 0pt)$) {\scriptsize \( \LaplaceDist(1) \)};
	\end{tikzpicture}
	\caption{%
		Qualitative results for imputation using the learned denoiser.
		Rows: Jump distributions.
		For each jump distribution, the \gls{mmse} estimates obtained by the different \gls{dps} algorithms and the gold-standard Gibbs methods are shown on top of the corresponding index-wise marginal variances.
		Columns: Different measurements.
	}%
	\label{fig:results imputation learned}
\end{figure*}
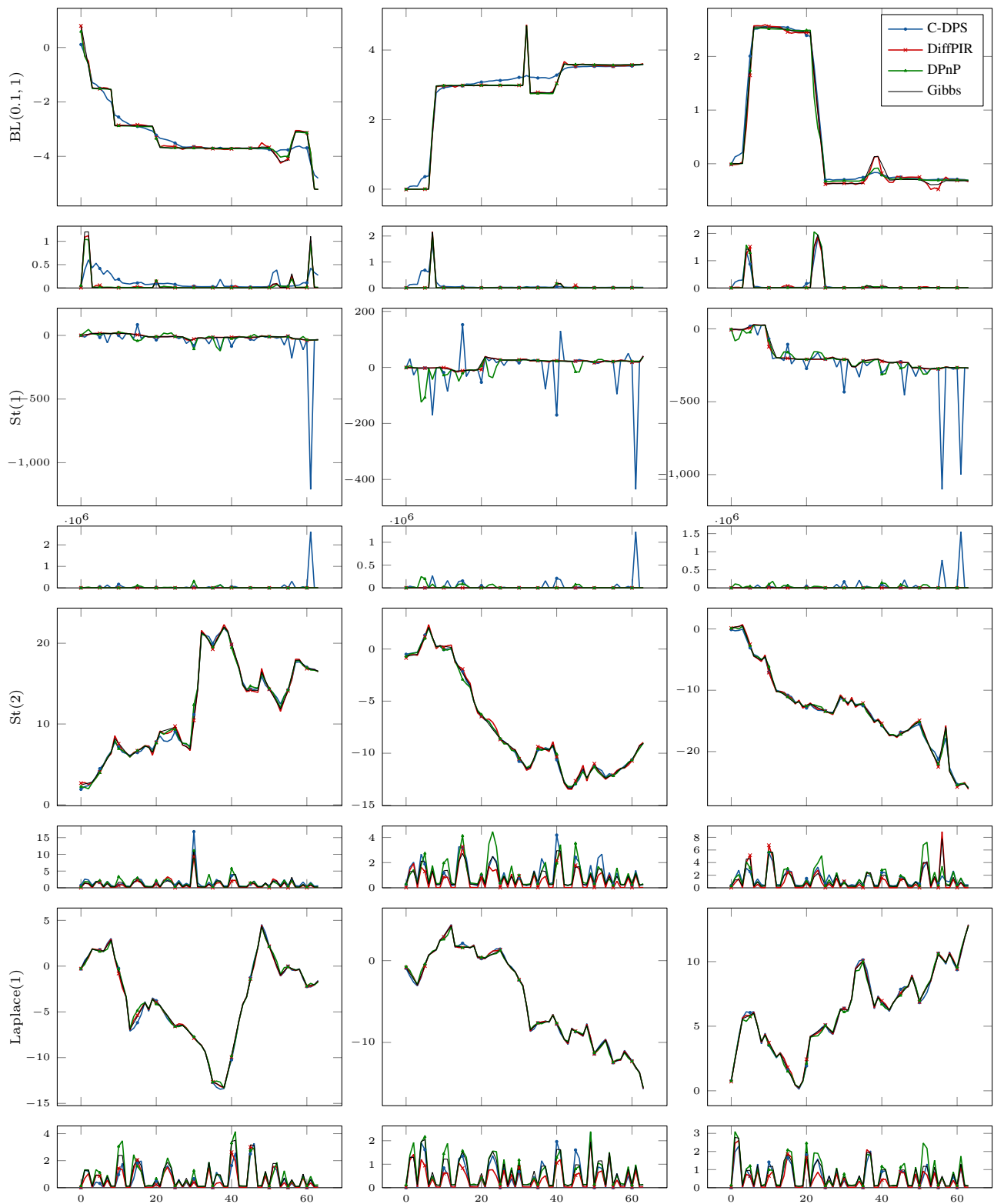

\subsection{Latent Distributions and Notation}
Some of the distributions that we rely on in this work have multiple competing parametrizations and we provide a summary of our definitions in \cref{tab:all distributions} to avoid any ambiguities.
\begin{table*}
	\centering
	\begin{threeparttable}
		\begin{tabular}{lllll}
			\toprule
			Name & Distribution & Parameter(s) & Support & Notation \\
			\midrule	
			Gaussian & $\frac{1}{\sqrt{2 \pi \sigma^2}}  \exp \Bigl(-\frac{(x - \mu)^2}{\sigma^2} \Bigr)$ & $\mu \in \R, \sigma^2 \in \R_{>0}$ & $\R$ & $\GaussianDist$ \\
			Exponential & $\lambda  \exp(-\lambda x)$ & $\lambda \in \R_{>0}$ & $\R_{\geq 0}$ & $\ExpDist$ \\
			Laplace & $\frac{1}{2b}  \exp\Bigl(-\frac{\abs{x}}{b}\Bigr)$ & $b \in \R_{>0}$ & $\R$ & $\LaplaceDist$ \\
			Student-t & $\frac{\Gamma\bigl(\frac{\nu + 1}{2}\bigr)}{\sqrt{\pi \nu}  \Gamma\bigl(\frac{\nu}{2}\bigr)}  \bigl(1 + \frac{x^2}{\nu}\bigr)^{-\frac{\nu + 1}{2}}$ & $\nu \in \R_{>0}$ & $\R$ & $\tDist$ \\
			Gamma & $\frac{\beta^\alpha}{\Gamma(\alpha)}  x^{\alpha - 1}  \exp(-\beta x)$ & $\alpha, \beta \in \R_{>0}$ & $\R_{> 0}$ & $\GammaDist$ \\ 
			Generalized inverse Gaussian & $\frac{(\frac{a}{b})^{\frac{p}{2}}}{2 K_p(\sqrt{ab})}  x^{p - 1}  \exp\bigl(-\frac{ax + b/x}{2}\bigr)$ & $a, b \in \R_{>0}, p \in \R$ & $\R_{>0}$ & $\GIGDist$ \\
			Bernoulli--Laplace & $\lambda \delta(x) + (1 - \lambda)\frac{1}{2b}\exp(-\frac{|x|}{b})$ & $\lambda\in[0,1], b\in\R_{>0}$ & $\R$ & $\BLDist$ \\
			\bottomrule
		\end{tabular}
		\begin{tablenotes}
		\item $\Gamma$ denotes the gamma function defined as $\Gamma(x) = \int^{\infty}_{0} t^{x - 1} \exp(-t) \,\mathrm{d}x$ for any $x \in \R_{>0}$.
		\item $K_{\nu}$ denotes the modified Bessel function of the second kind with parameter $\nu$.
		\end{tablenotes}
	\end{threeparttable}
	\caption{Summary of univariate distributions used throughout this work.}%
	\label{tab:all distributions}
\end{table*}
\Cref{tab:distributions} lists the latent maps and conditional latent distributions that are needed for the \gls{glm} for the distributions in this work.
\begin{table*}
	\centering
	\begin{tabular}{lllll}
		\toprule
		Type & Distribution $\phi_i$ & Latent distribution $f_i$ & Latent maps & Conditional latent distribution $p_{Z_i|X=(\mathbf{K}\mathbf{x})_i}$ \\
		\midrule
		Gauss & \( \GaussianDist(\mu, \sigma^2) \) & \( \delta(0) \) & \( \mu_i(z_i) = \mu,\ \sigma_i^2(z) = \sigma^2 \) & \( \delta(0) \) \\
		Laplace & $\LaplaceDist(b)$ & $\ExpDist\left(\frac{1}{2b^2}\right)$ & $\mu_i(z_i) = 0,\ \sigma_i^2(z_i) = z_i$ & $\GIGDist\left(\frac{1}{b^2}, (\mathbf{Kx})_i^2, \frac{1}{2}\right)$ \\
		Student-t & $\tDist(\nu)$ & $\GammaDist\left(\frac{\nu}{2}, \frac{\nu}{2}\right)$ & $\mu_i(z_i) = 0,\ \sigma_i^2(z_i) = \frac{1}{z_i}$ & $\GammaDist\left(\frac{\nu+1}{2}, \frac{\nu + (\mathbf{Kx})_i^2}{2}\right)$ \\
		\bottomrule
	\end{tabular}
	\caption{%
		Latent variable representations and conditional distributions for common distributions.
	}%
	\label{tab:distributions}
\end{table*}
\end{document}